\documentclass[aip,amsmath,amssymb, twocolumn, 10pt, times]{revtex4-1}
\usepackage{graphicx,graphics}
\usepackage{dcolumn}
\usepackage{bm}
\usepackage{eqnarray,amsmath}
\usepackage{mathptmx}
\usepackage{etoolbox}
\usepackage{verbatim}
\usepackage{color}
\usepackage{ulem}
\usepackage[table]{xcolor}
\usepackage{booktabs}
\usepackage{float}
\usepackage[colorlinks=true]{hyperref}
\makeatletter
\def\@email#1#2{%
\endgroup
\patchcmd{\titleblock@produce} {\frontmatter@RRAPformat}
{\frontmatter@RRAPformat{\produce@RRAP{*#1\href{mailto:#2}{#2}}}\frontmatter@RRAPformat}
  {}{}
}%

\makeatother
\begin{document}
\title[]{A proposal for skyrmion-based diode-like device in antiferromagnetic nanostripe}

\author{R. C. Silva}
\email{rodrigo.c.silva@ufes.br}
\affiliation{Departamento de Ci\^{e}ncias Naturais, Universidade Federal do Esp\'{i}rito Santo, Rodovia Governador M\'{a}rio Covas, Km 60, S\~{a}o Mateus, ES, 29932-540, Brazil.}

\author{R. L. Silva}
\affiliation{Departamento de Ci\^{e}ncias Naturais, Universidade Federal do Esp\'{i}rito Santo, Rodovia Governador M\'{a}rio Covas, Km 60, S\~{a}o Mateus, ES, 29932-540, Brazil.}

\date{\today}

\begin{abstract}
Micromagnetic simulations were employed to investigate the dynamics of a single skyrmion within an antiferromagnetic nanostripe with spatially engineered magnetic properties. This study investigates skyrmion motion within an antiferromagnetic nanostripe engineered with trapezoidal regions of enhanced magnetic anisotropy, enabling diode-like functionality by selectively directing skyrmion movement. Our findings demonstrate that skyrmions can cross these barriers in one direction while being obstructed in the reverse direction, mimicking diode behavior. A detailed analysis is presented on how geometric parameters, such as the inclination angle of the trapezoidal barriers, impact skyrmion motion and device efficacy. Additionally, we reveal that an optimal combination of current density and anisotropy is essential to facilitate efficient skyrmion transport through the nanostripe without reverse movement or annihilation. This work advances the development of skyrmion-based devices for spintronic applications. It provides valuable insights into designing structures that harness controlled topological dynamics.
\end{abstract}

\maketitle

\section{INTRODUCTION AND MOTIVATION}

Topological excitations in magnetic systems have become prominent topics in both fundamental and applied physics, attracting wide interest mainly due to their potential to create new spintronic technologies. Notable examples of these topological spin textures include magnetic vortex~\cite{TShinjo_2000, KYGuslienko_2008, HJung_2012}, skyrmions~\cite{AFert, NNagaosa_2013}, bobbers~\cite{FNRybakov_2015}, hopfions~\cite{PSutcliffe_2018, FNRybakov_2022}, merons~\cite{AABelavin_1975, XZYu_2018, JXia_2022}, and chiral solitons~\cite{JIKishine_2015, YTogawa_2016}. Among these, magnetic skyrmions have recently been the target of intense experimental and theoretical study, revealing an incredible diversity of technological applications.

In a two-dimensional system, a magnetic skyrmion is a quasi-particle whose magnetization in its center points in the opposite direction to its border~\cite{Everschor_2018}, corresponding to mapping the internal spin sphere onto the physical space. The number of times the magnetization wraps the spin sphere defines an integer topological invariant, commonly denominated topological charge. This characteristic provides topological protection~\cite{NNagaosa_2013, AOLeonov_2017} to skyrmions by imposing a high energy barrier that inhibits them from decaying to the ground state and assures their stability against thermal fluctuations. Furthermore, their nucleation arises from the competition among exchange, anisotropy, and Dzyaloshinskii-Moriya (DM) interactions~\cite{ABogdanov1989, Dzyaloshinsky_1958, Moriya_1960, ABogdanov1994, Robler_2006, Bogdanov_2020}. Since their first experimental observation~\cite{Muhlbauer_2009}, skyrmions have been found in a multitude of materials, including ferro (FM)-~\cite{ANeubauer2009, Karube_2016, SMuhlbauer_2016, Yu_2010, Yu_2011, Moreau_Luchaire_2016, RWiesendanger} and ferrimagnetic (FI)~\cite{TOgasawara_2009, MFinazzi_2013, ZLi_2021, RStreubel_2018, TXu_2023, CLuo_2023, KChen_2020, SVelez_2022} media.

The remarkable experimental advances in skyrmion detection across diverse materials, even at room temperature, underscore their immense potential for developing a brand-new spintronic technology. These applications include highly mobile, low-power consumption magnetic data storage devices~\cite{AFert, Sampaio_2013, XZhang_2015, Tomasello_2014}, skyrmion-based computation~\cite{TDohi_2023}, and other spintronic applications~\cite{NSisodia_2022, SLuo_2018, YFeng_2019, GSanchez_2016}. Such innovative advances become possible due to the skyrmion's reduced size~\cite{SHeinze, ASoumyanarayanan} and the ability to print~\cite{Buttner_2017, SFinizio_2019, Zhang_2020, VigoCotrina_2020}, manipulate~\cite{WJiang_2017, KLitzius_2017, Kasai_2019}, and annihilate~\cite{NMathur_2021, RIshikawa_2022} them using electronic methods. However, a troublesome issue in realizing such devices is the motion of skyrmions. In FM (or FI) systems, skyrmions are unable to move in the same direction along the applied spin-polarized current due to an additional transverse motion caused by the Magnus force~\cite{NNagaosa_2013, WJiang_2017, KLitzius_2017, Chen_2017}. Such phenomenon, known as the skyrmion Hall effect (SkHE), imposes a severe obstacle for practical applications necessitating the direct, unimpeded movement of skyrmions along the same direction of the applied current.

\begin{figure}[bt]
     \centering
     \includegraphics[width=10.0cm]{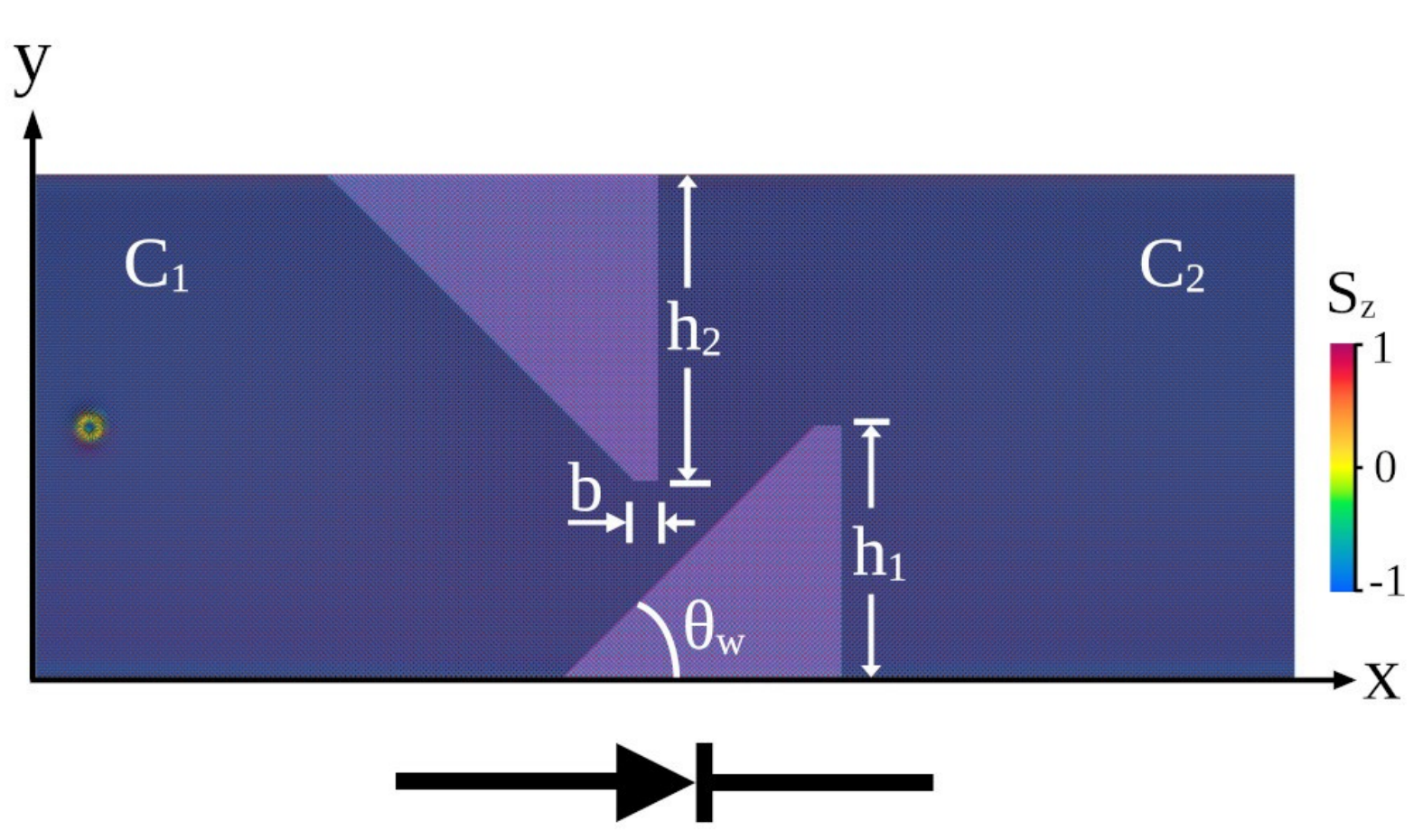}
    \caption{(Color online) Top view of the AFM racetrack with the skyrmion in its initial position $\vec{r}_{sk} = (40,160)$ nm. The racetrack has a length $L_{x}$ of 1000 nm and a width $L_{y}$ of 400 nm. The bright purple trapezoidal-shaped wedges in the central region have an easy-axis anisotropy constant higher than that in the rest of the nanostripe. This configuration directs skyrmion motion from region $C_{1}$ to $C_{2}$ while blocking reverse movement, achieving diode-like behavior.}
    \label{fig:device}
\end{figure}

An appealing alternative is the investigation of skyrmions in antiferromagnetic (AFM) systems. In fact, AFM materials are widely abundant in nature, encompassing metals consisting of Mn-based alloys, as well as various insulators and semiconductors~\cite{Barker_2016}. Hence, AFM skyrmions are able to move in the same direction as the applied electrical current\cite{Barker_2016, RLSilva_2019, RLSilva_2020}, exhibiting different dynamics compared to their FM analogs. This movement, without SkHE, is feasible because the AFM skyrmion texture can be understood as the fusion of two FM skyrmions with opposite spin orientations. Consequently, the topological charges of the sub-skyrmions cancel each other out, resulting in a null Magnus force. There are several other compelling advantages to considering skyrmions in AFM media, such as (i) their velocity is significantly higher, being dozens of times greater than that of FM skyrmions under the same electric current density~\cite{XZhang_SR_2016}; (ii) the minimum current density required to initiate motion in AFM skyrmions is approximately a hundred times lower than that needed for their FM counterparts~\cite{CJin_2016}; (iii) external magnetic fields do not affect the sample, rendering the AFM system more robust against magnetic perturbations~\cite{Barker_2016}. Due to the current limitations of real-space imaging techniques in addressing textures with vanishing magnetization, the direct visualization of AFM skyrmions remains challenging. Despite these experimental tasks, there has been indirect evidence reported for the nucleation of skyrmions in insulating antiferromagnetic La$_{2}$Cu$_{0.97}$Li$_{0.03}$O$_{4}$~\cite{IRaicevic_2011}, fractional AFM skyrmion lattice in MnSc$_{2}$S$_{4}$~\cite{SGao_2020} and synthetic AFM skyrmions and their motion have been experimentally realized~\cite{Dohi_2019, RJuge_2022, WLegrand_2020, RChen_2020}. He \textit{et al.} ~\cite{BHe_2024} reported the room-temperature observation of AFM skyrmions in IrMn, regardless of their experimental validation relied on indirect methods, using the uncompensated moments in IrMn layers and their interaction with adjacent ferromagnetic CoFeB layers. Indeed, detecting AFM skyrmions remains challenging, as they are invisible to real-space techniques such as MFM or LTEM. A promising approach for overcoming this limitation is the topological spin Hall effect (TSHE)~\cite{PMBuhl, BGobel, CAAkosa}, which has been proposed both as a probe to detect AFM skyrmions and as a mechanism to generate a spin current. Consequently, AFM systems show significant potential for advancing spintronics, offering a novel strategy for designing skyrmionic devices without the need for SkHE~\cite{LSmejkal_2018}.

In this work, we propose a new design for a skyrmion-based diode-like device. Although there are numerous theoretical proposals for FM skyrmion-based diodes~\cite{LZhao_2020, JWang_2020, DHJung_2021, LSong_2021, YFeng_2022, MXu_2023, YShu_2023}, the situation is quite different for their AFM counterparts, where the literature is less developed~\cite{MinXu_2023, Bindal_2023, CZhao_2024}. Our device contains two trapezoidal wedges with increased easy-axis anisotropy within the nanostripe. There are several strategies that can be used to create a region with improved magnetic anisotropy. One effective method is focused ion beam (FIB) irradiation~\cite{Chappert_98, Fassbender,Bera_2023}, which can locally modify anisotropy by either damaging the crystal lattice or altering the local composition of the material. Alternatively, using an antiferromagnetic bilayer with an insulating spacer between the layers offers another approach. In this configuration, the magnetic material is deposited solely in the wedge region in the new layer. Considering interlayer ferromagnetic interactions, the wedge region in the original layer will effectively exhibit an enhanced anisotropy. It is worth mentioning that many experimental studies have demonstrated the ability to selectively manipulate the interlayer interaction~\cite{SSPParkin_1990, GChen_2013, SHYang_2015, BDupe_2016, JMatsuno_2016, Zachary_2020}, wherein the coupling, antiferromagnetic or ferromagnetic, can be controlled by inserting a magnetic spacer with a precisely tailored chemical composition. The proposed design enables skyrmions to move in one direction while preventing movement in the opposite direction, regardless of their initial position along the racetrack, effectively mimicking the operation of a diode. Furthermore, the skyrmion consistently maintains the same vertical position within the racetrack when passing through the wedge region. We conducted extensive studies on the geometric design of the wedges, their anisotropy values, and the applied current to optimize the configuration for maximum device performance.

\section{MODEL AND METHODS}

The nanostripe under consideration is described by a Hamiltonian that incorporates isotropic Heisenberg, DM, and easy-axis anisotropy interactions that read:
\begin{equation}\label{eq:Hamiltoniano}
         \mathcal{H} = -J_{ex} \displaystyle{\sum_{\langle ij \rangle}} \vec{S}_{i} \cdot \vec{S}_{j} -\displaystyle{\sum_{\langle ij \rangle}} \vec{D}_{ij} \cdot \left(\vec{S}_{i} \times \vec{S}_{j} \right) - \displaystyle{\sum_{i}} K_{i}\left(\hat{z}\cdot \vec{S}_{i}\right)^{2},
\end{equation}
\noindent
where $\vec{S}_{k} = \left(S_{k}^{x},\, S_{k}^{y},\, S_{k}^{z}\right)$ represents the dimensionless unit magnetic moment situated at a lattice site \textit{k}. The symbol $\langle ij \rangle$ indicates the summation is taken over the nearest magnetic moment pairs, due to the short range of the exchange and DM interactions.

In our micromagnetic approach, we employed typical parameters for the KMnF$_{3}$~\cite{Barker_2016, XZhang_SR_2016, KSainki_1972}: exchange stiffness constant $A_{ex} = -6.59 \times 10^{-12}$ J/m,  the DM constant $D = 8.0 \times 10^{-4}$ J/m$^{2}$, the magnetocrystalline anisotropy constant $K = 1.16 \times 10^{5}$ J/m$^{3}$ and the saturation magnetization $M_{s} = 3.76 \times 10^{5}$ A/m. An accurate choice of the lattice spacing $a_{0}$ must consider the characteristic lengths of the system's interactions. These characteristic lengths are expressed as functions of the parameters of the magnetic material. Hence, the exchange length $\lambda_{ex} = \displaystyle{\sqrt{\frac{2| A_{ex}|}{\mu_{0} M_{s}^{2}}}} \approx 8.61$ nm, the wall width parameter $\lambda_{ww} = \displaystyle{\sqrt{\frac{| A_{ex}|}{K}}} \approx 7.54$ nm, and the characteristic length associated with DM interaction is $\lambda_{DM} = \displaystyle{\frac{2 |A_{ex}|}{D}} \approx 16.48$ nm. Therefore, a suitable choice for the lattice spacing is $a_{0} = 2$ nm which is less than $\lambda_{ww}$, the smallest characteristic length. Consequently, we partitioned the nanostripe into a square array of cubic cells, each one of a volume of $a_{0}^{3}$. Moreover, $J_{ex} = 2 a_{0} A_{ex} = -2.636 \times 10^{-20}$ J characterizes the strength of the exchange interaction, $\vec{D}_{ij} = D' \left( \hat{z} \times \hat{r}_{ij} \right)$ is the DM vector, where $D' = D a_{0}^{2} = 3.2 \times 10^{-21}$ J represents its strength, $\hat{z}$ is the versor perpendicular to the sample's plane, and $\hat{r}_{ij}$ is the normalized vector joining the lattice sites \textit{i} and \textit{j}~\cite{SRohart_2013, HYang_2015}. Such a DM vector favors the nucleation of interfacial N\'{e}el skyrmions (hedgehog-type skyrmion profile). In the last summation in Eq.\ref{eq:Hamiltoniano}, $K_{i} = K_{0} = K a_{0}^{3} = 9.28 \times 10^{-22}$ J depicts the easy-axis anisotropy constant. Note that the magnitude of all magnetic couplings is uniformly expressed in energy units, ensuring consistent dimensions.

We examined nanostripes with the following parameters: a horizontal length ($L_{x}$) of 1000 nm, a vertical width ($L_{y}$) of 400 nm, and a thickness ($L_{z}$) of 2 nm. In Figure~\ref{fig:device}, one can see the proposed racetrack, where the bright purple trapezoidal-shaped wedges represent the regions with an increased easy-axis anisotropy constant, denoted as $K_{w}$. Thus, for every lattice site \textit{i} within these trapezoidal regions, $K_{i} = K_{w}$. All other magnetic couplings remain unchanged. The key geometric attributes of the wedges include their heights, the inclination denoted by the angle $\theta_{w}$, the size $b$ of the smallest base of the trapezoid, and the distance d between them.

The magnetization dynamics are determined by solving the dimensionless Landau-Lifshitz-Gilbert (LLG) equation~\cite{Landau_1935, TGilbert}, incorporating the Zhang-Li spin transfer torque terms~\cite{Zhang_Li}:

\begin{equation}\label{eq:LLGZL}
    \begin{array}{ccc}
        \left(1+\alpha^{2}\right) \frac{\partial \vec{S}_{i}}{\partial \tau} = -\vec{S}_{i} \times \vec{B}_{i}^{ef}-\alpha \vec{S}_{i} \times \left(\vec{S}_{i} \times \vec{B}_{i}^{ef} \right) \\
        \\
        -\nu \left[\left(\beta - \alpha \right) \vec{S}_{i} \times \frac{\partial \vec{S}_{i}}{\partial x'}+\left(1+\alpha \beta \right)\vec{S}_{i} \times \left(\vec{S}_{i} \times \frac{\partial \vec{S}_{i}}{\partial x'} \right)\right],
    \end{array}
\end{equation}
\noindent
where $\alpha = 0.15$ is the Gilbert damping constant, $\vec{B}_{i}^{ef} = -\frac{1}{J_{ex}}\frac{\partial \mathcal{H}}{\partial \vec{S}_{i}}$ represents the dimensionless local effective magnetic field acting at lattice site \textit{i}. Moreover, the dimensionless time, denoted by $\tau$, is linked to real time  $t$ by the relationship $\Delta t = \eta \Delta \tau$, where $\eta = \frac{M_{s} a_{0}^{3}}{\gamma |J_{ex}|} \approx 0.65 \times 10^{-12}$ s. Additionally, the transformation between real spatial coordinates ($\Delta x$) and their dimensionless counterparts ($\Delta x'$) is given by $\Delta x = a_{0} \Delta x'$. The final two terms in Eq.~\ref{eq:LLGZL} represent the Zhang-Li spin-transfer torque, with $\nu = \frac{P j_{e} \mu_{B} a_{0}^{2}}{|e J_{ex}|\gamma \left(1+\beta^{2}\right)}$. Here, $P$ denotes the degree of polarization of the spin current density, $\mu_{B}$ stands for the Bohr magneton, $e$ represents the electronic charge, $\beta = 0.10 \alpha$ characterizes the non-adiabaticity coefficient, $j_{e}$ represents the magnitude of the electric current density, and $\gamma$ denotes the gyromagnetic ratio. The LLG equation has been integrated using a fourth-order Runge-Kutta method with a dimensionless time step of $\Delta \tau = 10^{-4}$.  Here, we set $\vec{j}_{e} =-j_{e} \,\, \hat{x}$ in order to drive the skyrmion from the left to the right and $\partial_{x'} \vec{S}$ is calculated in each sublattice~\cite{Barker_2016}. In addition, $j_{e}$ is given in terms of $j_{0} =  \frac{\gamma |e J_{ex}|}{\mu_{B} a_{0}^{2}} \approx 2.0 \times 10^{13}$ A/m$^{2}$.

The initial profile of the AFM skyrmion is derived from an analytical solution representing a single N\'{e}el skyrmion with a radius $R_{sk}$, positioned at coordinates $\vec{r}_{sk}=\left(x_{sk},y_{sk}\right)$. This solution is characterized by two scalar fields: the polar angle, $\theta$, and the azimuth angle, $\phi$, which define the orientation of the internal spin sphere at each lattice site \textit{i} with coordinates (x$_{i}$, y$_{i}$). Specifically, the spin vector $\vec{S}_{i}$ is expressed as $\vec{S}_{i} = (-1)^{x_{i}+y_{i}}\left(\sin \theta_{i} \cos \phi_{i}, \sin \theta_{i} \sin \phi_{i}, \cos \theta_{i} \right)$. Additionally,

\begin{equation} \label{eq:Nsky}
    \left\{
	\begin{array}{l}
	    \theta_{i} = \arccos{\left( \frac{R^{2}_{sk}-\rho^{2}_{i}}{R^{2}_{sk}+\rho^{2}_{i}}\right)} \\
	    \\
	    \phi_{i} = \arctan{\left(\frac{y_{i}-y_{sk}}{x_{i}-x_{sk}}\right)},
	\end{array}
	\right.
\end{equation}
\noindent
where $\rho_{i} = \sqrt{\left(x_{i}-x_{sk}\right)^{2}+\left(y_{i}-y_{sk}\right)^{2}}$. In order to enable the adjustment of the skyrmion's radius and its in-plane magnetization in the lattice, we integrate the LLG equation without the Zhang-Li term, so that the system achieves ground state. At the end of the relaxation process, the skyrmion radius was estimated to be approximately $R_{sk} \approx 8.7$ nm.

The position of the skyrmion can be determined by tracing the topological charge density, as outlined in ref.~\cite{CMoutafis_2009}, expressed as $\displaystyle{\sigma_{k} = \frac{1}{2} \varepsilon_{ij} \vec{\zeta}_{k} \cdot \left(\partial_{j} \vec{\zeta}_{k} \times \partial_{i} \vec{\zeta}_{k}\right)}$, where \textit{i} and \textit{j} represent summation over the horizontal and vertical directions of the racetrack. Here, $\vec{\zeta}_{k} = \frac{\vec{S}_{2k}-\vec{S}_{2k+1}}{2}$ denotes the N\'{e}el vector. Consequently, the skyrmion's mass center position is determined by~\cite{LShen_2020, MStier_2021}:
\begin{equation}
    \label{eq:sky_track}
    \displaystyle{\vec{r}_{sk} = \frac{\int \vec{r} \sigma d^{2}r}{Q}},
\end{equation}
\noindent
where $\vec{r} = \left(r_{x}, r_{y}\right)$ is the lattice site coordinates and $Q = \int \sigma d^{2}r$ is the N{\'{e}}el topological number. Skyrmion velocity is readily obtained by $\vec{V}_{sk} =d\vec{r}_{sk}/dt = \left(V_{x}, V_{y}\right)$, and its radius is evaluated like below:
\begin{equation}
	\displaystyle{R_{sk} = \sqrt{\int \left|\vec{r}-\vec{r}_{sk}\right|^{2}\frac{\sigma}{Q} d^{2}r}}.
 \end{equation}

\section{NUMERICAL RESULTS}

\subsection{SKYRMION-WEDGES INTERACTION POTENTIAL}

\begin{figure}[bt]
    \centering
    \includegraphics[width=\hsize]{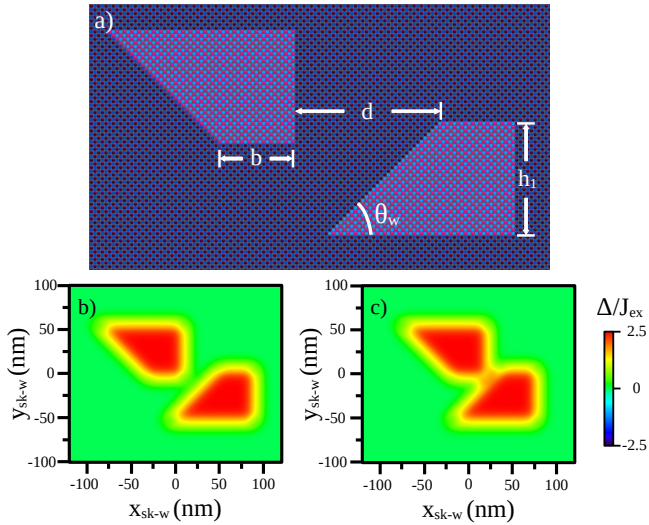}
     \caption{(Color online) a) Schematic top view of the identical reduced trapezoidal wedges used to analyze the skyrmion-wedges interaction potential. The geometric parameters for the wedges were fixed as follows: $h_{1} = 60$ nm, $b = 40$ nm, $\theta_{w} = 45^{\circ}$ and $d = 80$ nm. b) Skyrmion-wedge interaction potential $\Delta$ as a function of their center of mass distance $\vec{r}_{sk-w}$ for $d = 80$ nm and in (c) for $d = 16$ nm. In (b), there is a pathway between the trapezoids where $\Delta$ is null, facilitating the skyrmion's passage to reach the $C_{2}$ region. However, this pathway disappears when the trapezoids are close enough, making it more difficult for the skyrmion to cross the barriers.}
     \label{fig:potentials}
\end{figure}

In the following, we provide the numerical results of solving the LLG equation, incorporating current-induced torques within an engineered racetrack. First, it is crucial to understand how the trapezoidal wedges affect the skyrmion dynamic. The skyrmion-wedge interaction potential can supply valuable insights. However, the trapezoids shown in Fig.~\ref{fig:device} are pretty large, nearly matching the nanotrack width. Then, it is sufficient to consider smaller trapezoidal wedges for a more straightforward understanding of the skyrmion-affected dynamics. In fact, Fig.~\ref{fig:potentials}(a) shows two identical trapezoidal barriers with the following geometric parameters: $h_{1} = 60$ nm, $b = 40$ nm, $\theta_{w} = 45^{\circ}$ and $d = 80$ nm. Under these assumptions, to evaluate such an interaction, we consider a single skyrmion whose center is located at $\vec{r}_{sk}=\left(x_{sk},y_{sk} \right)$ in the middle of the racetrack. Then, by positioning the center of the wedges at $\vec{r}_{w} = \left(x_{w},y_{w}  \right)$, the total system’s energy is calculated for this spin configuration. By varying the position of the increased anisotropy obstacles, $\vec{r}_{w}$, the energy can be determined as a function of the center-to-center distance between the skyrmion and the wedges, $\vec{r}_{sk-w}=\left(x_{sk-w}, y_{sk-w}\right)=\vec{r}_{sk}-\vec{r}_{w}$. This energy, denoted as $E \left(\vec{r}_{sk-w}\right)$, depends solely on the relative coordinate $\vec{r}_{sk-w}$, as our simulation box is sufficiently large to minimize boundary effects. Additionally, note that when the skyrmion-wedges distance becomes sufficiently large, $|\vec{r}_{sk-w}| \rightarrow \infty$, the interaction between them vanishes. Thus, we define the skyrmion-wedges interaction potential as $\Delta = E \left(\vec{r}_{sk-w}\right) - E\left(\infty\right)$.

Figures~\ref{fig:potentials}(b) and~\ref{fig:potentials}(c) shows the potential landscape profile for $K_{w} = 1.50 K_{0}$. In Fig.~\ref{fig:potentials}(b), the separation between the two trapezoids is $d = 40$ nm, while in Fig.~\ref{fig:potentials}(c) this distance is reduced to $d = 16$ nm. The skyrmion-wedge interaction is repulsive in both cases since $\Delta > 0$ and its profile is closely related to the trapezoidal geometry of the wedges. This repulsion is significant only when the skyrmion is in close proximity to the wedges, with $\Delta$ becoming non-negligible only when the distance between the skyrmion and any wedge border is less than 30 nm. The findings are consistent with those presented in Refs.~\cite{DToscano_2020, RCSilva_2022, RCSilva_2024}, where the authors investigated the local variations in material properties of single or bilayer AFM/ FI nanostripes. They observed attractive and repulsive potentials resulting from changes in local magnetic coupling interactions.

The transition of the skyrmion from $C_{1}$ to $C_{2}$ regions should ideally occur with minimal interference from the wedges. This is possible when the trapezoids are sufficiently far apart, creating a pathway between them where the potential is null, as demonstrated in Fig.~\ref{fig:potentials}(b). In this case, the skyrmion can pass through the trapezoidal obstacles, interacting with only one wedge at a time. However, this free-interaction pathway disappears when the trapezoids are positioned closer together (Fig.~\ref{fig:potentials}(c)): the skyrmion must overcome an energy barrier from simultaneous interaction with both trapezoids to reach $C_{2}$ sector. This energy barrier increases as the separation between the wedges decreases. Consequently, a higher spin-polarized current density will be required for the skyrmion to bypass the wedges region. Depending on the separation between the trapezoids as well as the value of $K_{w}$, the skyrmion may be unable to overcome the barrier, staying trapped, or even be annihilated. On the other hand, when considering the movement of the skyrmion from $C_{2}$ to $C_{1}$ regions, it is noteworthy that within the area defined by the trapezoids, the potential is repulsive along the entire vertical extent of the barrier. Since the wedges span the full vertical extension of the nanotrack, there is no non-interacting path for the skyrmion to reverse its trajectory from $C_{2}$ to $C_{1}$. Our calculations have shown that $d = \frac{8R_{sk}}{\sin\left(\theta_{w}\right)}+\frac{\Delta h}{\tan\left(\theta_{w}\right)}$ is sufficiently large for smooth skyrmion transition. Here, $R_{sk}$ denotes the skyrmion radius and $\Delta h = h_{2}+h_{1}-L_{y}$.

\begin{figure}[bt]
    \centering
    \includegraphics[width=8.0cm]{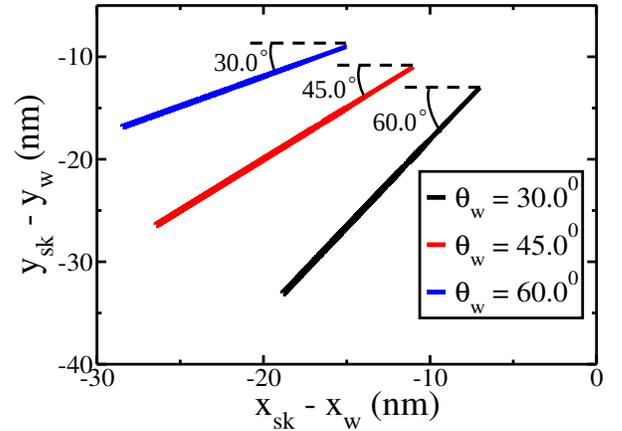}
     \caption{(Color online) Skyrmion motion is induced by its interaction with the trapezoid barrier. We considered wedges with three distinct angles ($\theta_{w} = 30.0^{\circ}$, $45.0^{\circ}$, and $60.0^{\circ}$). The trapezoid repels the skyrmion, which moves along a rectilinear path that forms a right angle with the inclined section of the trapezoid. Distances are measured relative to the inclined section, and the slope of the curves corresponds to the complementary angle of the trapezoid's inclination, confirming that the repulsive force is perpendicular to the inclined section.}
     \label{fig:push}
\end{figure}

Another helpful case for examining the interaction between a skyrmion and a trapezoidal wedge occurs when the skyrmion is initially positioned near the inclined section of the upper trapezoid. Similar results are observed when the skyrmion is near the lower one. In this scenario, we exclude the effects of spin-polarized electric currents, focusing solely on the skyrmion's movement as influenced by its interaction with the trapezoidal barrier. This analysis integrates the LLG equation (Eq.~\ref{eq:LLGZL}), excluding the spin-transfer torque terms. Our findings indicate that the trapezoidal barrier propels the skyrmion away from the it, exerting an outward force perpendicular to its inclined section, causing the skyrmion to move leftward and downward. This behavior aligns with the potential analysis. Insofar as the skyrmion departs sufficiently far from the upper trapezoid, its interaction with the wedge weakens, bringing it to rest. Figure~\ref{fig:push} illustrates this outcome for wedges with three distinct angles: $\theta_{w} = 30.0^{\circ}$, $45.0^{\circ}$, and $60.0^{\circ}$, with distances measured relative to the inclined section of the trapezoidal obstacle. The curve slopes, relative to the horizontal axis, match the complementary angle of the trapezoid's inclination. Consequently, the skyrmion, repelled by the trapezoid, follows a rectilinear path at a right angle to the inclined section of the wedge. These results confirm that the repulsive skyrmion-wedge force is perpendicular to the trapezoid's inclined section.

\subsection{SKYRMION DYNAMICS FROM $C_{1}$ TO $C_{2}$ REGIONS}

To investigate our prospect for a skyrmion-based diode device, we initially consider a single skyrmion located at the initial position ($x_{sk}$, $y_{sk}$) in the $C_{1}$ region, as depicted in Fig.~\ref{fig:device}. At this stage, we will examine the skyrmion’s passage from the $C_{1}$ to the $C_{2}$ region through the wedges. In these simulations, the skyrmion is created in the $C_{1}$ region at a fixed horizontal position of $x_{sk} = 40$ nm, with its vertical position in the range of 80 nm $\leq y_{sk} \leq 320$ nm to avoid interactions between the skyrmion and the borders of the nanostripe. In fact, for $y_{sk} < 80$ nm or $y_{sk} > 320$ nm, the skyrmion-edge repulsion is too strong, causing the skyrmion to acquire additional movement in the vertical direction (perpendicular to the applied current density), pushing it toward the center of the racetrack. This behavior is consistent with findings from ferromagnetic analogs, where similar skyrmion-edge repulsion has been observed~\cite{XZhang2_2015}.

\begin{figure}[bt]
    \centering
    \includegraphics[width=\hsize]{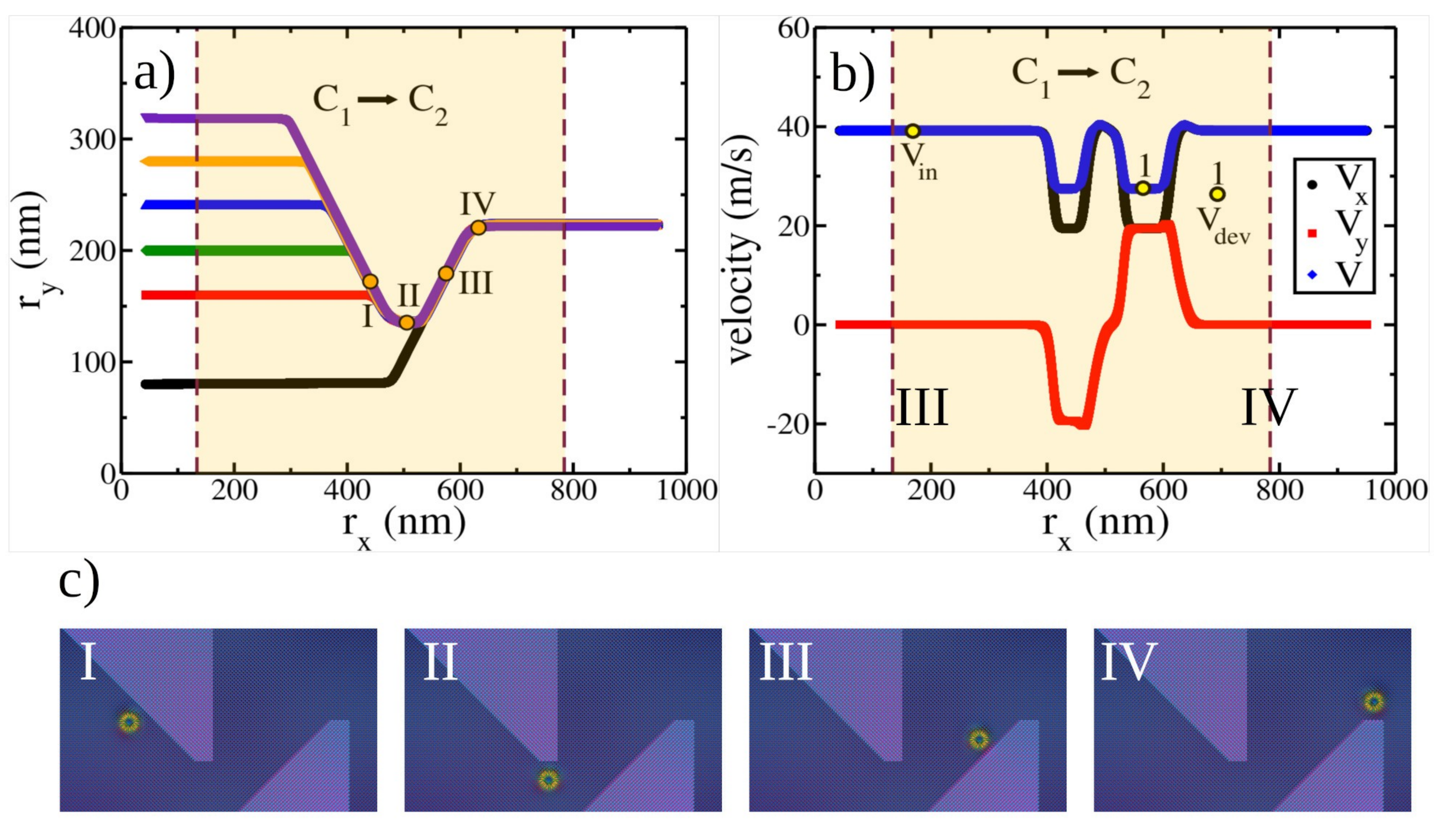}
     \caption{(Color online) a) The trajectory of the skyrmion from the $C_{1}$ to $C_{2}$ regions in the nanostripe. All curves start at the same initial horizontal coordinate but at different vertical positions. The graph's light-yellow region denotes the wedges' horizontal extent at the central position of the nanotrack. A spin-polarized current density of $j_{e} = 0.10 \,\, j_{0} \approx 2.0 \times 10^{12}$ A/m$^{2}$ is applied throughout the system to induce skyrmion motion. Notably, in all cases, the skyrmion passes through the device and arrives at region $C_{2}$ at the same vertical height. b) Skyrmion velocity along the racetrack. Black, red, and blue curves represent the horizontal ($V_{x}$), vertical ($V_{y}$) components, and the magnitude of the velocity $V=\sqrt{V_{x}^{2}+V_{y}^{2}}$, respectively. While moving in the $C_{1}$ or $C_{2}$ regions, the skyrmion travels in a straight line. The two yellow points indicate the magnitude of the skyrmion velocity in the $C_{1}$ region ($V_{in}$) and in the wedges sector ($V_{dev}$). Additionally, $v_{y}$ assumes non-zero values as the skyrmion bypasses the wedges. c) Panels I, II, III, and IV show snapshots, highlighted in the trajectory curve, of the skyrmion dynamics at different stages of its motion in the barriers region.}
    \label{fig:Sktraj}
\end{figure}

Fig.~\ref{fig:Sktraj}(a) shows the six skyrmion trajectories along the tailor-designed nanostripe. In these cases, the skyrmion dynamics are induced by applying a spin-polarized current density of $j_{e} = 0.10 \,\, j_{0}\approx 2.0 \times 10^{12}$ A/m$^{2}$ throughout the system. The results were obtained by considering the following geometric parameters of the wedges: $h_{1} = 200$ nm, $h_{2} = 240$ nm, $\theta_{w} = 45^{\circ}$, and $b = 20$ nm. Additionally, the easy-axis anisotropy constant of the wedges was fixed at $K_{w} = 1.50 K_{0}$.

In all cases, the skyrmion's initial motion aligns with the applied current, without exhibiting SkHE. For example, consider the black curve, where the skyrmion starts its dynamics at the coordinate (40, 80) nm. Indeed, the spin-polarized electric current generates a drag force that propels the skyrmion to the right. As the skyrmion approaches the trapezoid, it experiences a repulsive force perpendicular to the inclined border of the wedge. This force grows in magnitude as the skyrmion moves closer to the wedge. When the skyrmion is sufficiently near the trapezoid, the horizontal component of the repulsive force counterbalances the drag force. Simultaneously, its vertical component generates an upward thrust, causing the skyrmion to deviate from its initial straight horizontal path, moving parallel to the inclined edge of the barrier. For other initial conditions with higher vertical coordinates, the skyrmion's trajectory first encounters the upper trapezoid obstacle (as seen in the green curve of Fig.~\ref{fig:Sktraj}(a)). The interaction with the upper wedge causes the skyrmion to move downward, following the wedge's inclined path. After leaving the upper wedge, the skyrmion briefly travels straight before approaching and bypassing the lower obstacle, eventually arriving in the $C_{2}$ region. Moreover, the vertical position at which the skyrmion moves within region $C_{2}$ is controlled, on demand, by the height $h_{1}$ of the lower trapezoid. Notably, regardless of its initial vertical position in the $C_{1}$ region, the skyrmion maintains a consistent vertical height while traversing the wedges. This consistent behavior allows the device to function effectively as a skyrmion collimator, adding features to its capabilities.

\begin{figure}[bt]
    \centering
    \includegraphics[width=\hsize]{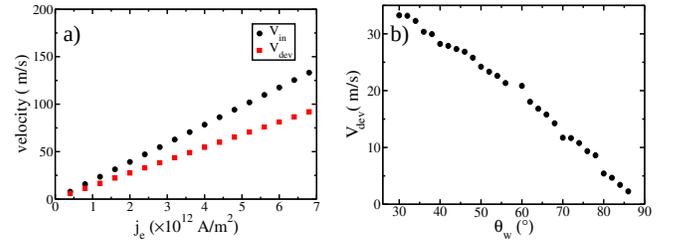}
     \caption{(Color online) a) Skyrmion velocity as a function of the spin-polarized current density magnitude $j_{e}$. The black dots represent the skyrmion velocity $V_{in}$ while moving through the $C_{1}$ and $C_{2}$ regions. Conversely, the red squares illustrate the skyrmion velocity $V_{dev}$ as it traverses the wedges region. It is important to note that in the wedge region, the skyrmion moves more slowly than in regions $C_{1}$ and $C_{2}$. b) $V_{dev}$ as a function of the angle $\theta_{w}$ of the inclination of the trapezoid. $V_{dev}$ decreases as $\theta_{w}$ increases, eventually vanishing at $\theta_{w} = 87^{\circ}$. At this angle, the trapezoids behave almost like a vertical barriers. }
    \label{fig:velocities}
\end{figure}

The skyrmion velocity as a function of its horizontal position in the racetrack is depicted in Fig.~\ref{fig:Sktraj}(b), obtained by considering the skyrmion's initial position at $(40,200)$ nm (the green curve in Fig.~\ref{fig:Sktraj}(a)). Black, red, and blue dots respectively represent the horizontal ($V_{x}$), vertical ($V_{y}$) components, and the magnitude $V = \sqrt{V_{x}^{2}+V_{y}^{2}}$ of the skyrmion's velocity. In the $C_{1}$ and $C_{2}$ regions, the skyrmion moves in a rectilinear trajectory. As the skyrmion approaches the upper trapezoid, their interaction becomes more prominent, resulting in additional downward vertical motion. Consequently, $V_{y}$ assumes negative values. Similarly, when the skyrmion approaches the lower obstacle, their interaction causes it to move upward along the ramp, resulting in $V_{y} > 0$. Although the skyrmion's velocity remains constant while traversing the wedges, its magnitude $V$ decreases, as shown by the blue curve. Let us define $V_{in}$ and $V_{dev}$ as the magnitudes of the skyrmion velocities in region $C_{1}$ (which is identical to the velocity in region $C_{2}$) and within the trapezoidal wedges, respectively. In this way, initially, the skyrmion moves with $V_{in} = 39.20$ m/s. Nonetheless, while moving through the wedges, the velocity drops to $V_{dev} \approx 27.30$ m/s (note that $V_{in}$ and $V_{dev}$ are detached by the yellow points in Fig.~\ref{fig:Sktraj}(b)). Additionally, Fig.~\ref{fig:Sktraj}(c) displays snapshots of the skyrmion's dynamics along the trapezoid barriers region at various stages. Panels I-IV correspond to the moments highlighted in trajectory analysis (Fig.~\ref{fig:Sktraj}(a)). Please refer to the movie mv1.mp4 in the supplementary material for additional details.

Another key feature to consider is the behavior of the skyrmion's velocity as a function of the magnitude of the applied current density, as shown in Fig.~\ref{fig:velocities}(a). The black dots represent the linear relationship between $V_{in}$ and $j_{e}$. Note that although the result depicts the skyrmion moving in the region $C_{1}$, such velocity is identical when it moves in $C_{2}$. The red squares, on the other hand, illustrate the linear dependence of $V_{dev}$ on the applied current density. When the skyrmion transits through the wedge region, its velocity is significantly lower than its movement outside. This difference highlights the influence of the wedge structure on skyrmion dynamics. Moreover, the difference between $V_{in}$ and $V_{dev}$ increases as $j_{e}$ rises, indicating a stronger impact of the applied current density in the wedge region. In both cases, the linear dependence of the velocity with the applied current is consistent with Thiele analysis~\cite{Barker_2016}. In addition, the relationship between $V_{dev}$ and the trapezoid tilt angle $\theta_{w}$ is shown in Fig.~\ref{fig:velocities}(b). The dependence of $V_{dev}$ on $\theta_{w}$ is nearly linear, with the velocity decreasing as $\theta_{w}$ increases, eventually reaching zero at $\theta_{w} = 87^{\circ}$. The skyrmion cannot cross the wedge region at this critical angle, causing the device to cease functioning as a diode. This angle is notably high, close to $90^{\circ}$, which makes the wedge resemble a vertical rectangular barrier. Thus, to optimize the device performance, it is desirable to build wedges where $30^{\circ} \leq \theta_{w} \leq 70^{\circ}$. For $\theta_{w} < 30^{\circ}$, the size of the wedge region increases considerably, becoming almost the size of the nanostripe, which is, therefore, an unfavorable factor. On the other hand, despite being able to move through the wedges at high angles, the skyrmion's movement is quite slow, with typical velocities around $V_{dev} \approx 1.5$ m/s. Therefore, the extended time required for the skyrmion to transfer from $C_{1}$ to $C_{2}$ regions significantly reduces the device's efficiency when $\theta_{w} > 70.0^{\circ}$.

\begin{figure}[H]
    \centering
    \includegraphics[width=\hsize]{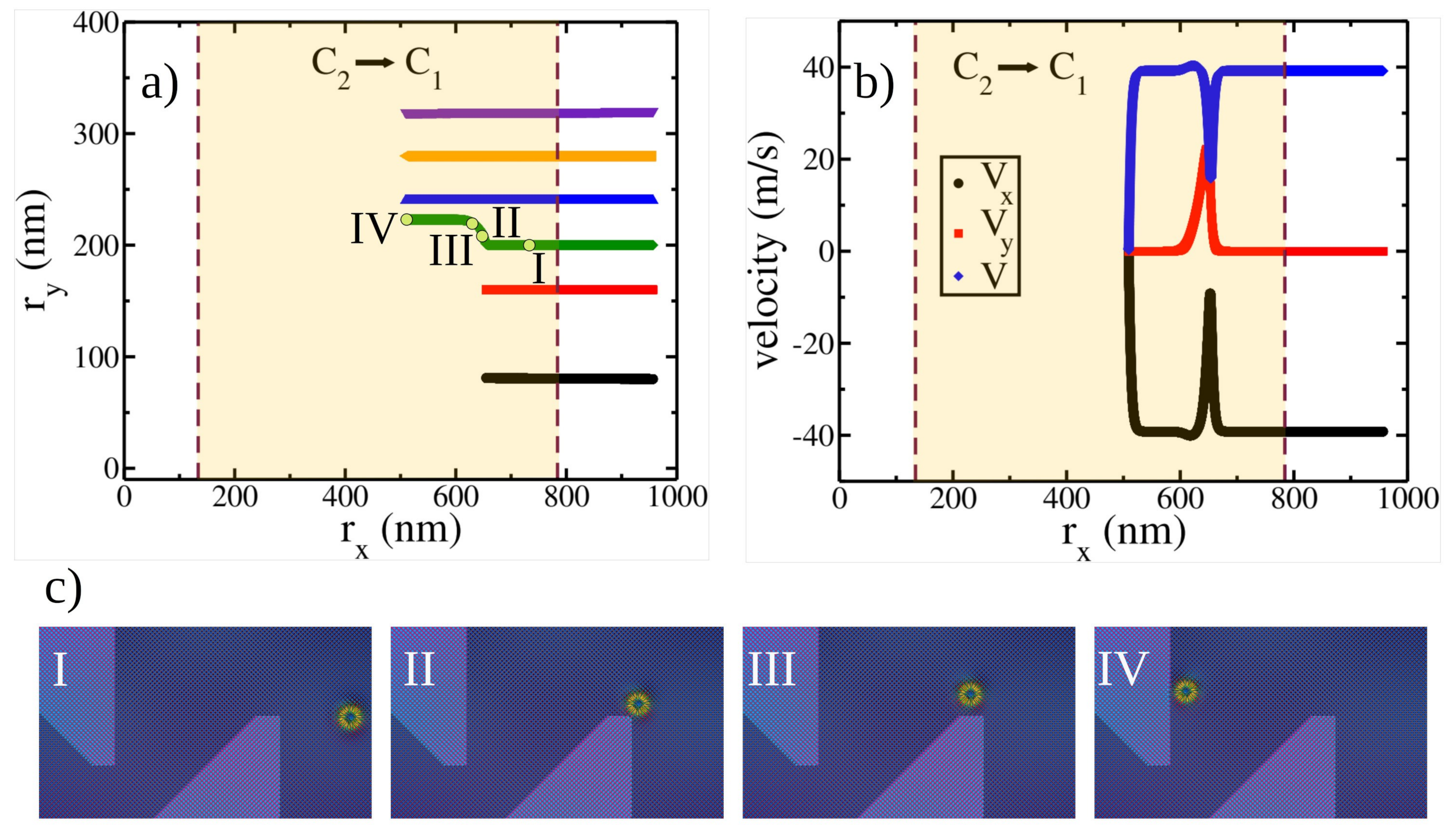}
     \caption{(Color online) a) The trajectories of skyrmions from different starting points show straight-line motion until they interact with the left side of the trapezoidal barrier. The green curve highlights the skyrmion trajectory with a vertical height close to $h_{1}$. In this case, the skyrmion overcomes the lower trapezoid. b) The velocity components of the skyrmion, $V_{x}$ (black) and $V_{y}$ (red), as well as the velocity modulus $V$ (blue) are plotted as a function of its horizontal position. These results were obtained by analyzing the trajectory represented by the green curve. c) Sequential snapshots capture the skyrmion's traversal of the lower trapezoid and its eventual halt at the upper barrier. Panels I-IV are marked in trajectory analysis.}
    \label{fig:reverse}
\end{figure}

\subsection{CONSTRAINTS ON SKYRMION REVERSE MOTION BETWEEN $C_{2}$ AND $C_{1}$ REGIONS}

We explore the reverse motion of the skyrmion, transitioning from the $C_{2}$ to $C_{1}$ sectors. In this analysis, the skyrmion is initially positioned at $x_{sk} =  960$ nm at different vertical heights $y_{sk}$, in which $80$ nm $\leq y_{sk} \leq 320$ nm. Herein, we have used $\vec{j}_{e} = 0.10\, j_{0}\,\, \hat{x}$ to drive the skyrmion from right to left across the system. Figure~\ref{fig:reverse}(a) displays its trajectories for different vertical starting points. Regardless of its initial position, the skyrmion exhibits a straight-line pathway until it draws near to the left side of the wedges. In most cases, as the skyrmion approaches the trapezoidal barrier, its movement is constrained, causing it to stay trapped near the trapezoid while the current is kept on. If the current is turned off, the skyrmion's interaction with the barrier pushes it to the right until $\Delta$ becomes negligible. Nonetheless, if the skyrmion's vertical coordinate $y_{sk}$ is close to $h_{1}$ (the height of the lower trapezoid), it overcomes the obstacle, as illustrated by the green curve in Fig.~\ref{fig:reverse}(a). Additionally, as the skyrmion moves horizontally, only its longitudinal velocity component ($V_{x}$) remains non-zero until its motion is interrupted by the interaction with the trapezoid. An exception occurs when the skyrmion hits near the upper extremity of the lower wedge. While circumventing the obstacle, the skyrmion briefly acquires a transverse velocity component ($V_{y}$). This behavior is shown in Fig.~\ref{fig:reverse}(b), where the black, red, and blue curves represent $V_{x}$, $V_{y}$, and $V$ (velocity modulus), respectively. Notably, the interaction with the barrier allows the skyrmion to bypass it, albeit at a reduced velocity. Fig.~\ref{fig:reverse}(c) presents a temporal sequence of snapshots of the skyrmion overcoming the lower trapezoid and halting its motion upon reaching the upper barrier. Panels I-IV correspond to specific points marked in trajectory analysis (please see the movie mv2.mp4 for more details).

\begin{figure}[bt]
    \centering
    \includegraphics[width=\hsize]{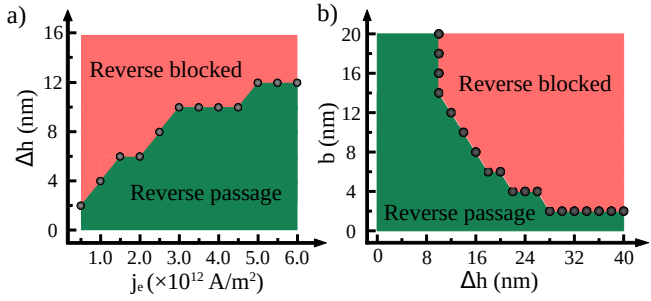}
     \caption{(Color online) a) Optimal values of $\Delta h$ and $j_{e}$ for effective skyrmion control in the diode-like device, with fixed parameters $b=20$ nm and $h_{1} = 200$ nm. The green region indicates $\left(\Delta h,j_{e}\right)$ values where the skyrmion moves unimpeded, while the red-shaded area denotes the range preventing its return to the $C_{1}$ region. b) Device response to varying $b$ of the trapezoidal wedge, highlighting the required $\Delta h$ values to block skyrmion return at specific $b$ values.}
    \label{fig:C2C1}
\end{figure}

The possibility of bypassing the lower wedge when approaching heights near the ends of the trapezoid supports the choice of $h_{2} > L_{y}-h_{1}$ for the upper trapezoid. To assess the impact of $h_{2}$ on the device's operation, we define $h_{2} = L_{y}-h_{1}+\Delta h$, where $\Delta h$ represents the required length of $h_{2}$ below the top vertex of the lower trapezoid. Fig.~\ref{fig:C2C1}(a)  exhibits the optimal values of $\Delta h$ and the applied current density $j_{e}$ to achieve proper operation of the diode-like device. These findings were obtained by fixing $b = 20$ nm and $h_{1} = 200$ nm (where $h_{1}$ denotes the height of the right wedge; refer to Fig.~\ref{fig:device} for details). The green region delineates the $(j_{e},\Delta h)$ values for which the skyrmion is unimpeded during its $C_{2} \rightarrow C_{1}$ motion within the nanostripe. This area should be avoided. Notably, if $\Delta h = 0$, so $h_{2} = L_{y}-h_{1} = 200$ nm, the skyrmion will cross the wedge region regardless of the $j_{e}$ value. As $j_{e}$ increases, $\Delta h$ also rises, ultimately saturating at $j_{e} = 5.0 \times 10^{12}$ A/m$^{2}$, with $\Delta h = 12$ nm. The red-shaded area designates the operational range where the device reliably prevents the skyrmion from returning to the $C_{1}$ region, which is the desired outcome for suitable functioning of the skyrmion-based diode. On the other hand, Fig.~\ref{fig:C2C1}(b) shows the device’s behavior with variations in the parameter $b$ of the trapezoidal wedge, with results obtained by setting $j_{e}=6.0 \times 10^{12}$ A/m$^{2}$. If $b= 0$, the wedges assume a triangular shape, with a smaller area near its ends, which is unable to prevent the skyrmion’s return to the $C_{1}$ region, even at higher $\Delta h$ values. In this case, the skyrmion experiences an asymmetric interaction potential during its reverse movement, making it easier to circumvent or traverse the wedge region, depending on $K_{w}$. Indeed, the entire green area represents $\left(\Delta h,b\right)$ values that fail to prevent the skyrmion’s return. Thus, to mitigate the effect of potential asymmetry from  $C_{2}$ to $C_{1}$, the parameter $b$ is relevant. Our findings indicate that the device blocks the skyrmion when $\Delta h > 12$ nm with $b \geq 14$ nm. For smaller values of $b$, a further increase in $\Delta h$ is required, with $\Delta h \geq 28 $ nm as the minimum height necessary for $b= 2$ nm, as observed in the simulations.

\begin{figure*}[bt]
    \centering
    \includegraphics[width=12.0cm]{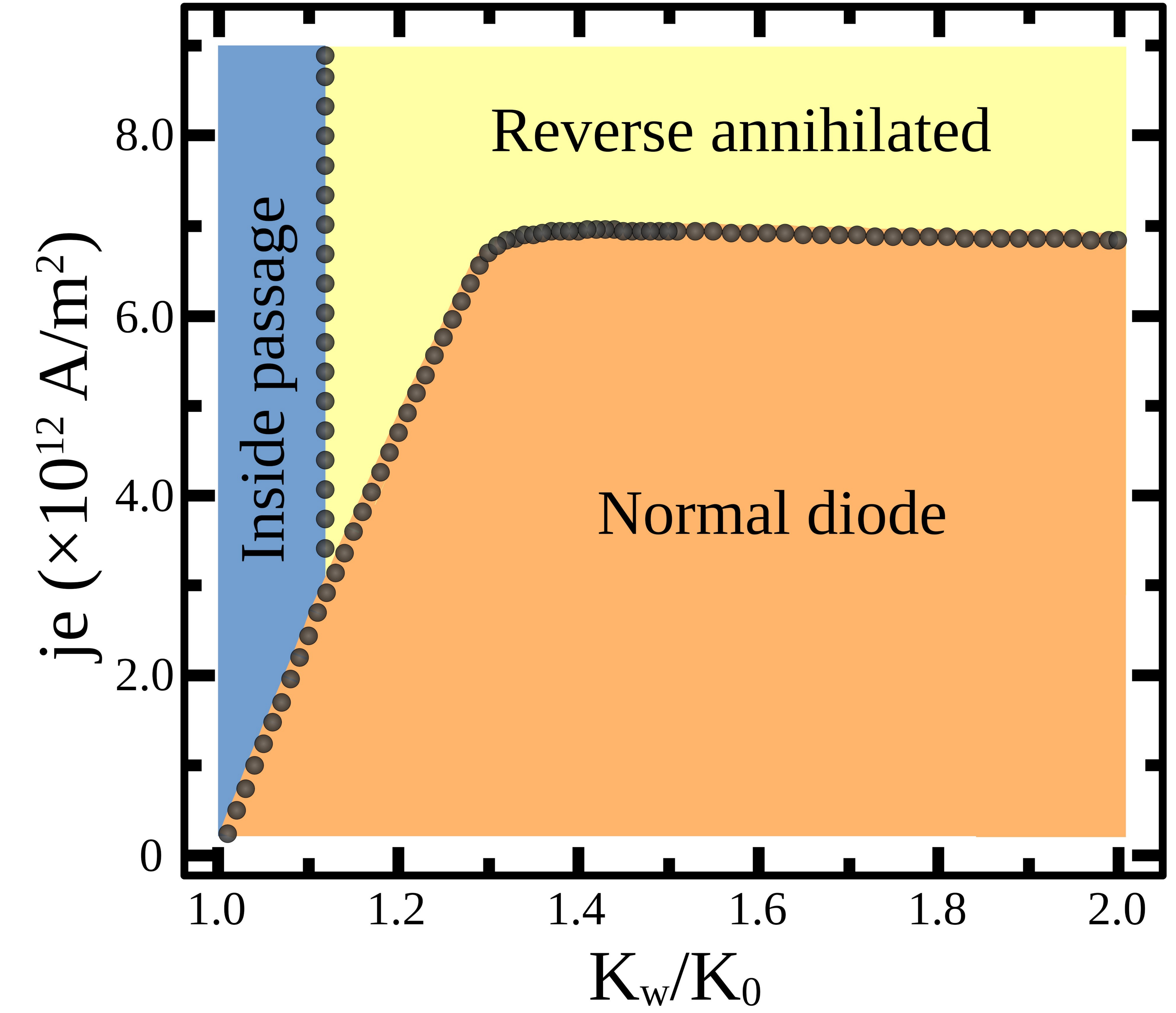}
     \caption{(Color online) Phase diagram showing the device performance based on current density $j_{e}$ and easy-axis anisotropy constant $K_{w}$. The orange region indicates optimal operation, with skyrmion movement from $C_{1}$ to $C_{2}$ and effective prevention of reverse motion. The blue region shows conditions where the device does not function as a diode, allowing skyrmion traverses within wedges areas in both directions. The yellow area highlights excessive current density, leading to skyrmion annihilation and information loss.}
    \label{fig:phase_d}
\end{figure*}

Another crucial aspect of the investigation focuses on the performance of the proposed device as the current density $j_{e}$ and the easy-axis anisotropy constant $K_{w}$ of the trapezoids are modified. The calculations were made by fixing $h_{1} = 200 $ nm, $h_{2} = 240 $ nm, $\theta_{w} = 45^{\circ}$, $b = 40$ nm, and $d = 80$ nm. Figure~\ref{fig:phase_d} decipts this behavior. The orange region represents the parameter ranges where the skyrmion-based diode functions optimally, allowing the skyrmion to move from the $C_{1}$ to $C_{2}$ sector. At the same time, the wedges effectively prevent its reverse motion. The blue region indicates values of $\left(K_{w}, j_{e}\right)$ where the skyrmion enters the trapezoid region from both directions ($C_{1} \rightarrow C_{2}$ and vice-versa). Therefore, the device does not operate as a diode in these conditions. On the other hand, the yellow region highlights scenarios where the skyrmion is annihilated during its reverse motion ($C_{2} \rightarrow C_{1}$) due to excessively high current density, which forces the skyrmion against the trapezoids.  Our studies reveal that the skyrmion lacks energetic stability for $K_{w} > 1.125 K_{0}$, blocking it from accessing in regions inside the trapezoids. Thereby, the intense current reduces the skyrmion's radius until it becomes too small to be stabilized in the lattice, leading to information loss. Such a regime should be avoided to ensure reliable device operation.

\section{CONCLUSION AND PROSPECTS}

In this study, we have investigated the dynamics of skyrmions within an engineered antiferromagnetic nanostripe. Our proposed racetrack comprises two right trapezoidal-shaped regions with an enhanced easy-plane anisotropy. This state of affairs enables the skyrmion passage from $C_{1}$ to $C_{2}$ regions while preventing its reverse motion, mimicking the operation of a diode. We have characterized the interaction potential between the skyrmion and the trapezoids, establishing a suitable separation between the wedges that enables efficient skyrmion transfer through the trapezoidal barriers. Additionally, the skyrmion transition for different factors, such as its initial position, spin-polarized current density $j_{e}$, and the inclination angle $\theta_{w}$ of the trapezoids, affect its movement. Remarkably, skyrmions can navigate through barriers with steep angles, below $\theta_{w} = 87.0^{\circ}$, though their velocity decreases in these wedge regions. Furthermore, our results show that in reverse motion from $C_{2}$ to $C_{1}$, skyrmion movement is strongly constrained with an appropriate choice of $h_{2}$ and $b$ (geometric parameters of the trapezoidal wedge) as well as the applied spin-polarized current density $j_{e}$. In addition, a phase diagram that maps the operational efficiency of the device as a function of $j_{e}$ and the easy-axis anisotropy constant $K_{w}$. In a forthcoming work, we will explore AFM skyrmion dynamics under temperature gradients and the influence of magnetic and non-magnetic impurities. Overall, our results contribute to the knowledge of skyrmion behavior in engineered nanostructures and offer a layout for creating next-generation spintronic devices. These findings provide experimental guidelines, underscoring the importance of precise control over trapezoidal geometry and current density to achieve optimal device performance. These insights may encourage further research and advance technological developments in skyrmion-based technology.

\section*{ACKNOWLEDGMENTS}
The authors acknowledge financial support from CNPq and the computational resources provided by the Sci-Com Lab.

\section*{SUPPLEMENTARY MATERIAL}
Please refer to the video files provided in the supplementary material. All movies depict a nanostripe with dimensions: length $L_{x} = 1000$ nm, width $L_{y} = 400$ nm, and thickness $L_{z} = 2$ nm. A spin-polarized electric current density $j_{e} = 0.10\, j_{0}$ is applied horizontally across the entire system. The racetrack contains two trapezoidal regions with an increased easy-axis constant, marked by the bright purple wedges, where we have fixed $K_{w} = 1.50 K_{0}$. In mv1.mp4, the current density $\vec{j}_{e} = -j_{e}\, \hat{x}$ drives the skyrmion from left to right. The skyrmion successfully traverses the obstacles and reaches the right end of the nanostripe. On the other hand, mv2.mp4 shows the skyrmion's reverse motion from right to left, driven by the current density $\vec{j}_{e} = j_{e} \, \hat{x}$. Initially, the skyrmion moves straight until it approaches the lower trapezoid. As its vertical position aligns closely with the height of the trapezoid, it passes through the lower barrier but eventually stops upon interacting with the upper trapezoid.

\bibliography{diode}

\begin{thebibliography}{105}%
\makeatletter
\providecommand \@ifxundefined [1]{%
 \@ifx{#1\undefined}
}%
\providecommand \@ifnum [1]{%
 \ifnum #1\expandafter \@firstoftwo
 \else \expandafter \@secondoftwo
 \fi
}%
\providecommand \@ifx [1]{%
 \ifx #1\expandafter \@firstoftwo
 \else \expandafter \@secondoftwo
 \fi
}%
\providecommand \natexlab [1]{#1}%
\providecommand \enquote  [1]{``#1''}%
\providecommand \bibnamefont  [1]{#1}%
\providecommand \bibfnamefont [1]{#1}%
\providecommand \citenamefont [1]{#1}%
\providecommand \href@noop [0]{\@secondoftwo}%
\providecommand \href [0]{\begingroup \@sanitize@url \@href}%
\providecommand \@href[1]{\@@startlink{#1}\@@href}%
\providecommand \@@href[1]{\endgroup#1\@@endlink}%
\providecommand \@sanitize@url [0]{\catcode `\\12\catcode `\$12\catcode
  `\&12\catcode `\#12\catcode `\^12\catcode `\_12\catcode `\%12\relax}%
\providecommand \@@startlink[1]{}%
\providecommand \@@endlink[0]{}%
\providecommand \url  [0]{\begingroup\@sanitize@url \@url }%
\providecommand \@url [1]{\endgroup\@href {#1}{\urlprefix }}%
\providecommand \urlprefix  [0]{URL }%
\providecommand \Eprint [0]{\href }%
\providecommand \doibase [0]{http://dx.doi.org/}%
\providecommand \selectlanguage [0]{\@gobble}%
\providecommand \bibinfo  [0]{\@secondoftwo}%
\providecommand \bibfield  [0]{\@secondoftwo}%
\providecommand \translation [1]{[#1]}%
\providecommand \BibitemOpen [0]{}%
\providecommand \bibitemStop [0]{}%
\providecommand \bibitemNoStop [0]{.\EOS\space}%
\providecommand \EOS [0]{\spacefactor3000\relax}%
\providecommand \BibitemShut  [1]{\csname bibitem#1\endcsname}%
\let\auto@bib@innerbib\@empty
\bibitem [{\citenamefont {Shinjo}\ \emph {et~al.}(2000)\citenamefont {Shinjo},
  \citenamefont {Okuno}, \citenamefont {Hassdorf}, \citenamefont {Shigeto},\
  and\ \citenamefont {Ono}}]{TShinjo_2000}%
  \BibitemOpen
  \bibfield  {author} {\bibinfo {author} {\bibfnamefont {T.}~\bibnamefont
  {Shinjo}}, \bibinfo {author} {\bibfnamefont {T.}~\bibnamefont {Okuno}},
  \bibinfo {author} {\bibfnamefont {R.}~\bibnamefont {Hassdorf}}, \bibinfo
  {author} {\bibfnamefont {K.}~\bibnamefont {Shigeto}}, \ and\ \bibinfo
  {author} {\bibfnamefont {T.}~\bibnamefont {Ono}},\ }\href {\doibase
  10.1126/science.289.5481.930} {\bibfield  {journal} {\bibinfo  {journal}
  {Sci.}\ }\textbf {\bibinfo {volume} {289}},\ \bibinfo {pages} {930} (\bibinfo
  {year} {2000})}\BibitemShut {NoStop}%
\bibitem [{\citenamefont {Guslienko}(2008)}]{KYGuslienko_2008}%
  \BibitemOpen
  \bibfield  {author} {\bibinfo {author} {\bibfnamefont {K.~Y.}\ \bibnamefont
  {Guslienko}},\ }\href {\doibase doi:10.1166/jnn.2008.18305} {\bibfield
  {journal} {\bibinfo  {journal} {J. Nanosci. Nanotechnol.}\ }\textbf {\bibinfo
  {volume} {8}},\ \bibinfo {pages} {2745} (\bibinfo {year} {2008})}\BibitemShut
  {NoStop}%
\bibitem [{\citenamefont {Jung}\ \emph {et~al.}(2012)\citenamefont {Jung},
  \citenamefont {Choi}, \citenamefont {Lee}, \citenamefont {Han}, \citenamefont
  {Yu}, \citenamefont {Im}, \citenamefont {Fischer},\ and\ \citenamefont
  {Kim}}]{HJung_2012}%
  \BibitemOpen
  \bibfield  {author} {\bibinfo {author} {\bibfnamefont {H.}~\bibnamefont
  {Jung}}, \bibinfo {author} {\bibfnamefont {Y.-S.}\ \bibnamefont {Choi}},
  \bibinfo {author} {\bibfnamefont {K.-S.}\ \bibnamefont {Lee}}, \bibinfo
  {author} {\bibfnamefont {D.-S.}\ \bibnamefont {Han}}, \bibinfo {author}
  {\bibfnamefont {Y.-S.}\ \bibnamefont {Yu}}, \bibinfo {author} {\bibfnamefont
  {M.-Y.}\ \bibnamefont {Im}}, \bibinfo {author} {\bibfnamefont
  {P.}~\bibnamefont {Fischer}}, \ and\ \bibinfo {author} {\bibfnamefont
  {S.-K.}\ \bibnamefont {Kim}},\ }\href {\doibase 10.1021/nn3000143} {\bibfield
   {journal} {\bibinfo  {journal} {ACS Nano}\ }\textbf {\bibinfo {volume}
  {6}},\ \bibinfo {pages} {3712} (\bibinfo {year} {2012})},\ \bibinfo {note}
  {pMID: 22533663}\BibitemShut {NoStop}%
\bibitem [{\citenamefont {Fert}\ \emph {et~al.}(2013)\citenamefont {Fert},
  \citenamefont {Cros},\ and\ \citenamefont {Sampaio}}]{AFert}%
  \BibitemOpen
  \bibfield  {author} {\bibinfo {author} {\bibfnamefont {A.}~\bibnamefont
  {Fert}}, \bibinfo {author} {\bibfnamefont {V.}~\bibnamefont {Cros}}, \ and\
  \bibinfo {author} {\bibfnamefont {J.}~\bibnamefont {Sampaio}},\ }\href
  {\doibase 10.1038/nnano.2013.29} {\bibfield  {journal} {\bibinfo  {journal}
  {Nat. Nanotech.}\ }\textbf {\bibinfo {volume} {8}},\ \bibinfo {pages} {152}
  (\bibinfo {year} {2013})}\BibitemShut {NoStop}%
\bibitem [{\citenamefont {Nagaosa}\ and\ \citenamefont
  {Tokura}(2013)}]{NNagaosa_2013}%
  \BibitemOpen
  \bibfield  {author} {\bibinfo {author} {\bibfnamefont {N.}~\bibnamefont
  {Nagaosa}}\ and\ \bibinfo {author} {\bibfnamefont {Y.}~\bibnamefont
  {Tokura}},\ }\href {\doibase 10.1038/nnano.2013.243} {\bibfield  {journal}
  {\bibinfo  {journal} {Nat. Nanotech.}\ }\textbf {\bibinfo {volume} {8}},\
  \bibinfo {pages} {899} (\bibinfo {year} {2013})}\BibitemShut {NoStop}%
\bibitem [{\citenamefont {Rybakov}\ \emph {et~al.}(2015)\citenamefont
  {Rybakov}, \citenamefont {Borisov}, \citenamefont {Bl{\"u}gel},\ and\
  \citenamefont {Kiselev}}]{FNRybakov_2015}%
  \BibitemOpen
  \bibfield  {author} {\bibinfo {author} {\bibfnamefont {F.~N.}\ \bibnamefont
  {Rybakov}}, \bibinfo {author} {\bibfnamefont {A.~B.}\ \bibnamefont
  {Borisov}}, \bibinfo {author} {\bibfnamefont {S.}~\bibnamefont {Bl{\"u}gel}},
  \ and\ \bibinfo {author} {\bibfnamefont {N.~S.}\ \bibnamefont {Kiselev}},\
  }\href {\doibase 10.1103/PhysRevLett.115.117201} {\bibfield  {journal}
  {\bibinfo  {journal} {Phys. Rev. Lett.}\ }\textbf {\bibinfo {volume} {115}},\
  \bibinfo {pages} {117201} (\bibinfo {year} {2015})}\BibitemShut {NoStop}%
\bibitem [{\citenamefont {Sutcliffe}(2018)}]{PSutcliffe_2018}%
  \BibitemOpen
  \bibfield  {author} {\bibinfo {author} {\bibfnamefont {P.}~\bibnamefont
  {Sutcliffe}},\ }\href {\doibase 10.1088/1751-8121/aad521} {\bibfield
  {journal} {\bibinfo  {journal} {J. Phys. A: Math. Theor.}\ }\textbf {\bibinfo
  {volume} {51}},\ \bibinfo {pages} {375401} (\bibinfo {year}
  {2018})}\BibitemShut {NoStop}%
\bibitem [{\citenamefont {Rybakov}\ \emph {et~al.}(2022)\citenamefont
  {Rybakov}, \citenamefont {Kiselev}, \citenamefont {Borisov}, \citenamefont
  {D{\"o}ring}, \citenamefont {Melcher},\ and\ \citenamefont
  {Bl{\"u}gel}}]{FNRybakov_2022}%
  \BibitemOpen
  \bibfield  {author} {\bibinfo {author} {\bibfnamefont {F.~N.}\ \bibnamefont
  {Rybakov}}, \bibinfo {author} {\bibfnamefont {N.~S.}\ \bibnamefont
  {Kiselev}}, \bibinfo {author} {\bibfnamefont {A.~B.}\ \bibnamefont
  {Borisov}}, \bibinfo {author} {\bibfnamefont {L.}~\bibnamefont {D{\"o}ring}},
  \bibinfo {author} {\bibfnamefont {C.}~\bibnamefont {Melcher}}, \ and\
  \bibinfo {author} {\bibfnamefont {S.}~\bibnamefont {Bl{\"u}gel}},\ }\href
  {\doibase 10.1063/5.0099942} {\bibfield  {journal} {\bibinfo  {journal} {APL
  Mater.}\ }\textbf {\bibinfo {volume} {10}},\ \bibinfo {pages} {111113}
  (\bibinfo {year} {2022})}\BibitemShut {NoStop}%
\bibitem [{\citenamefont {Belavin}\ and\ \citenamefont
  {Polyakov}(1975)}]{AABelavin_1975}%
  \BibitemOpen
  \bibfield  {author} {\bibinfo {author} {\bibfnamefont {A.~A.}\ \bibnamefont
  {Belavin}}\ and\ \bibinfo {author} {\bibfnamefont {A.~M.}\ \bibnamefont
  {Polyakov}},\ }\href {http://jetpletters.ru/ps/1529/article{$\_$}23383.pdf}
  {\bibfield  {journal} {\bibinfo  {journal} {JETP Lett.}\ }\textbf {\bibinfo
  {volume} {22}},\ \bibinfo {pages} {245} (\bibinfo {year} {1975})}\BibitemShut
  {NoStop}%
\bibitem [{\citenamefont {Yu}\ \emph {et~al.}(2018)\citenamefont {Yu},
  \citenamefont {Koshibae}, \citenamefont {Tokunaga}, \citenamefont {Shibata},
  \citenamefont {Taguchi}, \citenamefont {Nagaosa},\ and\ \citenamefont
  {Tokura}}]{XZYu_2018}%
  \BibitemOpen
  \bibfield  {author} {\bibinfo {author} {\bibfnamefont {X.~Z.}\ \bibnamefont
  {Yu}}, \bibinfo {author} {\bibfnamefont {W.}~\bibnamefont {Koshibae}},
  \bibinfo {author} {\bibfnamefont {Y.}~\bibnamefont {Tokunaga}}, \bibinfo
  {author} {\bibfnamefont {K.}~\bibnamefont {Shibata}}, \bibinfo {author}
  {\bibfnamefont {Y.}~\bibnamefont {Taguchi}}, \bibinfo {author} {\bibfnamefont
  {N.}~\bibnamefont {Nagaosa}}, \ and\ \bibinfo {author} {\bibfnamefont
  {Y.}~\bibnamefont {Tokura}},\ }\href {\doibase 10.1038/s41586-018-0745-3}
  {\bibfield  {journal} {\bibinfo  {journal} {Nature}\ }\textbf {\bibinfo
  {volume} {564}},\ \bibinfo {pages} {95} (\bibinfo {year} {2018})}\BibitemShut
  {NoStop}%
\bibitem [{\citenamefont {Xia}\ \emph {et~al.}(2022)\citenamefont {Xia},
  \citenamefont {Zhang}, \citenamefont {Liu}, \citenamefont {Zhou},\ and\
  \citenamefont {Ezawa}}]{JXia_2022}%
  \BibitemOpen
  \bibfield  {author} {\bibinfo {author} {\bibfnamefont {J.}~\bibnamefont
  {Xia}}, \bibinfo {author} {\bibfnamefont {X.}~\bibnamefont {Zhang}}, \bibinfo
  {author} {\bibfnamefont {X.}~\bibnamefont {Liu}}, \bibinfo {author}
  {\bibfnamefont {Y.}~\bibnamefont {Zhou}}, \ and\ \bibinfo {author}
  {\bibfnamefont {M.}~\bibnamefont {Ezawa}},\ }\href {\doibase
  10.1038/s43246-022-00311-w} {\bibfield  {journal} {\bibinfo  {journal}
  {Commun. Mater.}\ }\textbf {\bibinfo {volume} {3}},\ \bibinfo {pages} {1}
  (\bibinfo {year} {2022})}\BibitemShut {NoStop}%
\bibitem [{\citenamefont {Kishine}\ and\ \citenamefont
  {Ovchinnikov}(2015)}]{JIKishine_2015}%
  \BibitemOpen
  \bibfield  {author} {\bibinfo {author} {\bibfnamefont {J.-i.}\ \bibnamefont
  {Kishine}}\ and\ \bibinfo {author} {\bibfnamefont {A.~S.}\ \bibnamefont
  {Ovchinnikov}},\ }in\ \href {\doibase 10.1016/bs.ssp.2015.05.001} {\emph
  {\bibinfo {booktitle} {Solid {{State Physics}}}}},\ Vol.~\bibinfo {volume}
  {66},\ \bibinfo {editor} {edited by\ \bibinfo {editor} {\bibfnamefont
  {R.~E.}\ \bibnamefont {Camley}}\ and\ \bibinfo {editor} {\bibfnamefont
  {R.~L.}\ \bibnamefont {Stamps}}}\ (\bibinfo  {publisher} {{Academic Press}},\
  \bibinfo {address} {{San Diego}},\ \bibinfo {year} {2015})\ pp.\ \bibinfo
  {pages} {1--130}\BibitemShut {NoStop}%
\bibitem [{\citenamefont {Togawa}\ \emph {et~al.}(2016)\citenamefont {Togawa},
  \citenamefont {Kousaka}, \citenamefont {Inoue},\ and\ \citenamefont
  {Kishine}}]{YTogawa_2016}%
  \BibitemOpen
  \bibfield  {author} {\bibinfo {author} {\bibfnamefont {Y.}~\bibnamefont
  {Togawa}}, \bibinfo {author} {\bibfnamefont {Y.}~\bibnamefont {Kousaka}},
  \bibinfo {author} {\bibfnamefont {K.}~\bibnamefont {Inoue}}, \ and\ \bibinfo
  {author} {\bibfnamefont {J.}~\bibnamefont {Kishine}},\ }\href {\doibase
  10.7566/JPSJ.85.112001} {\bibfield  {journal} {\bibinfo  {journal} {J. Phys.
  Soc. Jpn.}\ }\textbf {\bibinfo {volume} {85}},\ \bibinfo {pages} {112001}
  (\bibinfo {year} {2016})}\BibitemShut {NoStop}%
\bibitem [{\citenamefont {Everschor-Sitte}\ \emph {et~al.}(2018)\citenamefont
  {Everschor-Sitte}, \citenamefont {Masell}, \citenamefont {Reeve},\ and\
  \citenamefont {Kl{\"a}ui}}]{Everschor_2018}%
  \BibitemOpen
  \bibfield  {author} {\bibinfo {author} {\bibfnamefont {K.}~\bibnamefont
  {Everschor-Sitte}}, \bibinfo {author} {\bibfnamefont {J.}~\bibnamefont
  {Masell}}, \bibinfo {author} {\bibfnamefont {R.~M.}\ \bibnamefont {Reeve}}, \
  and\ \bibinfo {author} {\bibfnamefont {M.}~\bibnamefont {Kl{\"a}ui}},\ }\href
  {\doibase 10.1063/1.5048972} {\bibfield  {journal} {\bibinfo  {journal} {J.
  Appl. Phys.}\ }\textbf {\bibinfo {volume} {124}},\ \bibinfo {pages} {240901}
  (\bibinfo {year} {2018})}\BibitemShut {NoStop}%
\bibitem [{\citenamefont {Leonov}\ and\ \citenamefont
  {K\'ezsm\'arki}(2017)}]{AOLeonov_2017}%
  \BibitemOpen
  \bibfield  {author} {\bibinfo {author} {\bibfnamefont {A.~O.}\ \bibnamefont
  {Leonov}}\ and\ \bibinfo {author} {\bibfnamefont {I.}~\bibnamefont
  {K\'ezsm\'arki}},\ }\href {\doibase 10.1103/PhysRevB.96.014423} {\bibfield
  {journal} {\bibinfo  {journal} {Phys. Rev. B}\ }\textbf {\bibinfo {volume}
  {96}},\ \bibinfo {pages} {014423} (\bibinfo {year} {2017})}\BibitemShut
  {NoStop}%
\bibitem [{\citenamefont {Bogdanov}\ and\ \citenamefont
  {Yablonskii}(1989)}]{ABogdanov1989}%
  \BibitemOpen
  \bibfield  {author} {\bibinfo {author} {\bibfnamefont {A.~N.}\ \bibnamefont
  {Bogdanov}}\ and\ \bibinfo {author} {\bibfnamefont {D.~A.}\ \bibnamefont
  {Yablonskii}},\ }\href {\doibase 10.1097/QAD.0b013e32830163c0} {\bibfield
  {journal} {\bibinfo  {journal} {Sov. Phys. JETP}\ }\textbf {\bibinfo {volume}
  {68}},\ \bibinfo {pages} {101} (\bibinfo {year} {1989})}\BibitemShut
  {NoStop}%
\bibitem [{\citenamefont {Dzyaloshinsky}(1958)}]{Dzyaloshinsky_1958}%
  \BibitemOpen
  \bibfield  {author} {\bibinfo {author} {\bibfnamefont {I.}~\bibnamefont
  {Dzyaloshinsky}},\ }\href {\doibase 10.1016/0022-3697(58)90076-3} {\bibfield
  {journal} {\bibinfo  {journal} {J. Phys. Chem. Solids}\ }\textbf {\bibinfo
  {volume} {4}},\ \bibinfo {pages} {241} (\bibinfo {year} {1958})}\BibitemShut
  {NoStop}%
\bibitem [{\citenamefont {Moriya}(1960)}]{Moriya_1960}%
  \BibitemOpen
  \bibfield  {author} {\bibinfo {author} {\bibfnamefont {T.}~\bibnamefont
  {Moriya}},\ }\href {\doibase 10.1103/PhysRev.120.91} {\bibfield  {journal}
  {\bibinfo  {journal} {Phys. Rev.}\ }\textbf {\bibinfo {volume} {120}},\
  \bibinfo {pages} {91} (\bibinfo {year} {1960})}\BibitemShut {NoStop}%
\bibitem [{\citenamefont {Bogdanov}\ and\ \citenamefont
  {Hubert}(1994)}]{ABogdanov1994}%
  \BibitemOpen
  \bibfield  {author} {\bibinfo {author} {\bibfnamefont {A.}~\bibnamefont
  {Bogdanov}}\ and\ \bibinfo {author} {\bibfnamefont {A.}~\bibnamefont
  {Hubert}},\ }\href {\doibase 10.1016/0304-8853(94)90046-9} {\bibfield
  {journal} {\bibinfo  {journal} {J. Magn. Magn. Mater.}\ }\textbf {\bibinfo
  {volume} {138}},\ \bibinfo {pages} {255} (\bibinfo {year}
  {1994})}\BibitemShut {NoStop}%
\bibitem [{\citenamefont {R{\"o}{\ss}ler}\ \emph {et~al.}(2006)\citenamefont
  {R{\"o}{\ss}ler}, \citenamefont {Bogdanov},\ and\ \citenamefont
  {Pfleiderer}}]{Robler_2006}%
  \BibitemOpen
  \bibfield  {author} {\bibinfo {author} {\bibfnamefont {U.~K.}\ \bibnamefont
  {R{\"o}{\ss}ler}}, \bibinfo {author} {\bibfnamefont {A.~N.}\ \bibnamefont
  {Bogdanov}}, \ and\ \bibinfo {author} {\bibfnamefont {C.}~\bibnamefont
  {Pfleiderer}},\ }\href {\doibase 10.1038/nature05056} {\bibfield  {journal}
  {\bibinfo  {journal} {Nature}\ }\textbf {\bibinfo {volume} {442}},\ \bibinfo
  {pages} {797} (\bibinfo {year} {2006})}\BibitemShut {NoStop}%
\bibitem [{\citenamefont {Bogdanov}\ and\ \citenamefont
  {Panagopoulos}(2020)}]{Bogdanov_2020}%
  \BibitemOpen
  \bibfield  {author} {\bibinfo {author} {\bibfnamefont {A.~N.}\ \bibnamefont
  {Bogdanov}}\ and\ \bibinfo {author} {\bibfnamefont {C.}~\bibnamefont
  {Panagopoulos}},\ }\href {\doibase 10.1038/s42254-020-0203-7} {\bibfield
  {journal} {\bibinfo  {journal} {Nat. Rev. Phys.}\ }\textbf {\bibinfo {volume}
  {2}},\ \bibinfo {pages} {492} (\bibinfo {year} {2020})}\BibitemShut {NoStop}%
\bibitem [{\citenamefont {M{\"u}hlbauer}\ \emph {et~al.}(2009)\citenamefont
  {M{\"u}hlbauer}, \citenamefont {Binz}, \citenamefont {Jonietz}, \citenamefont
  {Pfleiderer}, \citenamefont {Rosch}, \citenamefont {Neubauer}, \citenamefont
  {Georgii},\ and\ \citenamefont {B{\"o}ni}}]{Muhlbauer_2009}%
  \BibitemOpen
  \bibfield  {author} {\bibinfo {author} {\bibfnamefont {S.}~\bibnamefont
  {M{\"u}hlbauer}}, \bibinfo {author} {\bibfnamefont {B.}~\bibnamefont {Binz}},
  \bibinfo {author} {\bibfnamefont {F.}~\bibnamefont {Jonietz}}, \bibinfo
  {author} {\bibfnamefont {C.}~\bibnamefont {Pfleiderer}}, \bibinfo {author}
  {\bibfnamefont {A.}~\bibnamefont {Rosch}}, \bibinfo {author} {\bibfnamefont
  {A.}~\bibnamefont {Neubauer}}, \bibinfo {author} {\bibfnamefont
  {R.}~\bibnamefont {Georgii}}, \ and\ \bibinfo {author} {\bibfnamefont
  {P.}~\bibnamefont {B{\"o}ni}},\ }\href {\doibase 10.1126/science.1166767}
  {\bibfield  {journal} {\bibinfo  {journal} {Science}\ }\textbf {\bibinfo
  {volume} {323}},\ \bibinfo {pages} {915} (\bibinfo {year}
  {2009})}\BibitemShut {NoStop}%
\bibitem [{\citenamefont {Neubauer}\ \emph {et~al.}(2009)\citenamefont
  {Neubauer}, \citenamefont {Pfleiderer}, \citenamefont {Binz}, \citenamefont
  {Rosch}, \citenamefont {Ritz}, \citenamefont {Niklowitz},\ and\ \citenamefont
  {B{\"o}ni}}]{ANeubauer2009}%
  \BibitemOpen
  \bibfield  {author} {\bibinfo {author} {\bibfnamefont {A.}~\bibnamefont
  {Neubauer}}, \bibinfo {author} {\bibfnamefont {C.}~\bibnamefont
  {Pfleiderer}}, \bibinfo {author} {\bibfnamefont {B.}~\bibnamefont {Binz}},
  \bibinfo {author} {\bibfnamefont {A.}~\bibnamefont {Rosch}}, \bibinfo
  {author} {\bibfnamefont {R.}~\bibnamefont {Ritz}}, \bibinfo {author}
  {\bibfnamefont {P.~G.}\ \bibnamefont {Niklowitz}}, \ and\ \bibinfo {author}
  {\bibfnamefont {P.}~\bibnamefont {B{\"o}ni}},\ }\href {\doibase
  10.1103/PhysRevLett.102.186602} {\bibfield  {journal} {\bibinfo  {journal}
  {Phys. Rev. Lett.}\ }\textbf {\bibinfo {volume} {102}},\ \bibinfo {pages}
  {186602} (\bibinfo {year} {2009})}\BibitemShut {NoStop}%
\bibitem [{\citenamefont {Karube}\ \emph {et~al.}(2016)\citenamefont {Karube},
  \citenamefont {White}, \citenamefont {Reynolds}, \citenamefont {Gavilano},
  \citenamefont {Oike}, \citenamefont {Kikkawa}, \citenamefont {Kagawa},
  \citenamefont {Tokunaga}, \citenamefont {R{\o}nnow}, \citenamefont {Tokura},\
  and\ \citenamefont {Taguchi}}]{Karube_2016}%
  \BibitemOpen
  \bibfield  {author} {\bibinfo {author} {\bibfnamefont {K.}~\bibnamefont
  {Karube}}, \bibinfo {author} {\bibfnamefont {J.~S.}\ \bibnamefont {White}},
  \bibinfo {author} {\bibfnamefont {N.}~\bibnamefont {Reynolds}}, \bibinfo
  {author} {\bibfnamefont {J.~L.}\ \bibnamefont {Gavilano}}, \bibinfo {author}
  {\bibfnamefont {H.}~\bibnamefont {Oike}}, \bibinfo {author} {\bibfnamefont
  {A.}~\bibnamefont {Kikkawa}}, \bibinfo {author} {\bibfnamefont
  {F.}~\bibnamefont {Kagawa}}, \bibinfo {author} {\bibfnamefont
  {Y.}~\bibnamefont {Tokunaga}}, \bibinfo {author} {\bibfnamefont {H.~M.}\
  \bibnamefont {R{\o}nnow}}, \bibinfo {author} {\bibfnamefont {Y.}~\bibnamefont
  {Tokura}}, \ and\ \bibinfo {author} {\bibfnamefont {Y.}~\bibnamefont
  {Taguchi}},\ }\href {\doibase 10.1038/nmat4752} {\bibfield  {journal}
  {\bibinfo  {journal} {Nat. Mater.}\ }\textbf {\bibinfo {volume} {15}},\
  \bibinfo {pages} {1237} (\bibinfo {year} {2016})}\BibitemShut {NoStop}%
\bibitem [{\citenamefont {M{\"u}hlbauer}\ \emph {et~al.}(2016)\citenamefont
  {M{\"u}hlbauer}, \citenamefont {Kindervater}, \citenamefont {Adams},
  \citenamefont {Bauer}, \citenamefont {Keiderling},\ and\ \citenamefont
  {Pfleiderer}}]{SMuhlbauer_2016}%
  \BibitemOpen
  \bibfield  {author} {\bibinfo {author} {\bibfnamefont {S.}~\bibnamefont
  {M{\"u}hlbauer}}, \bibinfo {author} {\bibfnamefont {J.}~\bibnamefont
  {Kindervater}}, \bibinfo {author} {\bibfnamefont {T.}~\bibnamefont {Adams}},
  \bibinfo {author} {\bibfnamefont {A.}~\bibnamefont {Bauer}}, \bibinfo
  {author} {\bibfnamefont {U.}~\bibnamefont {Keiderling}}, \ and\ \bibinfo
  {author} {\bibfnamefont {C.}~\bibnamefont {Pfleiderer}},\ }\href {\doibase
  10.1088/1367-2630/18/7/075017} {\bibfield  {journal} {\bibinfo  {journal}
  {New J. Phys.}\ }\textbf {\bibinfo {volume} {18}},\ \bibinfo {pages} {075017}
  (\bibinfo {year} {2016})}\BibitemShut {NoStop}%
\bibitem [{\citenamefont {Yu}\ \emph {et~al.}(2010)\citenamefont {Yu},
  \citenamefont {Onose}, \citenamefont {Kanazawa}, \citenamefont {Park},
  \citenamefont {Han}, \citenamefont {Matsui}, \citenamefont {Nagaosa},\ and\
  \citenamefont {Tokura}}]{Yu_2010}%
  \BibitemOpen
  \bibfield  {author} {\bibinfo {author} {\bibfnamefont {X.~Z.}\ \bibnamefont
  {Yu}}, \bibinfo {author} {\bibfnamefont {Y.}~\bibnamefont {Onose}}, \bibinfo
  {author} {\bibfnamefont {N.}~\bibnamefont {Kanazawa}}, \bibinfo {author}
  {\bibfnamefont {J.~H.}\ \bibnamefont {Park}}, \bibinfo {author}
  {\bibfnamefont {J.~H.}\ \bibnamefont {Han}}, \bibinfo {author} {\bibfnamefont
  {Y.}~\bibnamefont {Matsui}}, \bibinfo {author} {\bibfnamefont
  {N.}~\bibnamefont {Nagaosa}}, \ and\ \bibinfo {author} {\bibfnamefont
  {Y.}~\bibnamefont {Tokura}},\ }\href {\doibase 10.1038/nature09124}
  {\bibfield  {journal} {\bibinfo  {journal} {Nature}\ }\textbf {\bibinfo
  {volume} {465}},\ \bibinfo {pages} {901} (\bibinfo {year}
  {2010})}\BibitemShut {NoStop}%
\bibitem [{\citenamefont {Yu}\ \emph {et~al.}(2011)\citenamefont {Yu},
  \citenamefont {Kanazawa}, \citenamefont {Onose}, \citenamefont {Kimoto},
  \citenamefont {Zhang}, \citenamefont {Ishiwata}, \citenamefont {Matsui},\
  and\ \citenamefont {Tokura}}]{Yu_2011}%
  \BibitemOpen
  \bibfield  {author} {\bibinfo {author} {\bibfnamefont {X.~Z.}\ \bibnamefont
  {Yu}}, \bibinfo {author} {\bibfnamefont {N.}~\bibnamefont {Kanazawa}},
  \bibinfo {author} {\bibfnamefont {Y.}~\bibnamefont {Onose}}, \bibinfo
  {author} {\bibfnamefont {K.}~\bibnamefont {Kimoto}}, \bibinfo {author}
  {\bibfnamefont {W.~Z.}\ \bibnamefont {Zhang}}, \bibinfo {author}
  {\bibfnamefont {S.}~\bibnamefont {Ishiwata}}, \bibinfo {author}
  {\bibfnamefont {Y.}~\bibnamefont {Matsui}}, \ and\ \bibinfo {author}
  {\bibfnamefont {Y.}~\bibnamefont {Tokura}},\ }\href {\doibase
  10.1038/nmat2916} {\bibfield  {journal} {\bibinfo  {journal} {Nat. Mater.}\
  }\textbf {\bibinfo {volume} {10}},\ \bibinfo {pages} {106} (\bibinfo {year}
  {2011})}\BibitemShut {NoStop}%
\bibitem [{\citenamefont {Moreau-Luchaire}\ \emph {et~al.}(2016)\citenamefont
  {Moreau-Luchaire}, \citenamefont {Moutafis}, \citenamefont {Reyren},
  \citenamefont {Sampaio}, \citenamefont {Vaz}, \citenamefont {Van~Horne},
  \citenamefont {Bouzehouane}, \citenamefont {Garcia}, \citenamefont
  {Deranlot}, \citenamefont {Warnicke}, \citenamefont {Wohlh{\"u}ter},
  \citenamefont {George}, \citenamefont {Weigand}, \citenamefont {Raabe},
  \citenamefont {Cros},\ and\ \citenamefont {Fert}}]{Moreau_Luchaire_2016}%
  \BibitemOpen
  \bibfield  {author} {\bibinfo {author} {\bibfnamefont {C.}~\bibnamefont
  {Moreau-Luchaire}}, \bibinfo {author} {\bibfnamefont {C.}~\bibnamefont
  {Moutafis}}, \bibinfo {author} {\bibfnamefont {N.}~\bibnamefont {Reyren}},
  \bibinfo {author} {\bibfnamefont {J.}~\bibnamefont {Sampaio}}, \bibinfo
  {author} {\bibfnamefont {C.~A.~F.}\ \bibnamefont {Vaz}}, \bibinfo {author}
  {\bibfnamefont {N.}~\bibnamefont {Van~Horne}}, \bibinfo {author}
  {\bibfnamefont {K.}~\bibnamefont {Bouzehouane}}, \bibinfo {author}
  {\bibfnamefont {K.}~\bibnamefont {Garcia}}, \bibinfo {author} {\bibfnamefont
  {C.}~\bibnamefont {Deranlot}}, \bibinfo {author} {\bibfnamefont
  {P.}~\bibnamefont {Warnicke}}, \bibinfo {author} {\bibfnamefont
  {P.}~\bibnamefont {Wohlh{\"u}ter}}, \bibinfo {author} {\bibfnamefont {J.-M.}\
  \bibnamefont {George}}, \bibinfo {author} {\bibfnamefont {M.}~\bibnamefont
  {Weigand}}, \bibinfo {author} {\bibfnamefont {J.}~\bibnamefont {Raabe}},
  \bibinfo {author} {\bibfnamefont {V.}~\bibnamefont {Cros}}, \ and\ \bibinfo
  {author} {\bibfnamefont {A.}~\bibnamefont {Fert}},\ }\href {\doibase
  10.1038/nnano.2015.313} {\bibfield  {journal} {\bibinfo  {journal} {Nat.
  Nanotech.}\ }\textbf {\bibinfo {volume} {11}},\ \bibinfo {pages} {444}
  (\bibinfo {year} {2016})}\BibitemShut {NoStop}%
\bibitem [{\citenamefont {Wiesendanger}(2016)}]{RWiesendanger}%
  \BibitemOpen
  \bibfield  {author} {\bibinfo {author} {\bibfnamefont {R.}~\bibnamefont
  {Wiesendanger}},\ }\href {\doibase 10.1038/natrevmats.2016.44} {\bibfield
  {journal} {\bibinfo  {journal} {Nat. Rev. Mater.}\ }\textbf {\bibinfo
  {volume} {1}},\ \bibinfo {pages} {16044} (\bibinfo {year}
  {2016})}\BibitemShut {NoStop}%
\bibitem [{\citenamefont {Ogasawara}\ \emph {et~al.}(2009)\citenamefont
  {Ogasawara}, \citenamefont {Iwata}, \citenamefont {Murakami}, \citenamefont
  {Okamoto},\ and\ \citenamefont {Tokura}}]{TOgasawara_2009}%
  \BibitemOpen
  \bibfield  {author} {\bibinfo {author} {\bibfnamefont {T.}~\bibnamefont
  {Ogasawara}}, \bibinfo {author} {\bibfnamefont {N.}~\bibnamefont {Iwata}},
  \bibinfo {author} {\bibfnamefont {Y.}~\bibnamefont {Murakami}}, \bibinfo
  {author} {\bibfnamefont {H.}~\bibnamefont {Okamoto}}, \ and\ \bibinfo
  {author} {\bibfnamefont {Y.}~\bibnamefont {Tokura}},\ }\href {\doibase
  10.1063/1.3123256} {\bibfield  {journal} {\bibinfo  {journal} {Appl. Phys.
  Lett.}\ }\textbf {\bibinfo {volume} {94}},\ \bibinfo {pages} {162507}
  (\bibinfo {year} {2009})}\BibitemShut {NoStop}%
\bibitem [{\citenamefont {Finazzi}\ \emph {et~al.}(2013)\citenamefont
  {Finazzi}, \citenamefont {Savoini}, \citenamefont {Khorsand}, \citenamefont
  {Tsukamoto}, \citenamefont {Itoh}, \citenamefont {Du{\`o}}, \citenamefont
  {Kirilyuk}, \citenamefont {Rasing},\ and\ \citenamefont
  {Ezawa}}]{MFinazzi_2013}%
  \BibitemOpen
  \bibfield  {author} {\bibinfo {author} {\bibfnamefont {M.}~\bibnamefont
  {Finazzi}}, \bibinfo {author} {\bibfnamefont {M.}~\bibnamefont {Savoini}},
  \bibinfo {author} {\bibfnamefont {A.~R.}\ \bibnamefont {Khorsand}}, \bibinfo
  {author} {\bibfnamefont {A.}~\bibnamefont {Tsukamoto}}, \bibinfo {author}
  {\bibfnamefont {A.}~\bibnamefont {Itoh}}, \bibinfo {author} {\bibfnamefont
  {L.}~\bibnamefont {Du{\`o}}}, \bibinfo {author} {\bibfnamefont
  {A.}~\bibnamefont {Kirilyuk}}, \bibinfo {author} {\bibfnamefont
  {T.}~\bibnamefont {Rasing}}, \ and\ \bibinfo {author} {\bibfnamefont
  {M.}~\bibnamefont {Ezawa}},\ }\href {\doibase 10.1103/PhysRevLett.110.177205}
  {\bibfield  {journal} {\bibinfo  {journal} {Phys. Rev. Lett.}\ }\textbf
  {\bibinfo {volume} {110}},\ \bibinfo {pages} {177205} (\bibinfo {year}
  {2013})}\BibitemShut {NoStop}%
\bibitem [{\citenamefont {Li}\ \emph {et~al.}(2021)\citenamefont {Li},
  \citenamefont {Su}, \citenamefont {Lin}, \citenamefont {Liu}, \citenamefont
  {Gao}, \citenamefont {Wang}, \citenamefont {Wei}, \citenamefont {Zhao},
  \citenamefont {Zhang}, \citenamefont {Cai},\ and\ \citenamefont
  {Shen}}]{ZLi_2021}%
  \BibitemOpen
  \bibfield  {author} {\bibinfo {author} {\bibfnamefont {Z.}~\bibnamefont
  {Li}}, \bibinfo {author} {\bibfnamefont {J.}~\bibnamefont {Su}}, \bibinfo
  {author} {\bibfnamefont {S.~Z.}\ \bibnamefont {Lin}}, \bibinfo {author}
  {\bibfnamefont {D.}~\bibnamefont {Liu}}, \bibinfo {author} {\bibfnamefont
  {Y.}~\bibnamefont {Gao}}, \bibinfo {author} {\bibfnamefont {S.}~\bibnamefont
  {Wang}}, \bibinfo {author} {\bibfnamefont {H.}~\bibnamefont {Wei}}, \bibinfo
  {author} {\bibfnamefont {T.}~\bibnamefont {Zhao}}, \bibinfo {author}
  {\bibfnamefont {Y.}~\bibnamefont {Zhang}}, \bibinfo {author} {\bibfnamefont
  {J.}~\bibnamefont {Cai}}, \ and\ \bibinfo {author} {\bibfnamefont
  {B.}~\bibnamefont {Shen}},\ }\href {\doibase 10.1038/s41467-021-25926-4}
  {\bibfield  {journal} {\bibinfo  {journal} {Nat. Commun.}\ }\textbf {\bibinfo
  {volume} {12}},\ \bibinfo {pages} {5604} (\bibinfo {year}
  {2021})}\BibitemShut {NoStop}%
\bibitem [{\citenamefont {Streubel}\ \emph {et~al.}(2018)\citenamefont
  {Streubel}, \citenamefont {Lambert}, \citenamefont {Kent}, \citenamefont
  {Ercius}, \citenamefont {N'Diaye}, \citenamefont {Ophus}, \citenamefont
  {Salahuddin},\ and\ \citenamefont {Fischer}}]{RStreubel_2018}%
  \BibitemOpen
  \bibfield  {author} {\bibinfo {author} {\bibfnamefont {R.}~\bibnamefont
  {Streubel}}, \bibinfo {author} {\bibfnamefont {C.-H.}\ \bibnamefont
  {Lambert}}, \bibinfo {author} {\bibfnamefont {N.}~\bibnamefont {Kent}},
  \bibinfo {author} {\bibfnamefont {P.}~\bibnamefont {Ercius}}, \bibinfo
  {author} {\bibfnamefont {A.~T.}\ \bibnamefont {N'Diaye}}, \bibinfo {author}
  {\bibfnamefont {C.}~\bibnamefont {Ophus}}, \bibinfo {author} {\bibfnamefont
  {S.}~\bibnamefont {Salahuddin}}, \ and\ \bibinfo {author} {\bibfnamefont
  {P.}~\bibnamefont {Fischer}},\ }\href {\doibase
  https://doi.org/10.1002/adma.201800199} {\bibfield  {journal} {\bibinfo
  {journal} {Adv. Mater.}\ }\textbf {\bibinfo {volume} {30}},\ \bibinfo {pages}
  {1800199} (\bibinfo {year} {2018})}\BibitemShut {NoStop}%
\bibitem [{\citenamefont {Xu}\ \emph {et~al.}(2023{\natexlab{a}})\citenamefont
  {Xu}, \citenamefont {Zhang}, \citenamefont {Wang}, \citenamefont {Bai},
  \citenamefont {Song}, \citenamefont {Liu}, \citenamefont {Zhou},
  \citenamefont {Je}, \citenamefont {N’Diaye}, \citenamefont {Im},
  \citenamefont {Yu}, \citenamefont {Chen},\ and\ \citenamefont
  {Jiang}}]{TXu_2023}%
  \BibitemOpen
  \bibfield  {author} {\bibinfo {author} {\bibfnamefont {T.}~\bibnamefont
  {Xu}}, \bibinfo {author} {\bibfnamefont {Y.}~\bibnamefont {Zhang}}, \bibinfo
  {author} {\bibfnamefont {Z.}~\bibnamefont {Wang}}, \bibinfo {author}
  {\bibfnamefont {H.}~\bibnamefont {Bai}}, \bibinfo {author} {\bibfnamefont
  {C.}~\bibnamefont {Song}}, \bibinfo {author} {\bibfnamefont {J.}~\bibnamefont
  {Liu}}, \bibinfo {author} {\bibfnamefont {Y.}~\bibnamefont {Zhou}}, \bibinfo
  {author} {\bibfnamefont {S.-G.}\ \bibnamefont {Je}}, \bibinfo {author}
  {\bibfnamefont {A.~T.}\ \bibnamefont {N’Diaye}}, \bibinfo {author}
  {\bibfnamefont {M.-Y.}\ \bibnamefont {Im}}, \bibinfo {author} {\bibfnamefont
  {R.}~\bibnamefont {Yu}}, \bibinfo {author} {\bibfnamefont {Z.}~\bibnamefont
  {Chen}}, \ and\ \bibinfo {author} {\bibfnamefont {W.}~\bibnamefont {Jiang}},\
  }\href {\doibase 10.1021/acsnano.3c02006} {\bibfield  {journal} {\bibinfo
  {journal} {ACS Nano}\ }\textbf {\bibinfo {volume} {17}},\ \bibinfo {pages}
  {7920} (\bibinfo {year} {2023}{\natexlab{a}})}\BibitemShut {NoStop}%
\bibitem [{\citenamefont {Luo}\ \emph {et~al.}(2023)\citenamefont {Luo},
  \citenamefont {Chen}, \citenamefont {Ukleev}, \citenamefont {Wintz},
  \citenamefont {Weigand}, \citenamefont {Abrudan}, \citenamefont
  {Proke{\v{s}}},\ and\ \citenamefont {Radu}}]{CLuo_2023}%
  \BibitemOpen
  \bibfield  {author} {\bibinfo {author} {\bibfnamefont {C.}~\bibnamefont
  {Luo}}, \bibinfo {author} {\bibfnamefont {K.}~\bibnamefont {Chen}}, \bibinfo
  {author} {\bibfnamefont {V.}~\bibnamefont {Ukleev}}, \bibinfo {author}
  {\bibfnamefont {S.}~\bibnamefont {Wintz}}, \bibinfo {author} {\bibfnamefont
  {M.}~\bibnamefont {Weigand}}, \bibinfo {author} {\bibfnamefont {R.-M.}\
  \bibnamefont {Abrudan}}, \bibinfo {author} {\bibfnamefont {K.}~\bibnamefont
  {Proke{\v{s}}}}, \ and\ \bibinfo {author} {\bibfnamefont {F.}~\bibnamefont
  {Radu}},\ }\href {\doibase 10.1038/s42005-023-01341-7} {\bibfield  {journal}
  {\bibinfo  {journal} {Commun. Phys.}\ }\textbf {\bibinfo {volume} {6}},\
  \bibinfo {pages} {218} (\bibinfo {year} {2023})}\BibitemShut {NoStop}%
\bibitem [{\citenamefont {Chen}\ \emph
  {et~al.}(2020{\natexlab{a}})\citenamefont {Chen}, \citenamefont {Lott},
  \citenamefont {Philippi-Kobs}, \citenamefont {Weigand}, \citenamefont {Luo},\
  and\ \citenamefont {Radu}}]{KChen_2020}%
  \BibitemOpen
  \bibfield  {author} {\bibinfo {author} {\bibfnamefont {K.}~\bibnamefont
  {Chen}}, \bibinfo {author} {\bibfnamefont {D.}~\bibnamefont {Lott}}, \bibinfo
  {author} {\bibfnamefont {A.}~\bibnamefont {Philippi-Kobs}}, \bibinfo {author}
  {\bibfnamefont {M.}~\bibnamefont {Weigand}}, \bibinfo {author} {\bibfnamefont
  {C.}~\bibnamefont {Luo}}, \ and\ \bibinfo {author} {\bibfnamefont
  {F.}~\bibnamefont {Radu}},\ }\href {\doibase 10.1039/D0NR02947E} {\bibfield
  {journal} {\bibinfo  {journal} {Nanoscale}\ }\textbf {\bibinfo {volume}
  {12}},\ \bibinfo {pages} {18137} (\bibinfo {year}
  {2020}{\natexlab{a}})}\BibitemShut {NoStop}%
\bibitem [{\citenamefont {V{\'e}lez}\ \emph {et~al.}(2022)\citenamefont
  {V{\'e}lez}, \citenamefont {Ruiz-G{\'o}mez}, \citenamefont {Schaab},
  \citenamefont {Gradauskaite}, \citenamefont {W{\"o}rnle}, \citenamefont
  {Welter}, \citenamefont {Jacot}, \citenamefont {Degen}, \citenamefont
  {Trassin}, \citenamefont {Fiebig},\ and\ \citenamefont
  {Gambardella}}]{SVelez_2022}%
  \BibitemOpen
  \bibfield  {author} {\bibinfo {author} {\bibfnamefont {S.}~\bibnamefont
  {V{\'e}lez}}, \bibinfo {author} {\bibfnamefont {S.}~\bibnamefont
  {Ruiz-G{\'o}mez}}, \bibinfo {author} {\bibfnamefont {J.}~\bibnamefont
  {Schaab}}, \bibinfo {author} {\bibfnamefont {E.}~\bibnamefont
  {Gradauskaite}}, \bibinfo {author} {\bibfnamefont {M.~S.}\ \bibnamefont
  {W{\"o}rnle}}, \bibinfo {author} {\bibfnamefont {P.}~\bibnamefont {Welter}},
  \bibinfo {author} {\bibfnamefont {B.~J.}\ \bibnamefont {Jacot}}, \bibinfo
  {author} {\bibfnamefont {C.~L.}\ \bibnamefont {Degen}}, \bibinfo {author}
  {\bibfnamefont {M.}~\bibnamefont {Trassin}}, \bibinfo {author} {\bibfnamefont
  {M.}~\bibnamefont {Fiebig}}, \ and\ \bibinfo {author} {\bibfnamefont
  {P.}~\bibnamefont {Gambardella}},\ }\href {\doibase
  10.1038/s41565-022-01144-x} {\bibfield  {journal} {\bibinfo  {journal} {Nat.
  Nanotechnol.}\ }\textbf {\bibinfo {volume} {17}},\ \bibinfo {pages} {834}
  (\bibinfo {year} {2022})}\BibitemShut {NoStop}%
\bibitem [{\citenamefont {Sampaio}\ \emph {et~al.}(2023)\citenamefont
  {Sampaio}, \citenamefont {Cros}, \citenamefont {Rohart}, \citenamefont
  {Thiaville},\ and\ \citenamefont {Fert}}]{Sampaio_2013}%
  \BibitemOpen
  \bibfield  {author} {\bibinfo {author} {\bibfnamefont {J.}~\bibnamefont
  {Sampaio}}, \bibinfo {author} {\bibfnamefont {V.}~\bibnamefont {Cros}},
  \bibinfo {author} {\bibfnamefont {S.}~\bibnamefont {Rohart}}, \bibinfo
  {author} {\bibfnamefont {A.}~\bibnamefont {Thiaville}}, \ and\ \bibinfo
  {author} {\bibfnamefont {A.}~\bibnamefont {Fert}},\ }\href {\doibase
  10.1038/nnano.2013.210} {\bibfield  {journal} {\bibinfo  {journal} {Nat.
  Nanotech.}\ }\textbf {\bibinfo {volume} {8}},\ \bibinfo {pages} {839}
  (\bibinfo {year} {2023})}\BibitemShut {NoStop}%
\bibitem [{\citenamefont {Zhang}\ \emph
  {et~al.}(2015{\natexlab{a}})\citenamefont {Zhang}, \citenamefont {Ezawa},\
  and\ \citenamefont {Zhou}}]{XZhang_2015}%
  \BibitemOpen
  \bibfield  {author} {\bibinfo {author} {\bibfnamefont {X.}~\bibnamefont
  {Zhang}}, \bibinfo {author} {\bibfnamefont {M.}~\bibnamefont {Ezawa}}, \ and\
  \bibinfo {author} {\bibfnamefont {Y.}~\bibnamefont {Zhou}},\ }\href {\doibase
  10.1038/srep09400} {\bibfield  {journal} {\bibinfo  {journal} {Sci. Rep.}\
  }\textbf {\bibinfo {volume} {5}},\ \bibinfo {pages} {9400} (\bibinfo {year}
  {2015}{\natexlab{a}})}\BibitemShut {NoStop}%
\bibitem [{\citenamefont {Tomasello}\ \emph {et~al.}(2014)\citenamefont
  {Tomasello}, \citenamefont {Martinez}, \citenamefont {Zivieri}, \citenamefont
  {Torres}, \citenamefont {Carpentieri},\ and\ \citenamefont
  {Finocchio}}]{Tomasello_2014}%
  \BibitemOpen
  \bibfield  {author} {\bibinfo {author} {\bibfnamefont {R.}~\bibnamefont
  {Tomasello}}, \bibinfo {author} {\bibfnamefont {E.}~\bibnamefont {Martinez}},
  \bibinfo {author} {\bibfnamefont {R.}~\bibnamefont {Zivieri}}, \bibinfo
  {author} {\bibfnamefont {L.}~\bibnamefont {Torres}}, \bibinfo {author}
  {\bibfnamefont {M.}~\bibnamefont {Carpentieri}}, \ and\ \bibinfo {author}
  {\bibfnamefont {G.}~\bibnamefont {Finocchio}},\ }\href {\doibase
  10.1038/srep06784} {\bibfield  {journal} {\bibinfo  {journal} {Sci. Rep.}\
  }\textbf {\bibinfo {volume} {4}},\ \bibinfo {pages} {6784} (\bibinfo {year}
  {2014})}\BibitemShut {NoStop}%
\bibitem [{\citenamefont {Dohi}\ \emph {et~al.}(2023)\citenamefont {Dohi},
  \citenamefont {Wei{\ss}enhofer}, \citenamefont {Kerber}, \citenamefont
  {Kammerbauer}, \citenamefont {Ge}, \citenamefont {Raab}, \citenamefont
  {Z{\'a}zvorka}, \citenamefont {Syskaki}, \citenamefont {Shahee},
  \citenamefont {Ruhwedel}, \citenamefont {B{\"o}ttcher}, \citenamefont
  {Pirro}, \citenamefont {Jakob}, \citenamefont {Nowak},\ and\ \citenamefont
  {Kl{\"a}ui}}]{TDohi_2023}%
  \BibitemOpen
  \bibfield  {author} {\bibinfo {author} {\bibfnamefont {T.}~\bibnamefont
  {Dohi}}, \bibinfo {author} {\bibfnamefont {M.}~\bibnamefont
  {Wei{\ss}enhofer}}, \bibinfo {author} {\bibfnamefont {N.}~\bibnamefont
  {Kerber}}, \bibinfo {author} {\bibfnamefont {F.}~\bibnamefont {Kammerbauer}},
  \bibinfo {author} {\bibfnamefont {Y.}~\bibnamefont {Ge}}, \bibinfo {author}
  {\bibfnamefont {K.}~\bibnamefont {Raab}}, \bibinfo {author} {\bibfnamefont
  {J.}~\bibnamefont {Z{\'a}zvorka}}, \bibinfo {author} {\bibfnamefont {M.-A.}\
  \bibnamefont {Syskaki}}, \bibinfo {author} {\bibfnamefont {A.}~\bibnamefont
  {Shahee}}, \bibinfo {author} {\bibfnamefont {M.}~\bibnamefont {Ruhwedel}},
  \bibinfo {author} {\bibfnamefont {T.}~\bibnamefont {B{\"o}ttcher}}, \bibinfo
  {author} {\bibfnamefont {P.}~\bibnamefont {Pirro}}, \bibinfo {author}
  {\bibfnamefont {G.}~\bibnamefont {Jakob}}, \bibinfo {author} {\bibfnamefont
  {U.}~\bibnamefont {Nowak}}, \ and\ \bibinfo {author} {\bibfnamefont
  {M.}~\bibnamefont {Kl{\"a}ui}},\ }\href {\doibase 10.1038/s41467-023-40720-0}
  {\bibfield  {journal} {\bibinfo  {journal} {Nat. Commun.}\ }\textbf {\bibinfo
  {volume} {14}},\ \bibinfo {pages} {5424} (\bibinfo {year}
  {2023})}\BibitemShut {NoStop}%
\bibitem [{\citenamefont {Sisodia}\ \emph {et~al.}(2022)\citenamefont
  {Sisodia}, \citenamefont {Pelloux-Prayer}, \citenamefont {Buda-Prejbeanu},
  \citenamefont {Anghel}, \citenamefont {Gaudin},\ and\ \citenamefont
  {Boulle}}]{NSisodia_2022}%
  \BibitemOpen
  \bibfield  {author} {\bibinfo {author} {\bibfnamefont {N.}~\bibnamefont
  {Sisodia}}, \bibinfo {author} {\bibfnamefont {J.}~\bibnamefont
  {Pelloux-Prayer}}, \bibinfo {author} {\bibfnamefont {L.~D.}\ \bibnamefont
  {Buda-Prejbeanu}}, \bibinfo {author} {\bibfnamefont {L.}~\bibnamefont
  {Anghel}}, \bibinfo {author} {\bibfnamefont {G.}~\bibnamefont {Gaudin}}, \
  and\ \bibinfo {author} {\bibfnamefont {O.}~\bibnamefont {Boulle}},\ }\href
  {\doibase 10.1103/PhysRevApplied.17.064035} {\bibfield  {journal} {\bibinfo
  {journal} {Phys. Rev. Appl.}\ }\textbf {\bibinfo {volume} {17}},\ \bibinfo
  {pages} {064035} (\bibinfo {year} {2022})}\BibitemShut {NoStop}%
\bibitem [{\citenamefont {Luo}\ \emph {et~al.}(2018)\citenamefont {Luo},
  \citenamefont {Song}, \citenamefont {Li}, \citenamefont {Zhang},
  \citenamefont {Hong}, \citenamefont {Yang}, \citenamefont {Zou},
  \citenamefont {Xu},\ and\ \citenamefont {You}}]{SLuo_2018}%
  \BibitemOpen
  \bibfield  {author} {\bibinfo {author} {\bibfnamefont {S.}~\bibnamefont
  {Luo}}, \bibinfo {author} {\bibfnamefont {M.}~\bibnamefont {Song}}, \bibinfo
  {author} {\bibfnamefont {X.}~\bibnamefont {Li}}, \bibinfo {author}
  {\bibfnamefont {Y.}~\bibnamefont {Zhang}}, \bibinfo {author} {\bibfnamefont
  {J.}~\bibnamefont {Hong}}, \bibinfo {author} {\bibfnamefont {X.}~\bibnamefont
  {Yang}}, \bibinfo {author} {\bibfnamefont {X.}~\bibnamefont {Zou}}, \bibinfo
  {author} {\bibfnamefont {N.}~\bibnamefont {Xu}}, \ and\ \bibinfo {author}
  {\bibfnamefont {L.}~\bibnamefont {You}},\ }\href {\doibase
  10.1021/acs.nanolett.7b04722} {\bibfield  {journal} {\bibinfo  {journal}
  {Nano Lett.}\ }\textbf {\bibinfo {volume} {18}},\ \bibinfo {pages} {1180}
  (\bibinfo {year} {2018})}\BibitemShut {NoStop}%
\bibitem [{\citenamefont {Feng}\ \emph {et~al.}(2019)\citenamefont {Feng},
  \citenamefont {Xia}, \citenamefont {Qiu}, \citenamefont {Cai}, \citenamefont
  {Shen}, \citenamefont {Morvan}, \citenamefont {Zhang}, \citenamefont {Zhou},\
  and\ \citenamefont {Zhao}}]{YFeng_2019}%
  \BibitemOpen
  \bibfield  {author} {\bibinfo {author} {\bibfnamefont {Y.}~\bibnamefont
  {Feng}}, \bibinfo {author} {\bibfnamefont {J.}~\bibnamefont {Xia}}, \bibinfo
  {author} {\bibfnamefont {L.}~\bibnamefont {Qiu}}, \bibinfo {author}
  {\bibfnamefont {X.}~\bibnamefont {Cai}}, \bibinfo {author} {\bibfnamefont
  {L.}~\bibnamefont {Shen}}, \bibinfo {author} {\bibfnamefont {F.~J.}\
  \bibnamefont {Morvan}}, \bibinfo {author} {\bibfnamefont {X.}~\bibnamefont
  {Zhang}}, \bibinfo {author} {\bibfnamefont {Y.}~\bibnamefont {Zhou}}, \ and\
  \bibinfo {author} {\bibfnamefont {G.}~\bibnamefont {Zhao}},\ }\href {\doibase
  10.1016/j.jmmm.2019.165610} {\bibfield  {journal} {\bibinfo  {journal} {J.
  Magn. Magn. Mater.}\ }\textbf {\bibinfo {volume} {491}},\ \bibinfo {pages}
  {165610} (\bibinfo {year} {2019})}\BibitemShut {NoStop}%
\bibitem [{\citenamefont {Garcia-Sanchez}\ \emph {et~al.}(2016)\citenamefont
  {Garcia-Sanchez}, \citenamefont {Sampaio}, \citenamefont {Reyren},
  \citenamefont {Cros},\ and\ \citenamefont {Kim}}]{GSanchez_2016}%
  \BibitemOpen
  \bibfield  {author} {\bibinfo {author} {\bibfnamefont {F.}~\bibnamefont
  {Garcia-Sanchez}}, \bibinfo {author} {\bibfnamefont {J.}~\bibnamefont
  {Sampaio}}, \bibinfo {author} {\bibfnamefont {N.}~\bibnamefont {Reyren}},
  \bibinfo {author} {\bibfnamefont {V.}~\bibnamefont {Cros}}, \ and\ \bibinfo
  {author} {\bibfnamefont {J.-V.}\ \bibnamefont {Kim}},\ }\href {\doibase
  10.1088/1367-2630/18/7/075011} {\bibfield  {journal} {\bibinfo  {journal}
  {New J. Phys.}\ }\textbf {\bibinfo {volume} {18}},\ \bibinfo {pages} {075011}
  (\bibinfo {year} {2016})}\BibitemShut {NoStop}%
\bibitem [{\citenamefont {Heinze}\ \emph {et~al.}(2011)\citenamefont {Heinze},
  \citenamefont {von Bergmann}, \citenamefont {Menzel}, \citenamefont {Brede},
  \citenamefont {Kubetzka}, \citenamefont {Wiesendanger}, \citenamefont
  {Bihlmayer},\ and\ \citenamefont {Bl{\"u}gel}}]{SHeinze}%
  \BibitemOpen
  \bibfield  {author} {\bibinfo {author} {\bibfnamefont {S.}~\bibnamefont
  {Heinze}}, \bibinfo {author} {\bibfnamefont {K.}~\bibnamefont {von
  Bergmann}}, \bibinfo {author} {\bibfnamefont {M.}~\bibnamefont {Menzel}},
  \bibinfo {author} {\bibfnamefont {J.}~\bibnamefont {Brede}}, \bibinfo
  {author} {\bibfnamefont {A.}~\bibnamefont {Kubetzka}}, \bibinfo {author}
  {\bibfnamefont {R.}~\bibnamefont {Wiesendanger}}, \bibinfo {author}
  {\bibfnamefont {G.}~\bibnamefont {Bihlmayer}}, \ and\ \bibinfo {author}
  {\bibfnamefont {S.}~\bibnamefont {Bl{\"u}gel}},\ }\href {\doibase
  10.1038/nphys2045} {\bibfield  {journal} {\bibinfo  {journal} {Nat. Phys.}\
  }\textbf {\bibinfo {volume} {7}},\ \bibinfo {pages} {713} (\bibinfo {year}
  {2011})}\BibitemShut {NoStop}%
\bibitem [{\citenamefont {Soumyanarayanan}\ \emph {et~al.}(2017)\citenamefont
  {Soumyanarayanan}, \citenamefont {Raju}, \citenamefont {Gonzalez~Oyarce},
  \citenamefont {Tan}, \citenamefont {Im}, \citenamefont {Petrovi{\'c}},
  \citenamefont {Ho}, \citenamefont {Khoo}, \citenamefont {Tran}, \citenamefont
  {Gan}, \citenamefont {Ernult},\ and\ \citenamefont
  {Panagopoulos}}]{ASoumyanarayanan}%
  \BibitemOpen
  \bibfield  {author} {\bibinfo {author} {\bibfnamefont {A.}~\bibnamefont
  {Soumyanarayanan}}, \bibinfo {author} {\bibfnamefont {M.}~\bibnamefont
  {Raju}}, \bibinfo {author} {\bibfnamefont {A.~L.}\ \bibnamefont
  {Gonzalez~Oyarce}}, \bibinfo {author} {\bibfnamefont {A.~K.~C.}\ \bibnamefont
  {Tan}}, \bibinfo {author} {\bibfnamefont {M.-Y.}\ \bibnamefont {Im}},
  \bibinfo {author} {\bibfnamefont {A.~P.}\ \bibnamefont {Petrovi{\'c}}},
  \bibinfo {author} {\bibfnamefont {P.}~\bibnamefont {Ho}}, \bibinfo {author}
  {\bibfnamefont {K.~H.}\ \bibnamefont {Khoo}}, \bibinfo {author}
  {\bibfnamefont {M.}~\bibnamefont {Tran}}, \bibinfo {author} {\bibfnamefont
  {C.~K.}\ \bibnamefont {Gan}}, \bibinfo {author} {\bibfnamefont
  {F.}~\bibnamefont {Ernult}}, \ and\ \bibinfo {author} {\bibfnamefont
  {C.}~\bibnamefont {Panagopoulos}},\ }\href {\doibase 10.1038/nmat4934}
  {\bibfield  {journal} {\bibinfo  {journal} {Nat. Mater.}\ }\textbf {\bibinfo
  {volume} {16}},\ \bibinfo {pages} {898} (\bibinfo {year} {2017})}\BibitemShut
  {NoStop}%
\bibitem [{\citenamefont {B{\"u}ttner}\ \emph {et~al.}(2017)\citenamefont
  {B{\"u}ttner}, \citenamefont {Lemesh}, \citenamefont {Schneider},
  \citenamefont {Pfau}, \citenamefont {G{\"u}nther}, \citenamefont {Hessing},
  \citenamefont {Geilhufe}, \citenamefont {Caretta}, \citenamefont {Engel},
  \citenamefont {Kr{\"u}ger}, \citenamefont {Viefhaus}, \citenamefont
  {Eisebitt},\ and\ \citenamefont {Beach}}]{Buttner_2017}%
  \BibitemOpen
  \bibfield  {author} {\bibinfo {author} {\bibfnamefont {F.}~\bibnamefont
  {B{\"u}ttner}}, \bibinfo {author} {\bibfnamefont {I.}~\bibnamefont {Lemesh}},
  \bibinfo {author} {\bibfnamefont {M.}~\bibnamefont {Schneider}}, \bibinfo
  {author} {\bibfnamefont {B.}~\bibnamefont {Pfau}}, \bibinfo {author}
  {\bibfnamefont {C.~M.}\ \bibnamefont {G{\"u}nther}}, \bibinfo {author}
  {\bibfnamefont {P.}~\bibnamefont {Hessing}}, \bibinfo {author} {\bibfnamefont
  {J.}~\bibnamefont {Geilhufe}}, \bibinfo {author} {\bibfnamefont
  {L.}~\bibnamefont {Caretta}}, \bibinfo {author} {\bibfnamefont
  {D.}~\bibnamefont {Engel}}, \bibinfo {author} {\bibfnamefont
  {B.}~\bibnamefont {Kr{\"u}ger}}, \bibinfo {author} {\bibfnamefont
  {J.}~\bibnamefont {Viefhaus}}, \bibinfo {author} {\bibfnamefont
  {S.}~\bibnamefont {Eisebitt}}, \ and\ \bibinfo {author} {\bibfnamefont
  {G.~S.~D.}\ \bibnamefont {Beach}},\ }\href {\doibase 10.1038/nnano.2017.178}
  {\bibfield  {journal} {\bibinfo  {journal} {Nat. Nanotech.}\ }\textbf
  {\bibinfo {volume} {12}},\ \bibinfo {pages} {1040} (\bibinfo {year}
  {2017})}\BibitemShut {NoStop}%
\bibitem [{\citenamefont {Finizio}\ \emph {et~al.}(2019)\citenamefont
  {Finizio}, \citenamefont {Zeissler}, \citenamefont {Wintz}, \citenamefont
  {Mayr}, \citenamefont {We{\ss}els}, \citenamefont {Huxtable}, \citenamefont
  {Burnell}, \citenamefont {Marrows},\ and\ \citenamefont
  {Raabe}}]{SFinizio_2019}%
  \BibitemOpen
  \bibfield  {author} {\bibinfo {author} {\bibfnamefont {S.}~\bibnamefont
  {Finizio}}, \bibinfo {author} {\bibfnamefont {K.}~\bibnamefont {Zeissler}},
  \bibinfo {author} {\bibfnamefont {S.}~\bibnamefont {Wintz}}, \bibinfo
  {author} {\bibfnamefont {S.}~\bibnamefont {Mayr}}, \bibinfo {author}
  {\bibfnamefont {T.}~\bibnamefont {We{\ss}els}}, \bibinfo {author}
  {\bibfnamefont {A.~J.}\ \bibnamefont {Huxtable}}, \bibinfo {author}
  {\bibfnamefont {G.}~\bibnamefont {Burnell}}, \bibinfo {author} {\bibfnamefont
  {C.~H.}\ \bibnamefont {Marrows}}, \ and\ \bibinfo {author} {\bibfnamefont
  {J.}~\bibnamefont {Raabe}},\ }\href {\doibase 10.1021/acs.nanolett.9b02840}
  {\bibfield  {journal} {\bibinfo  {journal} {Nano Lett.}\ }\textbf {\bibinfo
  {volume} {19}},\ \bibinfo {pages} {7246} (\bibinfo {year}
  {2019})}\BibitemShut {NoStop}%
\bibitem [{\citenamefont {Zhang}\ \emph {et~al.}(2020)\citenamefont {Zhang},
  \citenamefont {Zhou}, \citenamefont {Song}, \citenamefont {Park},
  \citenamefont {Xia}, \citenamefont {Ezawa}, \citenamefont {Liu},
  \citenamefont {Zhao}, \citenamefont {Zhao},\ and\ \citenamefont
  {Woo}}]{Zhang_2020}%
  \BibitemOpen
  \bibfield  {author} {\bibinfo {author} {\bibfnamefont {X.}~\bibnamefont
  {Zhang}}, \bibinfo {author} {\bibfnamefont {Y.}~\bibnamefont {Zhou}},
  \bibinfo {author} {\bibfnamefont {K.~S.}\ \bibnamefont {Song}}, \bibinfo
  {author} {\bibfnamefont {T.-E.}\ \bibnamefont {Park}}, \bibinfo {author}
  {\bibfnamefont {J.}~\bibnamefont {Xia}}, \bibinfo {author} {\bibfnamefont
  {M.}~\bibnamefont {Ezawa}}, \bibinfo {author} {\bibfnamefont
  {X.}~\bibnamefont {Liu}}, \bibinfo {author} {\bibfnamefont {W.}~\bibnamefont
  {Zhao}}, \bibinfo {author} {\bibfnamefont {G.}~\bibnamefont {Zhao}}, \ and\
  \bibinfo {author} {\bibfnamefont {S.}~\bibnamefont {Woo}},\ }\href {\doibase
  10.1088/1361-648X/ab5488} {\bibfield  {journal} {\bibinfo  {journal} {J.
  Phys.: Conden. Matter}\ }\textbf {\bibinfo {volume} {32}},\ \bibinfo {pages}
  {143001} (\bibinfo {year} {2020})}\BibitemShut {NoStop}%
\bibitem [{\citenamefont {Vigo-Cotrina}\ and\ \citenamefont
  {Guimar{\~a}es}(2020)}]{VigoCotrina_2020}%
  \BibitemOpen
  \bibfield  {author} {\bibinfo {author} {\bibfnamefont {H.}~\bibnamefont
  {Vigo-Cotrina}}\ and\ \bibinfo {author} {\bibfnamefont {A.~P.}\ \bibnamefont
  {Guimar{\~a}es}},\ }\href {\doibase 10.1016/j.jmmm.2020.166848} {\bibfield
  {journal} {\bibinfo  {journal} {J. Magn. Magn. Mater.}\ }\textbf {\bibinfo
  {volume} {507}},\ \bibinfo {pages} {166848} (\bibinfo {year}
  {2020})}\BibitemShut {NoStop}%
\bibitem [{\citenamefont {Jiang}\ \emph {et~al.}(2017)\citenamefont {Jiang},
  \citenamefont {Zhang}, \citenamefont {Yu}, \citenamefont {Zhang},
  \citenamefont {Wang}, \citenamefont {Benjamin~Jungfleisch}, \citenamefont
  {Pearson}, \citenamefont {Cheng}, \citenamefont {Heinonen}, \citenamefont
  {Wang}, \citenamefont {Zhou}, \citenamefont {Hoffmann},\ and\ \citenamefont
  {te~Velthuis}}]{WJiang_2017}%
  \BibitemOpen
  \bibfield  {author} {\bibinfo {author} {\bibfnamefont {W.}~\bibnamefont
  {Jiang}}, \bibinfo {author} {\bibfnamefont {X.}~\bibnamefont {Zhang}},
  \bibinfo {author} {\bibfnamefont {G.}~\bibnamefont {Yu}}, \bibinfo {author}
  {\bibfnamefont {W.}~\bibnamefont {Zhang}}, \bibinfo {author} {\bibfnamefont
  {X.}~\bibnamefont {Wang}}, \bibinfo {author} {\bibfnamefont {M.}~\bibnamefont
  {Benjamin~Jungfleisch}}, \bibinfo {author} {\bibfnamefont {J.~E.}\
  \bibnamefont {Pearson}}, \bibinfo {author} {\bibfnamefont {X.}~\bibnamefont
  {Cheng}}, \bibinfo {author} {\bibfnamefont {O.}~\bibnamefont {Heinonen}},
  \bibinfo {author} {\bibfnamefont {K.~L.}\ \bibnamefont {Wang}}, \bibinfo
  {author} {\bibfnamefont {Y.}~\bibnamefont {Zhou}}, \bibinfo {author}
  {\bibfnamefont {A.}~\bibnamefont {Hoffmann}}, \ and\ \bibinfo {author}
  {\bibfnamefont {S.~G.~E.}\ \bibnamefont {te~Velthuis}},\ }\href {\doibase
  10.1038/nphys3883} {\bibfield  {journal} {\bibinfo  {journal} {Nat. Phys.}\
  }\textbf {\bibinfo {volume} {13}},\ \bibinfo {pages} {162} (\bibinfo {year}
  {2017})}\BibitemShut {NoStop}%
\bibitem [{\citenamefont {Litzius}\ \emph {et~al.}(2017)\citenamefont
  {Litzius}, \citenamefont {Lemesh}, \citenamefont {Kr{\"u}ger}, \citenamefont
  {Bassirian}, \citenamefont {Caretta}, \citenamefont {Richter}, \citenamefont
  {B{\"u}ttner}, \citenamefont {Sato}, \citenamefont {Tretiakov}, \citenamefont
  {F{\"o}rster}, \citenamefont {Reeve}, \citenamefont {Weigand}, \citenamefont
  {Bykova}, \citenamefont {Stoll}, \citenamefont {Sch{\"u}tz}, \citenamefont
  {Beach},\ and\ \citenamefont {Kl{\"a}ui}}]{KLitzius_2017}%
  \BibitemOpen
  \bibfield  {author} {\bibinfo {author} {\bibfnamefont {K.}~\bibnamefont
  {Litzius}}, \bibinfo {author} {\bibfnamefont {I.}~\bibnamefont {Lemesh}},
  \bibinfo {author} {\bibfnamefont {B.}~\bibnamefont {Kr{\"u}ger}}, \bibinfo
  {author} {\bibfnamefont {P.}~\bibnamefont {Bassirian}}, \bibinfo {author}
  {\bibfnamefont {L.}~\bibnamefont {Caretta}}, \bibinfo {author} {\bibfnamefont
  {K.}~\bibnamefont {Richter}}, \bibinfo {author} {\bibfnamefont
  {F.}~\bibnamefont {B{\"u}ttner}}, \bibinfo {author} {\bibfnamefont
  {K.}~\bibnamefont {Sato}}, \bibinfo {author} {\bibfnamefont {O.~A.}\
  \bibnamefont {Tretiakov}}, \bibinfo {author} {\bibfnamefont {J.}~\bibnamefont
  {F{\"o}rster}}, \bibinfo {author} {\bibfnamefont {R.}~\bibnamefont {Reeve}},
  \bibinfo {author} {\bibfnamefont {M.}~\bibnamefont {Weigand}}, \bibinfo
  {author} {\bibfnamefont {I.}~\bibnamefont {Bykova}}, \bibinfo {author}
  {\bibfnamefont {H.}~\bibnamefont {Stoll}}, \bibinfo {author} {\bibfnamefont
  {G.}~\bibnamefont {Sch{\"u}tz}}, \bibinfo {author} {\bibfnamefont {G.~S.~D.}\
  \bibnamefont {Beach}}, \ and\ \bibinfo {author} {\bibfnamefont
  {M.}~\bibnamefont {Kl{\"a}ui}},\ }\href {\doibase 10.1038/nphys4000}
  {\bibfield  {journal} {\bibinfo  {journal} {Nat. Phys.}\ }\textbf {\bibinfo
  {volume} {13}},\ \bibinfo {pages} {170} (\bibinfo {year} {2017})}\BibitemShut
  {NoStop}%
\bibitem [{\citenamefont {Kasai}\ \emph {et~al.}(2019)\citenamefont {Kasai},
  \citenamefont {Sugimoto}, \citenamefont {Nakatani}, \citenamefont
  {Ishikawa},\ and\ \citenamefont {Takahashi}}]{Kasai_2019}%
  \BibitemOpen
  \bibfield  {author} {\bibinfo {author} {\bibfnamefont {S.}~\bibnamefont
  {Kasai}}, \bibinfo {author} {\bibfnamefont {S.}~\bibnamefont {Sugimoto}},
  \bibinfo {author} {\bibfnamefont {Y.}~\bibnamefont {Nakatani}}, \bibinfo
  {author} {\bibfnamefont {R.}~\bibnamefont {Ishikawa}}, \ and\ \bibinfo
  {author} {\bibfnamefont {Y.~K.}\ \bibnamefont {Takahashi}},\ }\href {\doibase
  10.7567/1882-0786/ab2baa} {\bibfield  {journal} {\bibinfo  {journal} {Appl.
  Phys. Express}\ }\textbf {\bibinfo {volume} {12}},\ \bibinfo {pages} {083001}
  (\bibinfo {year} {2019})}\BibitemShut {NoStop}%
\bibitem [{\citenamefont {Mathur}\ \emph {et~al.}(2021)\citenamefont {Mathur},
  \citenamefont {Yasin}, \citenamefont {Stolt}, \citenamefont {Nagai},
  \citenamefont {Kimoto}, \citenamefont {Du}, \citenamefont {Tian},
  \citenamefont {Tokura}, \citenamefont {Yu},\ and\ \citenamefont
  {Jin}}]{NMathur_2021}%
  \BibitemOpen
  \bibfield  {author} {\bibinfo {author} {\bibfnamefont {N.}~\bibnamefont
  {Mathur}}, \bibinfo {author} {\bibfnamefont {F.~S.}\ \bibnamefont {Yasin}},
  \bibinfo {author} {\bibfnamefont {M.~J.}\ \bibnamefont {Stolt}}, \bibinfo
  {author} {\bibfnamefont {T.}~\bibnamefont {Nagai}}, \bibinfo {author}
  {\bibfnamefont {K.}~\bibnamefont {Kimoto}}, \bibinfo {author} {\bibfnamefont
  {H.}~\bibnamefont {Du}}, \bibinfo {author} {\bibfnamefont {M.}~\bibnamefont
  {Tian}}, \bibinfo {author} {\bibfnamefont {Y.}~\bibnamefont {Tokura}},
  \bibinfo {author} {\bibfnamefont {X.}~\bibnamefont {Yu}}, \ and\ \bibinfo
  {author} {\bibfnamefont {S.}~\bibnamefont {Jin}},\ }\href {\doibase
  10.1002/adfm.202008521} {\bibfield  {journal} {\bibinfo  {journal} {Adv.
  Funct. Mater.}\ }\textbf {\bibinfo {volume} {31}},\ \bibinfo {pages}
  {2008521} (\bibinfo {year} {2021})}\BibitemShut {NoStop}%
\bibitem [{\citenamefont {Ishikawa}\ \emph {et~al.}(2022)\citenamefont
  {Ishikawa}, \citenamefont {Goto}, \citenamefont {Nomura},\ and\ \citenamefont
  {Suzuki}}]{RIshikawa_2022}%
  \BibitemOpen
  \bibfield  {author} {\bibinfo {author} {\bibfnamefont {R.}~\bibnamefont
  {Ishikawa}}, \bibinfo {author} {\bibfnamefont {M.}~\bibnamefont {Goto}},
  \bibinfo {author} {\bibfnamefont {H.}~\bibnamefont {Nomura}}, \ and\ \bibinfo
  {author} {\bibfnamefont {Y.}~\bibnamefont {Suzuki}},\ }\href {\doibase
  10.1063/5.0128385} {\bibfield  {journal} {\bibinfo  {journal} {Appl. Phys.
  Lett.}\ }\textbf {\bibinfo {volume} {121}},\ \bibinfo {pages} {252402}
  (\bibinfo {year} {2022})}\BibitemShut {NoStop}%
\bibitem [{\citenamefont {Chen}(2017)}]{Chen_2017}%
  \BibitemOpen
  \bibfield  {author} {\bibinfo {author} {\bibfnamefont {G.}~\bibnamefont
  {Chen}},\ }\href {\doibase 10.1038/nphys4030} {\bibfield  {journal} {\bibinfo
   {journal} {Nat. Phys.}\ }\textbf {\bibinfo {volume} {13}},\ \bibinfo {pages}
  {112} (\bibinfo {year} {2017})}\BibitemShut {NoStop}%
\bibitem [{\citenamefont {Barker}\ and\ \citenamefont
  {Tretiakov}(2016)}]{Barker_2016}%
  \BibitemOpen
  \bibfield  {author} {\bibinfo {author} {\bibfnamefont {J.}~\bibnamefont
  {Barker}}\ and\ \bibinfo {author} {\bibfnamefont {O.~A.}\ \bibnamefont
  {Tretiakov}},\ }\href {\doibase 10.1103/PhysRevLett.116.147203} {\bibfield
  {journal} {\bibinfo  {journal} {Phys. Rev. Lett.}\ }\textbf {\bibinfo
  {volume} {116}},\ \bibinfo {pages} {147203} (\bibinfo {year}
  {2016})}\BibitemShut {NoStop}%
\bibitem [{\citenamefont {Silva}\ \emph {et~al.}(2019)\citenamefont {Silva},
  \citenamefont {Silva}, \citenamefont {Pereira},\ and\ \citenamefont
  {Moura-Melo}}]{RLSilva_2019}%
  \BibitemOpen
  \bibfield  {author} {\bibinfo {author} {\bibfnamefont {R.~L.}\ \bibnamefont
  {Silva}}, \bibinfo {author} {\bibfnamefont {R.~C.}\ \bibnamefont {Silva}},
  \bibinfo {author} {\bibfnamefont {A.~R.}\ \bibnamefont {Pereira}}, \ and\
  \bibinfo {author} {\bibfnamefont {W.~A.}\ \bibnamefont {Moura-Melo}},\ }\href
  {\doibase 10.1088/1361-648X/ab0abd} {\bibfield  {journal} {\bibinfo
  {journal} {J. Phys.: Condens. Matter}\ }\textbf {\bibinfo {volume} {31}},\
  \bibinfo {pages} {225802} (\bibinfo {year} {2019})}\BibitemShut {NoStop}%
\bibitem [{\citenamefont {Silva}\ \emph {et~al.}(2020)\citenamefont {Silva},
  \citenamefont {Silva}, \citenamefont {Pereira},\ and\ \citenamefont
  {Moura-Melo}}]{RLSilva_2020}%
  \BibitemOpen
  \bibfield  {author} {\bibinfo {author} {\bibfnamefont {R.~L.}\ \bibnamefont
  {Silva}}, \bibinfo {author} {\bibfnamefont {R.~C.}\ \bibnamefont {Silva}},
  \bibinfo {author} {\bibfnamefont {A.~R.}\ \bibnamefont {Pereira}}, \ and\
  \bibinfo {author} {\bibfnamefont {W.~A.}\ \bibnamefont {Moura-Melo}},\ }\href
  {\doibase 10.1063/5.0024003} {\bibfield  {journal} {\bibinfo  {journal} {J.
  Appl. Phys.}\ }\textbf {\bibinfo {volume} {128}},\ \bibinfo {pages} {163902}
  (\bibinfo {year} {2020})}\BibitemShut {NoStop}%
\bibitem [{\citenamefont {Zhang}\ \emph {et~al.}(2016)\citenamefont {Zhang},
  \citenamefont {Zhou},\ and\ \citenamefont {Ezawa}}]{XZhang_SR_2016}%
  \BibitemOpen
  \bibfield  {author} {\bibinfo {author} {\bibfnamefont {X.}~\bibnamefont
  {Zhang}}, \bibinfo {author} {\bibfnamefont {Y.}~\bibnamefont {Zhou}}, \ and\
  \bibinfo {author} {\bibfnamefont {M.}~\bibnamefont {Ezawa}},\ }\href
  {\doibase 10.1038/srep24795} {\bibfield  {journal} {\bibinfo  {journal} {Sci.
  Rep.}\ }\textbf {\bibinfo {volume} {6}},\ \bibinfo {pages} {24795} (\bibinfo
  {year} {2016})}\BibitemShut {NoStop}%
\bibitem [{\citenamefont {Jin}\ \emph {et~al.}(2016)\citenamefont {Jin},
  \citenamefont {Song}, \citenamefont {Wang},\ and\ \citenamefont
  {Liu}}]{CJin_2016}%
  \BibitemOpen
  \bibfield  {author} {\bibinfo {author} {\bibfnamefont {C.}~\bibnamefont
  {Jin}}, \bibinfo {author} {\bibfnamefont {C.}~\bibnamefont {Song}}, \bibinfo
  {author} {\bibfnamefont {J.}~\bibnamefont {Wang}}, \ and\ \bibinfo {author}
  {\bibfnamefont {Q.}~\bibnamefont {Liu}},\ }\href {\doibase 10.1063/1.4967006}
  {\bibfield  {journal} {\bibinfo  {journal} {Appl. Phys. Lett.}\ }\textbf
  {\bibinfo {volume} {109}},\ \bibinfo {pages} {182404} (\bibinfo {year}
  {2016})}\BibitemShut {NoStop}%
\bibitem [{\citenamefont {Rai\ifmmode \check{c}\else
  \v{c}\fi{}evi\ifmmode~\acute{c}\else \'{c}\fi{}}\ \emph
  {et~al.}(2011)\citenamefont {Rai\ifmmode \check{c}\else
  \v{c}\fi{}evi\ifmmode~\acute{c}\else \'{c}\fi{}}, \citenamefont
  {Popovi\ifmmode~\acute{c}\else \'{c}\fi{}}, \citenamefont {Panagopoulos},
  \citenamefont {Benfatto}, \citenamefont {Silva~Neto}, \citenamefont {Choi},\
  and\ \citenamefont {Sasagawa}}]{IRaicevic_2011}%
  \BibitemOpen
  \bibfield  {author} {\bibinfo {author} {\bibfnamefont {I.}~\bibnamefont
  {Rai\ifmmode \check{c}\else \v{c}\fi{}evi\ifmmode~\acute{c}\else
  \'{c}\fi{}}}, \bibinfo {author} {\bibfnamefont {D.}~\bibnamefont
  {Popovi\ifmmode~\acute{c}\else \'{c}\fi{}}}, \bibinfo {author} {\bibfnamefont
  {C.}~\bibnamefont {Panagopoulos}}, \bibinfo {author} {\bibfnamefont
  {L.}~\bibnamefont {Benfatto}}, \bibinfo {author} {\bibfnamefont {M.~B.}\
  \bibnamefont {Silva~Neto}}, \bibinfo {author} {\bibfnamefont {E.~S.}\
  \bibnamefont {Choi}}, \ and\ \bibinfo {author} {\bibfnamefont
  {T.}~\bibnamefont {Sasagawa}},\ }\href {\doibase
  10.1103/PhysRevLett.106.227206} {\bibfield  {journal} {\bibinfo  {journal}
  {Phys. Rev. Lett.}\ }\textbf {\bibinfo {volume} {106}},\ \bibinfo {pages}
  {227206} (\bibinfo {year} {2011})}\BibitemShut {NoStop}%
\bibitem [{\citenamefont {Gao}\ \emph {et~al.}(2020)\citenamefont {Gao},
  \citenamefont {Rosales}, \citenamefont {G\'{o}mez~Albarrac\'{i}n},
  \citenamefont {Tsurkan}, \citenamefont {Kaur}, \citenamefont {Fennell},
  \citenamefont {Steffens}, \citenamefont {Boehm}, \citenamefont
  {\ifmmode~\check{c}\else \v{C}\fi{}erm\'{a}k}, \citenamefont {Schneidewind},
  \citenamefont {Ressouche}, \citenamefont {Cabra}, \citenamefont {R\"{u}egg},\
  and\ \citenamefont {Zaharko}}]{SGao_2020}%
  \BibitemOpen
  \bibfield  {author} {\bibinfo {author} {\bibfnamefont {S.}~\bibnamefont
  {Gao}}, \bibinfo {author} {\bibfnamefont {H.~D.}\ \bibnamefont {Rosales}},
  \bibinfo {author} {\bibfnamefont {F.~A.}\ \bibnamefont
  {G\'{o}mez~Albarrac\'{i}n}}, \bibinfo {author} {\bibfnamefont
  {V.}~\bibnamefont {Tsurkan}}, \bibinfo {author} {\bibfnamefont
  {G.}~\bibnamefont {Kaur}}, \bibinfo {author} {\bibfnamefont {T.}~\bibnamefont
  {Fennell}}, \bibinfo {author} {\bibfnamefont {P.}~\bibnamefont {Steffens}},
  \bibinfo {author} {\bibfnamefont {M.}~\bibnamefont {Boehm}}, \bibinfo
  {author} {\bibfnamefont {P.}~\bibnamefont {\ifmmode~\check{c}\else
  \v{C}\fi{}erm\'{a}k}}, \bibinfo {author} {\bibfnamefont {A.}~\bibnamefont
  {Schneidewind}}, \bibinfo {author} {\bibfnamefont {E.}~\bibnamefont
  {Ressouche}}, \bibinfo {author} {\bibfnamefont {D.~C.}\ \bibnamefont
  {Cabra}}, \bibinfo {author} {\bibfnamefont {C.}~\bibnamefont {R\"{u}egg}}, \
  and\ \bibinfo {author} {\bibfnamefont {O.}~\bibnamefont {Zaharko}},\ }\href
  {\doibase 10.1038/s41586-020-2716-8} {\bibfield  {journal} {\bibinfo
  {journal} {Nat.}\ }\textbf {\bibinfo {volume} {586}},\ \bibinfo {pages} {37}
  (\bibinfo {year} {2020})}\BibitemShut {NoStop}%
\bibitem [{\citenamefont {Dohi}\ \emph {et~al.}(2019)\citenamefont {Dohi},
  \citenamefont {DuttaGupta}, \citenamefont {Fukami},\ and\ \citenamefont
  {Ohno}}]{Dohi_2019}%
  \BibitemOpen
  \bibfield  {author} {\bibinfo {author} {\bibfnamefont {T.}~\bibnamefont
  {Dohi}}, \bibinfo {author} {\bibfnamefont {S.}~\bibnamefont {DuttaGupta}},
  \bibinfo {author} {\bibfnamefont {S.}~\bibnamefont {Fukami}}, \ and\ \bibinfo
  {author} {\bibfnamefont {H.}~\bibnamefont {Ohno}},\ }\href {\doibase
  10.1038/s41467-019-13182-6} {\bibfield  {journal} {\bibinfo  {journal} {Nat.
  Commun.}\ }\textbf {\bibinfo {volume} {10}},\ \bibinfo {pages} {5153}
  (\bibinfo {year} {2019})}\BibitemShut {NoStop}%
\bibitem [{\citenamefont {Juge}\ \emph {et~al.}(2022)\citenamefont {Juge},
  \citenamefont {Sisodia}, \citenamefont {Larra{\~n}aga}, \citenamefont
  {Zhang}, \citenamefont {Pham}, \citenamefont {Rana}, \citenamefont {Sarpi},
  \citenamefont {Mille}, \citenamefont {Stanescu}, \citenamefont {Belkhou},
  \citenamefont {Mawass}, \citenamefont {Novakovic-Marinkovic}, \citenamefont
  {Kronast}, \citenamefont {Weigand}, \citenamefont {Gr{\"a}fe}, \citenamefont
  {Wintz}, \citenamefont {Finizio}, \citenamefont {Raabe}, \citenamefont
  {Aballe}, \citenamefont {Foerster}, \citenamefont {Belmeguenai},
  \citenamefont {Buda-Prejbeanu}, \citenamefont {Pelloux-Prayer}, \citenamefont
  {Shaw}, \citenamefont {Nembach}, \citenamefont {Ranno}, \citenamefont
  {Gaudin},\ and\ \citenamefont {Boulle}}]{RJuge_2022}%
  \BibitemOpen
  \bibfield  {author} {\bibinfo {author} {\bibfnamefont {R.}~\bibnamefont
  {Juge}}, \bibinfo {author} {\bibfnamefont {N.}~\bibnamefont {Sisodia}},
  \bibinfo {author} {\bibfnamefont {J.~U.}\ \bibnamefont {Larra{\~n}aga}},
  \bibinfo {author} {\bibfnamefont {Q.}~\bibnamefont {Zhang}}, \bibinfo
  {author} {\bibfnamefont {V.~T.}\ \bibnamefont {Pham}}, \bibinfo {author}
  {\bibfnamefont {K.~G.}\ \bibnamefont {Rana}}, \bibinfo {author}
  {\bibfnamefont {B.}~\bibnamefont {Sarpi}}, \bibinfo {author} {\bibfnamefont
  {N.}~\bibnamefont {Mille}}, \bibinfo {author} {\bibfnamefont
  {S.}~\bibnamefont {Stanescu}}, \bibinfo {author} {\bibfnamefont
  {R.}~\bibnamefont {Belkhou}}, \bibinfo {author} {\bibfnamefont {M.-A.}\
  \bibnamefont {Mawass}}, \bibinfo {author} {\bibfnamefont {N.}~\bibnamefont
  {Novakovic-Marinkovic}}, \bibinfo {author} {\bibfnamefont {F.}~\bibnamefont
  {Kronast}}, \bibinfo {author} {\bibfnamefont {M.}~\bibnamefont {Weigand}},
  \bibinfo {author} {\bibfnamefont {J.}~\bibnamefont {Gr{\"a}fe}}, \bibinfo
  {author} {\bibfnamefont {S.}~\bibnamefont {Wintz}}, \bibinfo {author}
  {\bibfnamefont {S.}~\bibnamefont {Finizio}}, \bibinfo {author} {\bibfnamefont
  {J.}~\bibnamefont {Raabe}}, \bibinfo {author} {\bibfnamefont
  {L.}~\bibnamefont {Aballe}}, \bibinfo {author} {\bibfnamefont
  {M.}~\bibnamefont {Foerster}}, \bibinfo {author} {\bibfnamefont
  {M.}~\bibnamefont {Belmeguenai}}, \bibinfo {author} {\bibfnamefont {L.~D.}\
  \bibnamefont {Buda-Prejbeanu}}, \bibinfo {author} {\bibfnamefont
  {J.}~\bibnamefont {Pelloux-Prayer}}, \bibinfo {author} {\bibfnamefont
  {J.~M.}\ \bibnamefont {Shaw}}, \bibinfo {author} {\bibfnamefont {H.~T.}\
  \bibnamefont {Nembach}}, \bibinfo {author} {\bibfnamefont {L.}~\bibnamefont
  {Ranno}}, \bibinfo {author} {\bibfnamefont {G.}~\bibnamefont {Gaudin}}, \
  and\ \bibinfo {author} {\bibfnamefont {O.}~\bibnamefont {Boulle}},\ }\href
  {\doibase 10.1038/s41467-022-32525-4} {\bibfield  {journal} {\bibinfo
  {journal} {Nat. Commun.}\ }\textbf {\bibinfo {volume} {13}},\ \bibinfo
  {pages} {4807} (\bibinfo {year} {2022})}\BibitemShut {NoStop}%
\bibitem [{\citenamefont {Legrand}\ \emph {et~al.}(2020)\citenamefont
  {Legrand}, \citenamefont {Maccariello}, \citenamefont {Ajejas}, \citenamefont
  {Collin}, \citenamefont {Vecchiola}, \citenamefont {Bouzehouane},
  \citenamefont {Reyren}, \citenamefont {Cros},\ and\ \citenamefont
  {Fert}}]{WLegrand_2020}%
  \BibitemOpen
  \bibfield  {author} {\bibinfo {author} {\bibfnamefont {W.}~\bibnamefont
  {Legrand}}, \bibinfo {author} {\bibfnamefont {D.}~\bibnamefont
  {Maccariello}}, \bibinfo {author} {\bibfnamefont {F.}~\bibnamefont {Ajejas}},
  \bibinfo {author} {\bibfnamefont {S.}~\bibnamefont {Collin}}, \bibinfo
  {author} {\bibfnamefont {A.}~\bibnamefont {Vecchiola}}, \bibinfo {author}
  {\bibfnamefont {K.}~\bibnamefont {Bouzehouane}}, \bibinfo {author}
  {\bibfnamefont {N.}~\bibnamefont {Reyren}}, \bibinfo {author} {\bibfnamefont
  {V.}~\bibnamefont {Cros}}, \ and\ \bibinfo {author} {\bibfnamefont
  {A.}~\bibnamefont {Fert}},\ }\href {\doibase 10.1038/s41563-019-0468-3}
  {\bibfield  {journal} {\bibinfo  {journal} {Nat. Mater.}\ }\textbf {\bibinfo
  {volume} {19}},\ \bibinfo {pages} {34} (\bibinfo {year} {2020})}\BibitemShut
  {NoStop}%
\bibitem [{\citenamefont {Chen}\ \emph
  {et~al.}(2020{\natexlab{b}})\citenamefont {Chen}, \citenamefont {Gao},
  \citenamefont {Zhang}, \citenamefont {Zhang}, \citenamefont {Yin},
  \citenamefont {Chen}, \citenamefont {Zhou}, \citenamefont {Zhou},
  \citenamefont {Xia}, \citenamefont {Zhou}, \citenamefont {Wang},
  \citenamefont {Pan}, \citenamefont {Zhang},\ and\ \citenamefont
  {Song}}]{RChen_2020}%
  \BibitemOpen
  \bibfield  {author} {\bibinfo {author} {\bibfnamefont {R.}~\bibnamefont
  {Chen}}, \bibinfo {author} {\bibfnamefont {Y.}~\bibnamefont {Gao}}, \bibinfo
  {author} {\bibfnamefont {X.}~\bibnamefont {Zhang}}, \bibinfo {author}
  {\bibfnamefont {R.}~\bibnamefont {Zhang}}, \bibinfo {author} {\bibfnamefont
  {S.}~\bibnamefont {Yin}}, \bibinfo {author} {\bibfnamefont {X.}~\bibnamefont
  {Chen}}, \bibinfo {author} {\bibfnamefont {X.}~\bibnamefont {Zhou}}, \bibinfo
  {author} {\bibfnamefont {Y.}~\bibnamefont {Zhou}}, \bibinfo {author}
  {\bibfnamefont {J.}~\bibnamefont {Xia}}, \bibinfo {author} {\bibfnamefont
  {Y.}~\bibnamefont {Zhou}}, \bibinfo {author} {\bibfnamefont {S.}~\bibnamefont
  {Wang}}, \bibinfo {author} {\bibfnamefont {F.}~\bibnamefont {Pan}}, \bibinfo
  {author} {\bibfnamefont {Y.}~\bibnamefont {Zhang}}, \ and\ \bibinfo {author}
  {\bibfnamefont {C.}~\bibnamefont {Song}},\ }\href {\doibase
  10.1021/acs.nanolett.0c00116} {\bibfield  {journal} {\bibinfo  {journal}
  {Nano Lett.}\ }\textbf {\bibinfo {volume} {20}},\ \bibinfo {pages} {3299}
  (\bibinfo {year} {2020}{\natexlab{b}})}\BibitemShut {NoStop}%
\bibitem [{\citenamefont {He}\ \emph {et~al.}(2024)\citenamefont {He},
  \citenamefont {Jin}, \citenamefont {Zheng}, \citenamefont {Liu},
  \citenamefont {Li}, \citenamefont {Hu}, \citenamefont {Wang}, \citenamefont
  {Zhang}, \citenamefont {Peng}, \citenamefont {Wan}, \citenamefont {Zhu},
  \citenamefont {Han}, \citenamefont {Zhang},\ and\ \citenamefont
  {Yu}}]{BHe_2024}%
  \BibitemOpen
  \bibfield  {author} {\bibinfo {author} {\bibfnamefont {B.}~\bibnamefont
  {He}}, \bibinfo {author} {\bibfnamefont {H.}~\bibnamefont {Jin}}, \bibinfo
  {author} {\bibfnamefont {D.}~\bibnamefont {Zheng}}, \bibinfo {author}
  {\bibfnamefont {Y.}~\bibnamefont {Liu}}, \bibinfo {author} {\bibfnamefont
  {J.}~\bibnamefont {Li}}, \bibinfo {author} {\bibfnamefont {Y.}~\bibnamefont
  {Hu}}, \bibinfo {author} {\bibfnamefont {Y.}~\bibnamefont {Wang}}, \bibinfo
  {author} {\bibfnamefont {J.}~\bibnamefont {Zhang}}, \bibinfo {author}
  {\bibfnamefont {Y.}~\bibnamefont {Peng}}, \bibinfo {author} {\bibfnamefont
  {C.}~\bibnamefont {Wan}}, \bibinfo {author} {\bibfnamefont {T.}~\bibnamefont
  {Zhu}}, \bibinfo {author} {\bibfnamefont {X.}~\bibnamefont {Han}}, \bibinfo
  {author} {\bibfnamefont {S.}~\bibnamefont {Zhang}}, \ and\ \bibinfo {author}
  {\bibfnamefont {G.}~\bibnamefont {Yu}},\ }\href {\doibase
  10.1021/acs.nanolett.3c04221} {\bibfield  {journal} {\bibinfo  {journal}
  {Nano Letters}\ }\textbf {\bibinfo {volume} {24}},\ \bibinfo {pages} {2196}
  (\bibinfo {year} {2024})},\ \bibinfo {note} {pMID: 38329428}\BibitemShut
  {NoStop}%
\bibitem [{\citenamefont {Buhl}\ \emph {et~al.}(2017)\citenamefont {Buhl},
  \citenamefont {Freimuth}, \citenamefont {Blügel},\ and\ \citenamefont
  {Mokrousov}}]{PMBuhl}%
  \BibitemOpen
  \bibfield  {author} {\bibinfo {author} {\bibfnamefont {P.~M.}\ \bibnamefont
  {Buhl}}, \bibinfo {author} {\bibfnamefont {F.}~\bibnamefont {Freimuth}},
  \bibinfo {author} {\bibfnamefont {S.}~\bibnamefont {Blügel}}, \ and\
  \bibinfo {author} {\bibfnamefont {Y.}~\bibnamefont {Mokrousov}},\ }\href
  {\doibase 10.1002/pssr.201700007} {\bibfield  {journal} {\bibinfo  {journal}
  {Phys. Status Solidi RRL}\ }\textbf {\bibinfo {volume} {11}},\ \bibinfo
  {pages} {1700007} (\bibinfo {year} {2017})}\BibitemShut {NoStop}%
\bibitem [{\citenamefont {G{\"o}bel}\ \emph {et~al.}(2017)\citenamefont
  {G{\"o}bel}, \citenamefont {Mook}, \citenamefont {Henk},\ and\ \citenamefont
  {Mertig}}]{BGobel}%
  \BibitemOpen
  \bibfield  {author} {\bibinfo {author} {\bibfnamefont {B.}~\bibnamefont
  {G{\"o}bel}}, \bibinfo {author} {\bibfnamefont {A.}~\bibnamefont {Mook}},
  \bibinfo {author} {\bibfnamefont {J.}~\bibnamefont {Henk}}, \ and\ \bibinfo
  {author} {\bibfnamefont {I.}~\bibnamefont {Mertig}},\ }\href {\doibase
  10.1103/PhysRevB.96.060406} {\bibfield  {journal} {\bibinfo  {journal} {Phys.
  Rev. B}\ }\textbf {\bibinfo {volume} {96}},\ \bibinfo {pages} {060406}
  (\bibinfo {year} {2017})}\BibitemShut {NoStop}%
\bibitem [{\citenamefont {Akosa}\ \emph {et~al.}(2018)\citenamefont {Akosa},
  \citenamefont {Tretiakov}, \citenamefont {Tatara},\ and\ \citenamefont
  {Manchon}}]{CAAkosa}%
  \BibitemOpen
  \bibfield  {author} {\bibinfo {author} {\bibfnamefont {C.~A.}\ \bibnamefont
  {Akosa}}, \bibinfo {author} {\bibfnamefont {O.~A.}\ \bibnamefont
  {Tretiakov}}, \bibinfo {author} {\bibfnamefont {G.}~\bibnamefont {Tatara}}, \
  and\ \bibinfo {author} {\bibfnamefont {A.}~\bibnamefont {Manchon}},\ }\href
  {\doibase 10.1103/PhysRevLett.121.097204} {\bibfield  {journal} {\bibinfo
  {journal} {Phys. Rev. Lett.}\ }\textbf {\bibinfo {volume} {121}},\ \bibinfo
  {pages} {097204} (\bibinfo {year} {2018})}\BibitemShut {NoStop}%
\bibitem [{\citenamefont {\ifmmode~\check{c}\else \v{S}\fi{}mejkal}\ \emph
  {et~al.}(2018)\citenamefont {\ifmmode~\check{c}\else \v{S}\fi{}mejkal},
  \citenamefont {Mokrousov}, \citenamefont {Yan},\ and\ \citenamefont
  {MacDonald}}]{LSmejkal_2018}%
  \BibitemOpen
  \bibfield  {author} {\bibinfo {author} {\bibfnamefont {L.}~\bibnamefont
  {\ifmmode~\check{c}\else \v{S}\fi{}mejkal}}, \bibinfo {author} {\bibfnamefont
  {Y.}~\bibnamefont {Mokrousov}}, \bibinfo {author} {\bibfnamefont
  {B.}~\bibnamefont {Yan}}, \ and\ \bibinfo {author} {\bibfnamefont {A.~H.}\
  \bibnamefont {MacDonald}},\ }\href {\doibase 10.1038/s41567-018-0064-5}
  {\bibfield  {journal} {\bibinfo  {journal} {Nat. Phys.}\ }\textbf {\bibinfo
  {volume} {14}},\ \bibinfo {pages} {242} (\bibinfo {year} {2018})}\BibitemShut
  {NoStop}%
\bibitem [{\citenamefont {Zhao}\ \emph {et~al.}(2020)\citenamefont {Zhao},
  \citenamefont {Liang}, \citenamefont {Xia}, \citenamefont {Zhao},\ and\
  \citenamefont {Zhou}}]{LZhao_2020}%
  \BibitemOpen
  \bibfield  {author} {\bibinfo {author} {\bibfnamefont {L.}~\bibnamefont
  {Zhao}}, \bibinfo {author} {\bibfnamefont {X.}~\bibnamefont {Liang}},
  \bibinfo {author} {\bibfnamefont {J.}~\bibnamefont {Xia}}, \bibinfo {author}
  {\bibfnamefont {G.}~\bibnamefont {Zhao}}, \ and\ \bibinfo {author}
  {\bibfnamefont {Y.}~\bibnamefont {Zhou}},\ }\href {\doibase
  10.1039/C9NR10528J} {\bibfield  {journal} {\bibinfo  {journal} {Nanoscale}\
  }\textbf {\bibinfo {volume} {12}},\ \bibinfo {pages} {9507} (\bibinfo {year}
  {2020})}\BibitemShut {NoStop}%
\bibitem [{\citenamefont {Wang}\ \emph {et~al.}(2020)\citenamefont {Wang},
  \citenamefont {Xia}, \citenamefont {Zhang}, \citenamefont {Zheng},
  \citenamefont {Li}, \citenamefont {Chen}, \citenamefont {Zhou}, \citenamefont
  {Wu}, \citenamefont {Yin}, \citenamefont {Chantrell},\ and\ \citenamefont
  {Xu}}]{JWang_2020}%
  \BibitemOpen
  \bibfield  {author} {\bibinfo {author} {\bibfnamefont {J.}~\bibnamefont
  {Wang}}, \bibinfo {author} {\bibfnamefont {J.}~\bibnamefont {Xia}}, \bibinfo
  {author} {\bibfnamefont {X.}~\bibnamefont {Zhang}}, \bibinfo {author}
  {\bibfnamefont {X.}~\bibnamefont {Zheng}}, \bibinfo {author} {\bibfnamefont
  {G.}~\bibnamefont {Li}}, \bibinfo {author} {\bibfnamefont {L.}~\bibnamefont
  {Chen}}, \bibinfo {author} {\bibfnamefont {Y.}~\bibnamefont {Zhou}}, \bibinfo
  {author} {\bibfnamefont {J.}~\bibnamefont {Wu}}, \bibinfo {author}
  {\bibfnamefont {H.}~\bibnamefont {Yin}}, \bibinfo {author} {\bibfnamefont
  {R.}~\bibnamefont {Chantrell}}, \ and\ \bibinfo {author} {\bibfnamefont
  {Y.}~\bibnamefont {Xu}},\ }\href {\doibase 10.1063/5.0025124} {\bibfield
  {journal} {\bibinfo  {journal} {Appl. Phys. Lett.}\ }\textbf {\bibinfo
  {volume} {117}},\ \bibinfo {pages} {202401} (\bibinfo {year}
  {2020})}\BibitemShut {NoStop}%
\bibitem [{\citenamefont {Jung}\ \emph {et~al.}(2021)\citenamefont {Jung},
  \citenamefont {Han}, \citenamefont {Kim}, \citenamefont {Kim}, \citenamefont
  {Jeong}, \citenamefont {Lee}, \citenamefont {Kang}, \citenamefont {Im},\ and\
  \citenamefont {Lee}}]{DHJung_2021}%
  \BibitemOpen
  \bibfield  {author} {\bibinfo {author} {\bibfnamefont {D.-H.}\ \bibnamefont
  {Jung}}, \bibinfo {author} {\bibfnamefont {H.-S.}\ \bibnamefont {Han}},
  \bibinfo {author} {\bibfnamefont {N.}~\bibnamefont {Kim}}, \bibinfo {author}
  {\bibfnamefont {G.}~\bibnamefont {Kim}}, \bibinfo {author} {\bibfnamefont
  {S.}~\bibnamefont {Jeong}}, \bibinfo {author} {\bibfnamefont
  {S.}~\bibnamefont {Lee}}, \bibinfo {author} {\bibfnamefont {M.}~\bibnamefont
  {Kang}}, \bibinfo {author} {\bibfnamefont {M.-Y.}\ \bibnamefont {Im}}, \ and\
  \bibinfo {author} {\bibfnamefont {K.-S.}\ \bibnamefont {Lee}},\ }\href
  {\doibase 10.1103/PhysRevB.104.L060408} {\bibfield  {journal} {\bibinfo
  {journal} {Phys. Rev. B}\ }\textbf {\bibinfo {volume} {104}},\ \bibinfo
  {pages} {L060408} (\bibinfo {year} {2021})}\BibitemShut {NoStop}%
\bibitem [{\citenamefont {Song}\ \emph {et~al.}(2021)\citenamefont {Song},
  \citenamefont {Yang}, \citenamefont {Liu}, \citenamefont {Meng},
  \citenamefont {Cao},\ and\ \citenamefont {Yan}}]{LSong_2021}%
  \BibitemOpen
  \bibfield  {author} {\bibinfo {author} {\bibfnamefont {L.}~\bibnamefont
  {Song}}, \bibinfo {author} {\bibfnamefont {H.}~\bibnamefont {Yang}}, \bibinfo
  {author} {\bibfnamefont {B.}~\bibnamefont {Liu}}, \bibinfo {author}
  {\bibfnamefont {H.}~\bibnamefont {Meng}}, \bibinfo {author} {\bibfnamefont
  {Y.}~\bibnamefont {Cao}}, \ and\ \bibinfo {author} {\bibfnamefont
  {P.}~\bibnamefont {Yan}},\ }\href {\doibase 10.1016/j.jmmm.2021.167975}
  {\bibfield  {journal} {\bibinfo  {journal} {J. Magn. Magn. Mater.}\ }\textbf
  {\bibinfo {volume} {532}},\ \bibinfo {pages} {167975} (\bibinfo {year}
  {2021})}\BibitemShut {NoStop}%
\bibitem [{\citenamefont {Feng}\ \emph {et~al.}(2022)\citenamefont {Feng},
  \citenamefont {Zhang}, \citenamefont {Zhao},\ and\ \citenamefont
  {Xiang}}]{YFeng_2022}%
  \BibitemOpen
  \bibfield  {author} {\bibinfo {author} {\bibfnamefont {Y.}~\bibnamefont
  {Feng}}, \bibinfo {author} {\bibfnamefont {X.}~\bibnamefont {Zhang}},
  \bibinfo {author} {\bibfnamefont {G.}~\bibnamefont {Zhao}}, \ and\ \bibinfo
  {author} {\bibfnamefont {G.}~\bibnamefont {Xiang}},\ }\href {\doibase
  10.1109/TED.2021.3138837} {\bibfield  {journal} {\bibinfo  {journal} {IEEE
  Trans. Electron. Devices}\ }\textbf {\bibinfo {volume} {69}},\ \bibinfo
  {pages} {1293} (\bibinfo {year} {2022})}\BibitemShut {NoStop}%
\bibitem [{\citenamefont {Xu}\ \emph {et~al.}(2023{\natexlab{b}})\citenamefont
  {Xu}, \citenamefont {Chen}, \citenamefont {Chen}, \citenamefont {Hu},
  \citenamefont {Zhang}, \citenamefont {Jiang},\ and\ \citenamefont
  {Zhang}}]{MXu_2023}%
  \BibitemOpen
  \bibfield  {author} {\bibinfo {author} {\bibfnamefont {M.}~\bibnamefont
  {Xu}}, \bibinfo {author} {\bibfnamefont {W.}~\bibnamefont {Chen}}, \bibinfo
  {author} {\bibfnamefont {Y.}~\bibnamefont {Chen}}, \bibinfo {author}
  {\bibfnamefont {C.}~\bibnamefont {Hu}}, \bibinfo {author} {\bibfnamefont
  {Z.}~\bibnamefont {Zhang}}, \bibinfo {author} {\bibfnamefont
  {G.}~\bibnamefont {Jiang}}, \ and\ \bibinfo {author} {\bibfnamefont
  {J.}~\bibnamefont {Zhang}},\ }\href {\doibase 10.1088/1361-648X/ace6ea}
  {\bibfield  {journal} {\bibinfo  {journal} {J. Phys.: Conden. Matter}\
  }\textbf {\bibinfo {volume} {35}},\ \bibinfo {pages} {425801} (\bibinfo
  {year} {2023}{\natexlab{b}})}\BibitemShut {NoStop}%
\bibitem [{\citenamefont {Shu}\ \emph {et~al.}(2023)\citenamefont {Shu},
  \citenamefont {Li}, \citenamefont {Xia}, \citenamefont {Lai}, \citenamefont
  {Zhao}, \citenamefont {Zhou}, \citenamefont {Liu},\ and\ \citenamefont
  {Zhao}}]{YShu_2023}%
  \BibitemOpen
  \bibfield  {author} {\bibinfo {author} {\bibfnamefont {Y.}~\bibnamefont
  {Shu}}, \bibinfo {author} {\bibfnamefont {Q.}~\bibnamefont {Li}}, \bibinfo
  {author} {\bibfnamefont {J.}~\bibnamefont {Xia}}, \bibinfo {author}
  {\bibfnamefont {P.}~\bibnamefont {Lai}}, \bibinfo {author} {\bibfnamefont
  {Y.}~\bibnamefont {Zhao}}, \bibinfo {author} {\bibfnamefont {Y.}~\bibnamefont
  {Zhou}}, \bibinfo {author} {\bibfnamefont {X.}~\bibnamefont {Liu}}, \ and\
  \bibinfo {author} {\bibfnamefont {G.}~\bibnamefont {Zhao}},\ }\href {\doibase
  10.1016/j.jmmm.2023.170387} {\bibfield  {journal} {\bibinfo  {journal} {J.
  Magn. Magn. Mater.}\ }\textbf {\bibinfo {volume} {568}},\ \bibinfo {pages}
  {170387} (\bibinfo {year} {2023})}\BibitemShut {NoStop}%
\bibitem [{\citenamefont {Xu}\ \emph {et~al.}(2023{\natexlab{c}})\citenamefont
  {Xu}, \citenamefont {Zhang}, \citenamefont {Zhang}, \citenamefont {Jiang},
  \citenamefont {Chen}, \citenamefont {Chen},\ and\ \citenamefont
  {Hu}}]{MinXu_2023}%
  \BibitemOpen
  \bibfield  {author} {\bibinfo {author} {\bibfnamefont {M.}~\bibnamefont
  {Xu}}, \bibinfo {author} {\bibfnamefont {Z.}~\bibnamefont {Zhang}}, \bibinfo
  {author} {\bibfnamefont {J.}~\bibnamefont {Zhang}}, \bibinfo {author}
  {\bibfnamefont {G.}~\bibnamefont {Jiang}}, \bibinfo {author} {\bibfnamefont
  {Y.}~\bibnamefont {Chen}}, \bibinfo {author} {\bibfnamefont {W.}~\bibnamefont
  {Chen}}, \ and\ \bibinfo {author} {\bibfnamefont {C.}~\bibnamefont {Hu}},\
  }\href {\doibase 10.1063/5.0142460} {\bibfield  {journal} {\bibinfo
  {journal} {Appl. Phys. Lett.}\ }\textbf {\bibinfo {volume} {122}},\ \bibinfo
  {pages} {152404} (\bibinfo {year} {2023}{\natexlab{c}})}\BibitemShut
  {NoStop}%
\bibitem [{\citenamefont {Bindal}\ \emph {et~al.}(2023)\citenamefont {Bindal},
  \citenamefont {Raj},\ and\ \citenamefont {Kaushik}}]{Bindal_2023}%
  \BibitemOpen
  \bibfield  {author} {\bibinfo {author} {\bibfnamefont {N.}~\bibnamefont
  {Bindal}}, \bibinfo {author} {\bibfnamefont {R.~K.}\ \bibnamefont {Raj}}, \
  and\ \bibinfo {author} {\bibfnamefont {B.~K.}\ \bibnamefont {Kaushik}},\
  }\href {\doibase 10.1039/D2NA00748G} {\bibfield  {journal} {\bibinfo
  {journal} {Nanoscale Adv.}\ }\textbf {\bibinfo {volume} {5}},\ \bibinfo
  {pages} {450} (\bibinfo {year} {2023})}\BibitemShut {NoStop}%
\bibitem [{\citenamefont {Zhao}\ \emph {et~al.}(2024)\citenamefont {Zhao},
  \citenamefont {Shen}, \citenamefont {Cui}, \citenamefont {He}, \citenamefont
  {Wu}, \citenamefont {Li}, \citenamefont {Guang}, \citenamefont {Nie},
  \citenamefont {Zheng}, \citenamefont {Yin}, \citenamefont {Shen},
  \citenamefont {Wang}, \citenamefont {Liang}, \citenamefont {Zhou},
  \citenamefont {Han},\ and\ \citenamefont {Yu}}]{CZhao_2024}%
  \BibitemOpen
  \bibfield  {author} {\bibinfo {author} {\bibfnamefont {C.}~\bibnamefont
  {Zhao}}, \bibinfo {author} {\bibfnamefont {L.}~\bibnamefont {Shen}}, \bibinfo
  {author} {\bibfnamefont {B.}~\bibnamefont {Cui}}, \bibinfo {author}
  {\bibfnamefont {B.}~\bibnamefont {He}}, \bibinfo {author} {\bibfnamefont
  {C.}~\bibnamefont {Wu}}, \bibinfo {author} {\bibfnamefont {J.}~\bibnamefont
  {Li}}, \bibinfo {author} {\bibfnamefont {Y.}~\bibnamefont {Guang}}, \bibinfo
  {author} {\bibfnamefont {Z.}~\bibnamefont {Nie}}, \bibinfo {author}
  {\bibfnamefont {X.}~\bibnamefont {Zheng}}, \bibinfo {author} {\bibfnamefont
  {G.}~\bibnamefont {Yin}}, \bibinfo {author} {\bibfnamefont {K.}~\bibnamefont
  {Shen}}, \bibinfo {author} {\bibfnamefont {H.}~\bibnamefont {Wang}}, \bibinfo
  {author} {\bibfnamefont {S.}~\bibnamefont {Liang}}, \bibinfo {author}
  {\bibfnamefont {Y.}~\bibnamefont {Zhou}}, \bibinfo {author} {\bibfnamefont
  {X.}~\bibnamefont {Han}}, \ and\ \bibinfo {author} {\bibfnamefont
  {G.}~\bibnamefont {Yu}},\ }\href {\doibase 10.1002/adfm.202405296} {\bibfield
   {journal} {\bibinfo  {journal} {Adv. Funct. Mater.}\ }\textbf {\bibinfo
  {volume} {n/a}},\ \bibinfo {pages} {2405296} (\bibinfo {year}
  {2024})}\BibitemShut {NoStop}%
\bibitem [{\citenamefont {Chappert}\ \emph {et~al.}(1998)\citenamefont
  {Chappert}, \citenamefont {Bernas}, \citenamefont {Ferré}, \citenamefont
  {Kottler}, \citenamefont {Jamet}, \citenamefont {Chen}, \citenamefont
  {Cambril}, \citenamefont {Devolder}, \citenamefont {Rousseaux}, \citenamefont
  {Mathet},\ and\ \citenamefont {Launois}}]{Chappert_98}%
  \BibitemOpen
  \bibfield  {author} {\bibinfo {author} {\bibfnamefont {C.}~\bibnamefont
  {Chappert}}, \bibinfo {author} {\bibfnamefont {H.}~\bibnamefont {Bernas}},
  \bibinfo {author} {\bibfnamefont {J.}~\bibnamefont {Ferré}}, \bibinfo
  {author} {\bibfnamefont {V.}~\bibnamefont {Kottler}}, \bibinfo {author}
  {\bibfnamefont {J.-P.}\ \bibnamefont {Jamet}}, \bibinfo {author}
  {\bibfnamefont {Y.}~\bibnamefont {Chen}}, \bibinfo {author} {\bibfnamefont
  {E.}~\bibnamefont {Cambril}}, \bibinfo {author} {\bibfnamefont
  {T.}~\bibnamefont {Devolder}}, \bibinfo {author} {\bibfnamefont
  {F.}~\bibnamefont {Rousseaux}}, \bibinfo {author} {\bibfnamefont
  {V.}~\bibnamefont {Mathet}}, \ and\ \bibinfo {author} {\bibfnamefont
  {H.}~\bibnamefont {Launois}},\ }\href {\doibase
  10.1126/science.280.5371.1919} {\bibfield  {journal} {\bibinfo  {journal}
  {Science}\ }\textbf {\bibinfo {volume} {280}},\ \bibinfo {pages} {1919}
  (\bibinfo {year} {1998})}\BibitemShut {NoStop}%
\bibitem [{\citenamefont {Fassbender}\ and\ \citenamefont
  {McCord}(2008)}]{Fassbender}%
  \BibitemOpen
  \bibfield  {author} {\bibinfo {author} {\bibfnamefont {J.}~\bibnamefont
  {Fassbender}}\ and\ \bibinfo {author} {\bibfnamefont {J.}~\bibnamefont
  {McCord}},\ }\href {\doibase 10.1016/j.jmmm.2007.07.032} {\bibfield
  {journal} {\bibinfo  {journal} {J. Magn. Magn. Mater.}\ }\textbf {\bibinfo
  {volume} {320}},\ \bibinfo {pages} {579} (\bibinfo {year}
  {2008})}\BibitemShut {NoStop}%
\bibitem [{\citenamefont {Bera}\ \emph {et~al.}(2023)\citenamefont {Bera},
  \citenamefont {Dev},\ and\ \citenamefont {Kumar}}]{Bera_2023}%
  \BibitemOpen
  \bibfield  {author} {\bibinfo {author} {\bibfnamefont {A.~K.}\ \bibnamefont
  {Bera}}, \bibinfo {author} {\bibfnamefont {A.~S.}\ \bibnamefont {Dev}}, \
  and\ \bibinfo {author} {\bibfnamefont {D.}~\bibnamefont {Kumar}},\ }\href
  {\doibase 10.1063/5.0125851} {\bibfield  {journal} {\bibinfo  {journal}
  {Appl. Phys. Lett.}\ }\textbf {\bibinfo {volume} {122}},\ \bibinfo {pages}
  {022405} (\bibinfo {year} {2023})}\BibitemShut {NoStop}%
\bibitem [{\citenamefont {Parkin}\ \emph {et~al.}(1990)\citenamefont {Parkin},
  \citenamefont {More},\ and\ \citenamefont {Roche}}]{SSPParkin_1990}%
  \BibitemOpen
  \bibfield  {author} {\bibinfo {author} {\bibfnamefont {S.~S.~P.}\
  \bibnamefont {Parkin}}, \bibinfo {author} {\bibfnamefont {N.}~\bibnamefont
  {More}}, \ and\ \bibinfo {author} {\bibfnamefont {K.~P.}\ \bibnamefont
  {Roche}},\ }\href {\doibase 10.1103/PhysRevLett.64.2304} {\bibfield
  {journal} {\bibinfo  {journal} {Phys. Rev. Lett.}\ }\textbf {\bibinfo
  {volume} {64}},\ \bibinfo {pages} {2304} (\bibinfo {year}
  {1990})}\BibitemShut {NoStop}%
\bibitem [{\citenamefont {Chen}\ \emph {et~al.}(2013)\citenamefont {Chen},
  \citenamefont {Ma}, \citenamefont {N’Diaye}, \citenamefont {Kwon},
  \citenamefont {Won}, \citenamefont {Wu},\ and\ \citenamefont
  {Schmid}}]{GChen_2013}%
  \BibitemOpen
  \bibfield  {author} {\bibinfo {author} {\bibfnamefont {G.}~\bibnamefont
  {Chen}}, \bibinfo {author} {\bibfnamefont {T.}~\bibnamefont {Ma}}, \bibinfo
  {author} {\bibfnamefont {A.~T.}\ \bibnamefont {N’Diaye}}, \bibinfo {author}
  {\bibfnamefont {H.}~\bibnamefont {Kwon}}, \bibinfo {author} {\bibfnamefont
  {C.}~\bibnamefont {Won}}, \bibinfo {author} {\bibfnamefont {Y.}~\bibnamefont
  {Wu}}, \ and\ \bibinfo {author} {\bibfnamefont {A.~K.}\ \bibnamefont
  {Schmid}},\ }\href {\doibase 10.1038/ncomms3671} {\bibfield  {journal}
  {\bibinfo  {journal} {Nat. Commun.}\ }\textbf {\bibinfo {volume} {4}},\
  \bibinfo {pages} {2671} (\bibinfo {year} {2013})}\BibitemShut {NoStop}%
\bibitem [{\citenamefont {Yang}\ \emph
  {et~al.}(2015{\natexlab{a}})\citenamefont {Yang}, \citenamefont {Ryu},\ and\
  \citenamefont {Parkin}}]{SHYang_2015}%
  \BibitemOpen
  \bibfield  {author} {\bibinfo {author} {\bibfnamefont {S.-H.}\ \bibnamefont
  {Yang}}, \bibinfo {author} {\bibfnamefont {K.-S.}\ \bibnamefont {Ryu}}, \
  and\ \bibinfo {author} {\bibfnamefont {S.}~\bibnamefont {Parkin}},\ }\href
  {\doibase 10.1038/nnano.2014.324} {\bibfield  {journal} {\bibinfo  {journal}
  {Nat. Nanotech.}\ }\textbf {\bibinfo {volume} {10}},\ \bibinfo {pages} {221}
  (\bibinfo {year} {2015}{\natexlab{a}})}\BibitemShut {NoStop}%
\bibitem [{\citenamefont {Dup{\'e}}\ \emph {et~al.}(2016)\citenamefont
  {Dup{\'e}}, \citenamefont {Bihlmayer}, \citenamefont {Bl{\"u}gel},\ and\
  \citenamefont {Heinze}}]{BDupe_2016}%
  \BibitemOpen
  \bibfield  {author} {\bibinfo {author} {\bibfnamefont {B.}~\bibnamefont
  {Dup{\'e}}}, \bibinfo {author} {\bibfnamefont {M.}~\bibnamefont {Bihlmayer},
  \bibfnamefont {G.and~B{\"o}ttcher}}, \bibinfo {author} {\bibfnamefont
  {S.}~\bibnamefont {Bl{\"u}gel}}, \ and\ \bibinfo {author} {\bibfnamefont
  {S.}~\bibnamefont {Heinze}},\ }\href {\doibase 10.1038/ncomms11779}
  {\bibfield  {journal} {\bibinfo  {journal} {Nat. Commun.}\ }\textbf {\bibinfo
  {volume} {7}},\ \bibinfo {pages} {11779} (\bibinfo {year}
  {2016})}\BibitemShut {NoStop}%
\bibitem [{\citenamefont {Matsuno}\ \emph {et~al.}(2016)\citenamefont
  {Matsuno}, \citenamefont {Ogawa}, \citenamefont {Yasuda}, \citenamefont
  {Kagawa}, \citenamefont {Koshibae}, \citenamefont {Nagaosa}, \citenamefont
  {Tokura},\ and\ \citenamefont {Kawasaki}}]{JMatsuno_2016}%
  \BibitemOpen
  \bibfield  {author} {\bibinfo {author} {\bibfnamefont {J.}~\bibnamefont
  {Matsuno}}, \bibinfo {author} {\bibfnamefont {N.}~\bibnamefont {Ogawa}},
  \bibinfo {author} {\bibfnamefont {K.}~\bibnamefont {Yasuda}}, \bibinfo
  {author} {\bibfnamefont {F.}~\bibnamefont {Kagawa}}, \bibinfo {author}
  {\bibfnamefont {W.}~\bibnamefont {Koshibae}}, \bibinfo {author}
  {\bibfnamefont {N.}~\bibnamefont {Nagaosa}}, \bibinfo {author} {\bibfnamefont
  {Y.}~\bibnamefont {Tokura}}, \ and\ \bibinfo {author} {\bibfnamefont
  {M.}~\bibnamefont {Kawasaki}},\ }\href {\doibase 10.1126/sciadv.1600304}
  {\bibfield  {journal} {\bibinfo  {journal} {Sci. Adv.}\ }\textbf {\bibinfo
  {volume} {2}},\ \bibinfo {pages} {e1600304} (\bibinfo {year}
  {2016})}\BibitemShut {NoStop}%
\bibitem [{\citenamefont {Nunn}\ \emph {et~al.}(2020)\citenamefont {Nunn},
  \citenamefont {Abert}, \citenamefont {Suess},\ and\ \citenamefont
  {Girt}}]{Zachary_2020}%
  \BibitemOpen
  \bibfield  {author} {\bibinfo {author} {\bibfnamefont {Z.~R.}\ \bibnamefont
  {Nunn}}, \bibinfo {author} {\bibfnamefont {C.}~\bibnamefont {Abert}},
  \bibinfo {author} {\bibfnamefont {D.}~\bibnamefont {Suess}}, \ and\ \bibinfo
  {author} {\bibfnamefont {E.}~\bibnamefont {Girt}},\ }\href {\doibase
  10.1126/sciadv.abd8861} {\bibfield  {journal} {\bibinfo  {journal} {Sci.
  Adv.}\ }\textbf {\bibinfo {volume} {6}},\ \bibinfo {pages} {eabd8861}
  (\bibinfo {year} {2020})}\BibitemShut {NoStop}%
\bibitem [{\citenamefont {Saiki}(1972)}]{KSainki_1972}%
  \BibitemOpen
  \bibfield  {author} {\bibinfo {author} {\bibfnamefont {K.}~\bibnamefont
  {Saiki}},\ }\href {\doibase 10.1143/JPSJ.33.1284} {\bibfield  {journal}
  {\bibinfo  {journal} {J. Phys. Soc. Japan}\ }\textbf {\bibinfo {volume}
  {33}},\ \bibinfo {pages} {1284} (\bibinfo {year} {1972})}\BibitemShut
  {NoStop}%
\bibitem [{\citenamefont {Rohart}\ and\ \citenamefont
  {Thiaville}(2013)}]{SRohart_2013}%
  \BibitemOpen
  \bibfield  {author} {\bibinfo {author} {\bibfnamefont {S.}~\bibnamefont
  {Rohart}}\ and\ \bibinfo {author} {\bibfnamefont {A.}~\bibnamefont
  {Thiaville}},\ }\href {\doibase 10.1103/PhysRevB.88.184422} {\bibfield
  {journal} {\bibinfo  {journal} {Phys. Rev. B}\ }\textbf {\bibinfo {volume}
  {88}},\ \bibinfo {pages} {184422} (\bibinfo {year} {2013})}\BibitemShut
  {NoStop}%
\bibitem [{\citenamefont {Yang}\ \emph
  {et~al.}(2015{\natexlab{b}})\citenamefont {Yang}, \citenamefont {Thiaville},
  \citenamefont {Rohart}, \citenamefont {Fert},\ and\ \citenamefont
  {Chshiev}}]{HYang_2015}%
  \BibitemOpen
  \bibfield  {author} {\bibinfo {author} {\bibfnamefont {H.}~\bibnamefont
  {Yang}}, \bibinfo {author} {\bibfnamefont {A.}~\bibnamefont {Thiaville}},
  \bibinfo {author} {\bibfnamefont {S.}~\bibnamefont {Rohart}}, \bibinfo
  {author} {\bibfnamefont {A.}~\bibnamefont {Fert}}, \ and\ \bibinfo {author}
  {\bibfnamefont {M.}~\bibnamefont {Chshiev}},\ }\href {\doibase
  10.1103/PhysRevLett.115.267210} {\bibfield  {journal} {\bibinfo  {journal}
  {Phys. Rev. Lett.}\ }\textbf {\bibinfo {volume} {115}},\ \bibinfo {pages}
  {267210} (\bibinfo {year} {2015}{\natexlab{b}})}\BibitemShut {NoStop}%
\bibitem [{\citenamefont {Landau}\ and\ \citenamefont
  {Lifshitz}(1935)}]{Landau_1935}%
  \BibitemOpen
  \bibfield  {author} {\bibinfo {author} {\bibfnamefont {L.~D.}\ \bibnamefont
  {Landau}}\ and\ \bibinfo {author} {\bibfnamefont {E.}~\bibnamefont
  {Lifshitz}},\ }\href {http://cds.cern.ch/record/437299} {\bibfield  {journal}
  {\bibinfo  {journal} {Phys. Z. Sowjet.}\ }\textbf {\bibinfo {volume} {8}},\
  \bibinfo {pages} {153} (\bibinfo {year} {1935})}\BibitemShut {NoStop}%
\bibitem [{\citenamefont {Gilbert}(2004)}]{TGilbert}%
  \BibitemOpen
  \bibfield  {author} {\bibinfo {author} {\bibfnamefont {T.}~\bibnamefont
  {Gilbert}},\ }\href {\doibase 10.1109/TMAG.2004.836740} {\bibfield  {journal}
  {\bibinfo  {journal} {IEEE Trans. Magn.}\ }\textbf {\bibinfo {volume} {40}},\
  \bibinfo {pages} {3443} (\bibinfo {year} {2004})}\BibitemShut {NoStop}%
\bibitem [{\citenamefont {Zhang}\ and\ \citenamefont {Li}(2004)}]{Zhang_Li}%
  \BibitemOpen
  \bibfield  {author} {\bibinfo {author} {\bibfnamefont {S.}~\bibnamefont
  {Zhang}}\ and\ \bibinfo {author} {\bibfnamefont {Z.}~\bibnamefont {Li}},\
  }\href {\doibase 10.1103/PhysRevLett.93.127204} {\bibfield  {journal}
  {\bibinfo  {journal} {Phys. Rev. Lett.}\ }\textbf {\bibinfo {volume} {93}},\
  \bibinfo {pages} {127204} (\bibinfo {year} {2004})}\BibitemShut {NoStop}%
\bibitem [{\citenamefont {Moutafis}\ \emph {et~al.}(2009)\citenamefont
  {Moutafis}, \citenamefont {Komineas},\ and\ \citenamefont
  {Bland}}]{CMoutafis_2009}%
  \BibitemOpen
  \bibfield  {author} {\bibinfo {author} {\bibfnamefont {C.}~\bibnamefont
  {Moutafis}}, \bibinfo {author} {\bibfnamefont {S.}~\bibnamefont {Komineas}},
  \ and\ \bibinfo {author} {\bibfnamefont {J.~A.~C.}\ \bibnamefont {Bland}},\
  }\href {\doibase 10.1103/PhysRevB.79.224429} {\bibfield  {journal} {\bibinfo
  {journal} {Phys. Rev. B}\ }\textbf {\bibinfo {volume} {79}},\ \bibinfo
  {pages} {224429} (\bibinfo {year} {2009})}\BibitemShut {NoStop}%
\bibitem [{\citenamefont {Shen}\ \emph {et~al.}(2020)\citenamefont {Shen},
  \citenamefont {Xia}, \citenamefont {Zhang}, \citenamefont {Ezawa},
  \citenamefont {Tretiakov}, \citenamefont {Liu}, \citenamefont {Zhao},\ and\
  \citenamefont {Zhou}}]{LShen_2020}%
  \BibitemOpen
  \bibfield  {author} {\bibinfo {author} {\bibfnamefont {L.}~\bibnamefont
  {Shen}}, \bibinfo {author} {\bibfnamefont {J.}~\bibnamefont {Xia}}, \bibinfo
  {author} {\bibfnamefont {X.}~\bibnamefont {Zhang}}, \bibinfo {author}
  {\bibfnamefont {M.}~\bibnamefont {Ezawa}}, \bibinfo {author} {\bibfnamefont
  {O.~A.}\ \bibnamefont {Tretiakov}}, \bibinfo {author} {\bibfnamefont
  {X.}~\bibnamefont {Liu}}, \bibinfo {author} {\bibfnamefont {G.}~\bibnamefont
  {Zhao}}, \ and\ \bibinfo {author} {\bibfnamefont {Y.}~\bibnamefont {Zhou}},\
  }\href {\doibase 10.1103/PhysRevLett.124.037202} {\bibfield  {journal}
  {\bibinfo  {journal} {Phys. Rev. Lett.}\ }\textbf {\bibinfo {volume} {124}},\
  \bibinfo {pages} {037202} (\bibinfo {year} {2020})}\BibitemShut {NoStop}%
\bibitem [{\citenamefont {Stier}\ \emph {et~al.}(2021)\citenamefont {Stier},
  \citenamefont {Strobel}, \citenamefont {Krause}, \citenamefont {H\"ausler},\
  and\ \citenamefont {Thorwart}}]{MStier_2021}%
  \BibitemOpen
  \bibfield  {author} {\bibinfo {author} {\bibfnamefont {M.}~\bibnamefont
  {Stier}}, \bibinfo {author} {\bibfnamefont {R.}~\bibnamefont {Strobel}},
  \bibinfo {author} {\bibfnamefont {S.}~\bibnamefont {Krause}}, \bibinfo
  {author} {\bibfnamefont {W.}~\bibnamefont {H\"ausler}}, \ and\ \bibinfo
  {author} {\bibfnamefont {M.}~\bibnamefont {Thorwart}},\ }\href {\doibase
  10.1103/PhysRevB.103.054420} {\bibfield  {journal} {\bibinfo  {journal}
  {Phys. Rev. B}\ }\textbf {\bibinfo {volume} {103}},\ \bibinfo {pages}
  {054420} (\bibinfo {year} {2021})}\BibitemShut {NoStop}%
\bibitem [{\citenamefont {Toscano}\ \emph {et~al.}(2020)\citenamefont
  {Toscano}, \citenamefont {Santece}, \citenamefont {Guedes}, \citenamefont
  {Assis}, \citenamefont {Miranda}, \citenamefont {de~Araujo}, \citenamefont
  {Sato}, \citenamefont {Coura},\ and\ \citenamefont {Leonel}}]{DToscano_2020}%
  \BibitemOpen
  \bibfield  {author} {\bibinfo {author} {\bibfnamefont {D.}~\bibnamefont
  {Toscano}}, \bibinfo {author} {\bibfnamefont {I.~A.}\ \bibnamefont
  {Santece}}, \bibinfo {author} {\bibfnamefont {R.~C.~O.}\ \bibnamefont
  {Guedes}}, \bibinfo {author} {\bibfnamefont {H.~S.}\ \bibnamefont {Assis}},
  \bibinfo {author} {\bibfnamefont {A.~L.~S.}\ \bibnamefont {Miranda}},
  \bibinfo {author} {\bibfnamefont {C.~I.~L.}\ \bibnamefont {de~Araujo}},
  \bibinfo {author} {\bibfnamefont {F.}~\bibnamefont {Sato}}, \bibinfo {author}
  {\bibfnamefont {P.~Z.}\ \bibnamefont {Coura}}, \ and\ \bibinfo {author}
  {\bibfnamefont {S.~A.}\ \bibnamefont {Leonel}},\ }\href {\doibase
  10.1063/5.0006219} {\bibfield  {journal} {\bibinfo  {journal} {J. Appl.
  Phys.}\ }\textbf {\bibinfo {volume} {127}},\ \bibinfo {pages} {193902}
  (\bibinfo {year} {2020})}\BibitemShut {NoStop}%
\bibitem [{\citenamefont {Silva}\ \emph {et~al.}(2022)\citenamefont {Silva},
  \citenamefont {Silva}, \citenamefont {Carvalho-Santos}, \citenamefont
  {Moura-Melo},\ and\ \citenamefont {Pereira}}]{RCSilva_2022}%
  \BibitemOpen
  \bibfield  {author} {\bibinfo {author} {\bibfnamefont {R.~C.}\ \bibnamefont
  {Silva}}, \bibinfo {author} {\bibfnamefont {R.~L.}\ \bibnamefont {Silva}},
  \bibinfo {author} {\bibfnamefont {V.~L.}\ \bibnamefont {Carvalho-Santos}},
  \bibinfo {author} {\bibfnamefont {W.~A.}\ \bibnamefont {Moura-Melo}}, \ and\
  \bibinfo {author} {\bibfnamefont {A.~R.}\ \bibnamefont {Pereira}},\ }\href
  {\doibase 10.1016/j.jmmm.2021.168997} {\bibfield  {journal} {\bibinfo
  {journal} {J. Magn. Magn. Mater.}\ }\textbf {\bibinfo {volume} {549}},\
  \bibinfo {pages} {168997} (\bibinfo {year} {2022})}\BibitemShut {NoStop}%
\bibitem [{\citenamefont {Silva}\ \emph {et~al.}(2024)\citenamefont {Silva},
  \citenamefont {Silva}, \citenamefont {Moreira}, \citenamefont {Moura-Melo},\
  and\ \citenamefont {Pereira}}]{RCSilva_2024}%
  \BibitemOpen
  \bibfield  {author} {\bibinfo {author} {\bibfnamefont {R.~C.}\ \bibnamefont
  {Silva}}, \bibinfo {author} {\bibfnamefont {R.~L.}\ \bibnamefont {Silva}},
  \bibinfo {author} {\bibfnamefont {J.~C.}\ \bibnamefont {Moreira}}, \bibinfo
  {author} {\bibfnamefont {W.~A.}\ \bibnamefont {Moura-Melo}}, \ and\ \bibinfo
  {author} {\bibfnamefont {A.~R.}\ \bibnamefont {Pereira}},\ }\href {\doibase
  10.1063/5.0206403} {\bibfield  {journal} {\bibinfo  {journal} {J. Appl.
  Phys.}\ }\textbf {\bibinfo {volume} {135}},\ \bibinfo {pages} {183902}
  (\bibinfo {year} {2024})}\BibitemShut {NoStop}%
\bibitem [{\citenamefont {Zhang}\ \emph
  {et~al.}(2015{\natexlab{b}})\citenamefont {Zhang}, \citenamefont {Zhao},
  \citenamefont {Fangohr}, \citenamefont {Liu}, \citenamefont {Xia},
  \citenamefont {Xia},\ and\ \citenamefont {Morvan}}]{XZhang2_2015}%
  \BibitemOpen
  \bibfield  {author} {\bibinfo {author} {\bibfnamefont {X.}~\bibnamefont
  {Zhang}}, \bibinfo {author} {\bibfnamefont {G.~P.}\ \bibnamefont {Zhao}},
  \bibinfo {author} {\bibfnamefont {H.}~\bibnamefont {Fangohr}}, \bibinfo
  {author} {\bibfnamefont {J.~P.}\ \bibnamefont {Liu}}, \bibinfo {author}
  {\bibfnamefont {W.~X.}\ \bibnamefont {Xia}}, \bibinfo {author} {\bibfnamefont
  {J.}~\bibnamefont {Xia}}, \ and\ \bibinfo {author} {\bibfnamefont {F.~J.}\
  \bibnamefont {Morvan}},\ }\href {\doibase 10.1038/srep07643} {\bibfield
  {journal} {\bibinfo  {journal} {Sci. Rep.}\ }\textbf {\bibinfo {volume}
  {5}},\ \bibinfo {pages} {7643} (\bibinfo {year}
  {2015}{\natexlab{b}})}\BibitemShut {NoStop}%
\end{thebibliography}%

\end{document}